\documentclass[a4paper,fleqn,usenatbib]{mnras}

\usepackage{graphicx}
\usepackage{amsmath}
\usepackage{amssymb}
\usepackage[normalem]{ulem}

\newcommand{\fable}{{\sc fable}}
\newcommand{\overbar}[1]{\mkern 1.5mu\overline{\mkern-1.5mu#1\mkern-1.5mu}\mkern 1.5mu}

\title[The FABLE simulations]{The FABLE simulations: A feedback model for galaxies, groups and clusters}

\author[Henden et al.]{Nicholas A. Henden,$^{1}$\thanks{E-mail: n.henden@ast.cam.ac.uk}
Ewald Puchwein,$^{1,2}$
Sijing Shen$^{3}$
and Debora Sijacki$^{1,2}$
\\
$^{1}$Institute of Astronomy, University of Cambridge, Madingley Road, Cambridge, CB3 0HA, UK\\
$^{2}$Kavli Institute for Cosmology, University of Cambridge, Madingley Road, Cambridge CB3 0HA, UK\\
$^{3}$Institute of Theoretical Astrophysics, University of Oslo, P.O. Box 1029 Blindern, NO-0315 Oslo, Norway
}

\date{Accepted XXX. Received YYY; in original form ZZZ}

\pubyear{2018}

\begin{document}
\label{firstpage}
\pagerange{\pageref{firstpage}--\pageref{lastpage}}
\maketitle

\begin{abstract}
We present the \textit{Feedback Acting on Baryons in Large-scale Environments} (\fable) suite of cosmological hydrodynamical simulations of galaxies, groups and clusters.
The simulations use the {\sc arepo} moving-mesh code with a set of physical models for galaxy formation based on the successful Illustris simulation, but with updated AGN and supernovae feedback models. This allows us to simultaneously reproduce the observed redshift evolution of the galaxy stellar mass function together with the stellar and gas mass fractions of local groups and clusters across a wide range of halo masses.
Focusing on the properties of groups and clusters, we find very good agreement with a range of observed scaling relations, including the X-ray luminosity--total mass and gas mass relations as well as the total mass--temperature and Sunyaev-Zel'dovich flux--mass relations.
Careful comparison of our results with scaling relations based on X-ray hydrostatic masses as opposed to weak lensing-derived masses reveals some discrepancies, which hint towards a non-negligible X-ray mass bias in observed samples.
We further show that radial profiles of density, pressure and temperature of the simulated intracluster medium are in very good agreement with observations, in particular for $r > 0.3\,r_{500}$. In the innermost regions however we find too large entropy cores, which indicates that a more sophisticated modelling of the physics of AGN feedback may be required to accurately reproduce the observed populations of cool-core and non-cool-core clusters.
\end{abstract}


\section{Introduction}\label{sec:intro}
Galaxy clusters collapse from the rarest peaks in the matter density field and therefore provide large leverage to probe cosmic structure growth and cosmology.
Ongoing and forthcoming surveys are set to greatly extend the number of known galaxy clusters and groups.
In X-rays, the upcoming eROSITA telescope \citep{Merloni2012, Pillepich2012} is expected to detect as many as one hundred thousand groups and clusters of galaxies out to $z \sim 1$ while observations of the Sunyaev--Zel'dovich effect (SZ) with SPT-3G \citep{Benson2014} and ACTpol \citep{Henderson2016} extend the search to higher redshift.
Detailed follow-up observations will be possible at X-ray wavelengths with missions such as \textit{XARM} (replacement for \textit{Hitomi}; \citealt{Takahashi2010, Takahashi2016}) and \textit{Athena} \citep{Nandra2013} and at optical and near-infrared wavelengths with ground-based surveys such as the Dark Energy Survey \citep{DES2005} and the Large Synoptic Survey Telescope \citep{LSST2012}, as well as future space-based missions such as Euclid \citep{Laureijs2011}.

This wealth of data holds great potential to provide stringent constraints on cluster physics and cosmological parameters, such as the amplitude and slope of the matter power spectrum and the densities of baryons, dark matter and dark energy (see \citealt{Allen2011, Kravtsov2012, Planelles2015} for recent reviews).
However, the cosmological interpretation of observed cluster data is crucially dependent on the degree to which we can connect cluster observables to theoretical predictions for a given cosmological model. Simulations of cosmic structure formation play an important role in this process. At the most basic level, they provide predictions for the abundance of clusters as a function of their mass for a particular cosmology \citep[e.g.][]{Cohn2008, Tinker2008, Courtin2010, Crocce2010, Bhattacharya2011, Murray2013, Watson2013}.
Yet as larger cluster surveys reduce the statistical uncertainty, we are increasingly limited by our incomplete understanding of cluster physics and its impact on the relation between cluster observables (such as X-ray luminosity, temperature or SZ flux) and mass \citep[e.g.][]{Rudd2008, Semboloni2011, VanDaalen2011, Cui2014, Cusworth2014, Velliscig2014, Henson2017, Chisari2018}.

Cosmological hydrodynamical simulations can aid in understanding and constraining these relations, as well as explaining potential systematic biases in observations.
As an example, the most common method for measuring cluster masses is by analysing X-ray data under the assumption that the X-ray emitting gas is in hydrostatic equilibrium with the gravitational potential of the cluster. This method has been reported to systematically underestimate the true halo mass (e.g. \citealt{Miralda-Escude1995, Wu1997, Mahdavi2008, Richard2010, Mahdavi2013, Foex2017}), however, there is still no consensus on the exact magnitude of the effect or indeed whether such a bias exists at all \citep[e.g.][]{Zhang2010, Gruen2014, Applegate2016, PlanckXXIV2015, Rines2016}.
Furthermore, as cluster surveys expand in size and depth, it will be vital to understand potential selection biases associated with a given observable. For example, X-ray surveys may be biased toward cool-core clusters with strong central X-ray emission, the prevalence and evolution of which remains uncertain \citep[e.g.][]{Andrade-Santos2017, Rossetti2017}.
Cosmological hydrodynamical simulations are in a unique position to quantify such potential biases and therefore facilitate the use of clusters as precise cosmological probes.

The hydrodynamical modelling of cluster formation has progressed considerably in recent years, largely due to increases in computing power and improvements in the treatment of physical processes that act below the resolution scale of the simulation, commonly referred to as ``sub-grid'' models. In particular, feedback from Active Galactic Nuclei (AGN) has proven critical in explaining a range of cluster observables (e.g. \citealt{Sijacki2007, Puchwein2008, Fabjan2010, Puchwein2010, McCarthy2010, McCarthy2011, Gaspari2014}).
These improvements have facilitated hydrodynamical simulations aimed at reproducing various observational properties of clusters, such as the cool-core/non-cool-core dichotomy (e.g. \citealt{Rasia2015, Hahn2017}) and the X-ray and SZ scaling relations (e.g. \citealt{Pike2014, Planelles2014}), including their redshift evolution (e.g. \citealt{Fabjan2011, LeBrun2017, Truong2018}).
A small number of groups have also produced hydrodynamical simulations of statistical samples of massive haloes useful for cluster cosmology \citep{LeBrun2014, Dolag2016, MACSIS, McCarthy2016}. One of the limitations of these works however is that they lack the numerical resolution needed to resolve the detailed structure of the cluster galaxies. Given that galaxies are often used as observational tracers of large-scale structure, it is important to work towards reproducing the properties of the galaxy populations of clusters in addition to their global stellar, gas and halo properties.

As progress has been made in the modelling of cluster formation, so too has there been a marked increase in the realism of simulated field galaxy populations.
A number of groups have produced high-resolution cosmological hydrodynamical simulations of galaxy formation capable of reproducing a range of key observations of galaxies, such as their sizes, morphologies, passive fractions and the build-up of their stellar mass.
Notable examples include Illustris \citep{Vogelsberger2014}, {\sc eagle} \citep{Schaye2014}, Horizon-AGN \citep{Dubois2014} and MassiveBlack-II \citep{Khandai2015}.
The computational expense associated with the high resolution and complex sub-grid models of these simulations limits their total volume. As a result, few, if any, galaxy clusters exist within the simulation volume and it is difficult to assess the ability of the galaxy formation model to reproduce observations of massive collapsed structures.

Recently, the {\sc c-eagle} \citep{CEAGLE, Bahe2017} and IllustrisTNG \citep{Pillepich2018b} projects have combined these two approaches and produced high-resolution simulations that resolve the full dynamic range from galaxies to massive clusters. The {\sc c-eagle} project consists of ``zoom-in'' simulations of 30 galaxy clusters in the mass range $M_{200} = 10^{14} - 10^{15.4} M_{\odot}$\footnote{Throughout this paper, spherical-overdensity masses and radii use the critical density of the Universe as a reference point. Hence, $M_{200}$ is the total mass inside a sphere of radius $r_{200}$ within which the average density is 200 times the critical density of the Universe.} simulated with the same galaxy formation model as the {\sc eagle} simulation and at the same resolution. The {\sc c-eagle} clusters are a good match to a number of cluster observables, including the satellite stellar mass function, the total stellar and metal content and the scaling of X-ray spectroscopic temperature and SZ flux with total mass. Conversely, the clusters are too gas rich and possess too high central temperatures. Their results imply that improved agreement with observations will require revision of their model for AGN feedback (see discussion in \citealt{CEAGLE}).
The IllustrisTNG project consists of three large, uniformly-sampled cosmological volumes of approximately 50, 100 and 300 Mpc on a side simulated with magneto-hydrodynamics \citep{Pillepich2018a}. The largest volume contains numerous low-mass clusters, with a few objects reaching masses $M_{200} \sim 10^{15} M_{\odot}$ at $z=0$.
IllustrisTNG is a follow-up project of the Illustris simulation \citep{Vogelsberger2013, Vogelsberger2014, Genel2014} and improves upon Illustris by extending the explorable mass range of simulated haloes and revising the astrophysical modelling to address some of the shortcomings of Illustris. The Illustris simulation was successful in reproducing a broad range of observations of galaxy populations at various redshifts, including the observed range of galaxy morphologies and the evolution of galaxy specific star formation rates.
However, various tensions with observations remained, especially on the scale of galaxy groups and (low-mass) clusters. In particular, the AGN feedback in Illustris ejected too much gas from low-redshift, massive haloes ($M_{500} \approx 10^{13}-10^{14} M_{\odot}$; \citealt{Genel2014}).
IllustrisTNG introduce a new model for AGN feedback which is able to suppress star formation in massive haloes without removing too much gas \citep{Weinberger2017a}. Combined with changes to other aspects of the modelling, such as modified galactic winds and the introduction of magnetic fields \citep{Pillepich2018a}, this has allowed IllustrisTNG to reproduce a range of observables properties of galaxies, groups and clusters.

The \textit{Feedback Acting on Baryons in Large-scale Environments} (\fable) project presented here shares a similar motivation to IllustrisTNG. We have worked independently to build a suite of simulations based upon the framework of the successful Illustris project but improve upon the agreement with observations on scales larger than galaxies, e.g., with constraints on the gas content of massive haloes. By using zoom-in simulations we also extend the comparison with observations to massive clusters, which were not present in the original Illustris simulation.
Like IllustrisTNG we have updated the Illustris models for galactic winds and AGN feedback.
We find that our changes improve the realism of groups and clusters significantly.
The \fable\ simulations consist of a uniformly-sampled cosmological volume about 60 Mpc on a side and a series of zoom-in simulations of groups and clusters approximately uniformly spaced in logarithmic halo mass.
In this paper we focus on the $z = 0$ properties of \fable\ groups and clusters in comparison to observations and demonstrate good agreement in a number of key areas.

The outline of this paper is as follows. Section~\ref{sec:sims} describes the characteristics of the \fable\ simulations and Section~\ref{sec:comparison} describes our methods for comparing the simulations with observations. We then present the galaxy stellar mass function at different redshifts (Section~\ref{sec:galaxies}), the global properties of massive haloes (Section~\ref{sec:global}) and the hot gas profiles of groups and clusters (Section~\ref{sec:profiles}). We discuss our results  in Section~\ref{sec:discussion} and summarise our findings in Section~\ref{sec:conclusion}.

\section{Simulations}\label{sec:sims}
\subsection{Basic simulation properties} \label{subsec:sims}
We utilise the cosmological hydrodynamic moving-mesh code {\sc arepo} \citep{Arepo}, which solves the Euler equations of hydrodynamics on a quasi-Lagrangian moving Voronoi mesh.
On top of gravity, hydrodynamics and a spatially uniform ionizing background, the \fable\ simulations employ a set of sub-grid models for processes important for galaxy formation. The majority of our sub-grid models are unchanged from Illustris and are described in full detail in \cite{Vogelsberger2013} and \cite{Torrey2014}. These include models for star formation \citep{Springel2003, Springel2005}, radiative cooling \citep{Katz1996, Wiersma2009} and chemical enrichment \citep{Wiersma2009a}. The sub-grid models for feedback from stars \citep{Vogelsberger2013} and AGN \citep{DiMatteo2005, Springel2005, Sijacki2007} have been modified from the Illustris models and are described in Sections~\ref{subsec:SF} and \ref{subsec:BH}.
The parameters of the feedback models have been calibrated to reproduce observations of the local galaxy stellar mass function (Section~\ref{sec:galaxies}) and the gas mass fractions of massive haloes (Section~\ref{sec:global}).
This same calibration strategy was adopted for the {\sc bahamas} simulations \citep{McCarthy2016}, which successfully reproduce a broad range of observed hot gas and stellar properties of massive systems.
The calibration process involved a series of periodic boxes of length 40~$h^{-1}$~Mpc simulated with different parametrizations of stellar and AGN feedback. We note that, as this volume is relatively small, these simulations did not contain massive clusters and therefore gas fractions were initially matched to observations on group-scales ($M_{500} \lesssim 10^{14} M_{\odot}$). Results from a sample of these calibration simulations are presented in Appendix~\ref{A:models}.
The \fable\ suite of simulations consists of the calibration volume and a series of zoom-in simulations of individual galaxy groups and clusters, which apply our calibrated model to higher mass objects.

The 40~$h^{-1}$ (comoving) Mpc periodic box was evolved to $z=0$ from initial conditions based on a Planck cosmology \citep{PlanckXII2015} with cosmological parameters $\Omega_{\Lambda}=$~0.6935, $\Omega_{\rm M}=$~0.3065, $\Omega_{\rm b}=$~0.0483, $\sigma_8=$~0.8154, $n_s=$~0.9681 and $H_0=67.9$~km~s$^{-1}$~Mpc$^{-1}=h \times$100~km~s$^{-1}$~Mpc$^{-1}$. The simulation follows 512$^3$ dark matter particles of mass $m_{\rm DM}=3.4\times$10$^7$ $h^{-1}$~M$_{\odot}$ and approximately 512$^3$ baryonic resolution elements (gas cells and star/BH particles) of typical mass $\overline{m}_{\rm b}=6.4\times$10$^6$~$h^{-1}$~M$_{\rm \odot}$.
The gravitational softening length was fixed to $2.393$~$h^{-1}$~kpc in physical coordinates below $z=5$ and fixed in comoving coordinates at higher redshifts.
This choice is consistent with the $z=0$ optimal softening length for low-mass cluster galaxies according to the empirical rule determined by \cite{Power2003}\footnote{Note that this softening length is somewhat larger than used in Illustris.}.

For our series of zoom-in simulations, individual groups and clusters were chosen from a collisionless (dark matter-only) parent simulation and re-simulated at high resolution. The zoom-in regions were drawn from Millennium XXL, a periodic box of side length $3~h^{-1}$~Gpc \citep{Angulo2012}. Six systems were selected logarithmically-spaced in mass spanning the mass range from groups ($\sim 10^{13} M_{\odot}$) to massive clusters ($\sim 10^{15} M_{\odot}$). These systems were selected only by their total mass within the parent simulation, with no prior knowledge regarding, for example, their dynamical state.
The high-resolution regions were chosen such that they are free from lower resolution particles out to approximately $5\,r_{500}$ at $z=0$. Dark matter particles in the high-resolution region have a mass of $m_{\rm DM}=5.5\times$10$^7$ $h^{-1}$~M$_{\odot}$. The gravitational softening length was fixed at $2.8125$~$h^{-1}$~kpc in physical coordinates for $z \leq 5$ and fixed in comoving coordinates for $z > 5$. The softening lengths of boundary particles outside the high-resolution region were allowed to vary with their mass.
Mode amplitudes in the initial conditions were scaled to a Planck cosmology \citep{PlanckXII2015} with $\Omega_{\Lambda}=$~0.6911, $\Omega_{\rm M}=$~0.3089, $\Omega_{\rm b}=$~0.0486, $\sigma_8=$~0.8159, $n_s=$~0.9667 and $H_0=67.74$~km~s$^{-1}$~Mpc$^{-1}$.
In presenting our results we rescale the appropriate quantities to the cosmology of the periodic box described above, although we note that this is a very small effect due to the similarity between the cosmological parameters.

The simulations were processed on-the-fly with the friends-of-friends (FoF) and \mbox{\sc subfind} algorithms \citep{Davis1985, Springel2001, Dolag2009} to identify gravitationally bound groups of dark matter, stars and gas (``haloes''), using a linking length of $0.2$ times the mean dark matter inter-particle separation, and decompose them into self-bound substructures (``subhaloes'').

\begin{figure*}
	\includegraphics[width=\textwidth]{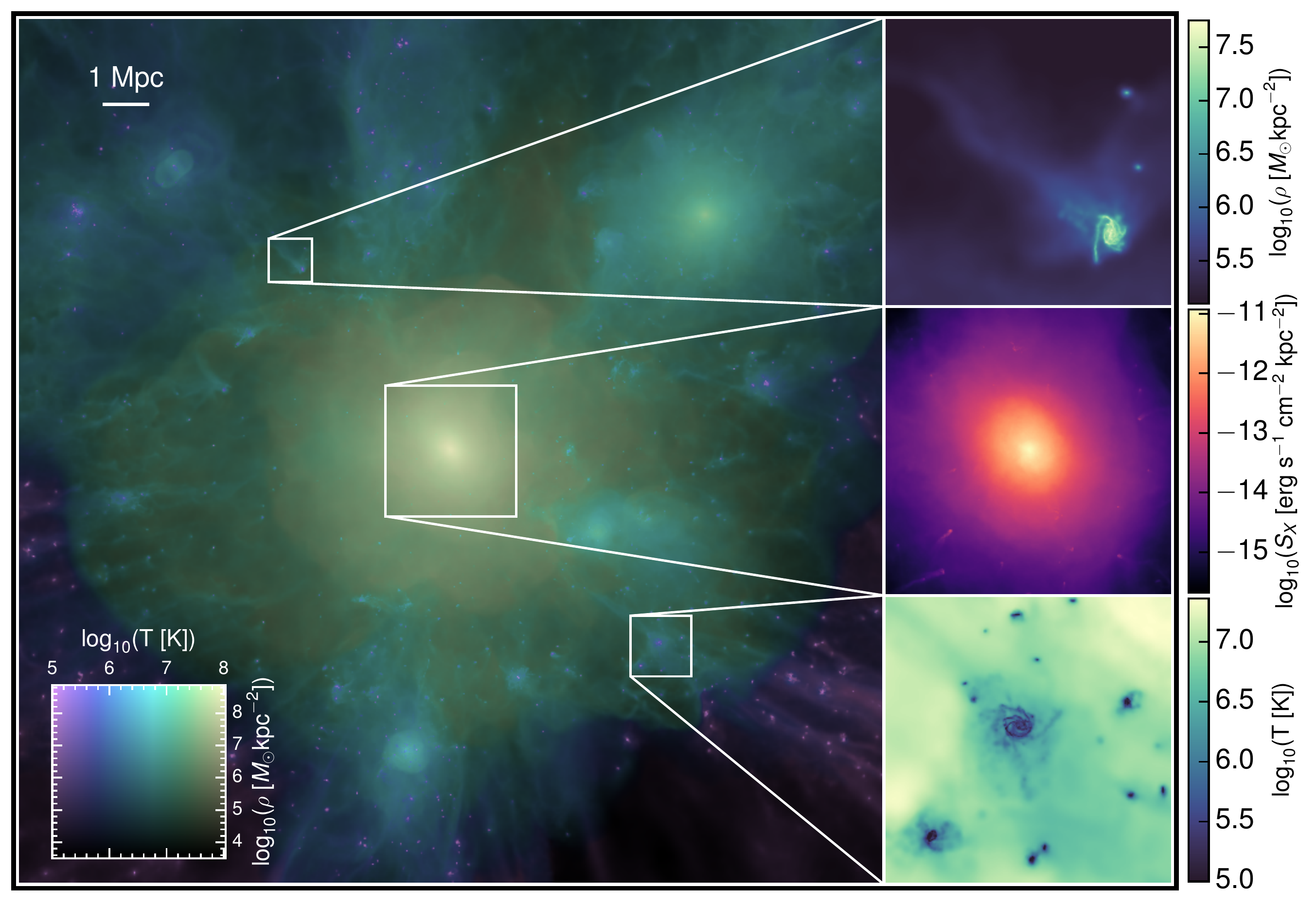}
        \caption{A visualization of the gas properties of a massive \fable\ cluster with $M_{500} = 9 \times 10^{14} M_{\odot}$ at $z=0$. The main panel shows a $20 \times 20 \times 5$~Mpc slice through the cluster with panels on the right hand side zooming in on individual galaxies and the cluster core. In the main panel the gas surface density is represented by the image brightness while the gas temperature corresponds to the hue/saturation of the image according to the inset colour scale.
          The bottom-right panel shows the gas temperature distribution of a group of galaxies of various sizes and morphologies. The top-right panel shows the surface density of gas surrounding an individual spiral galaxy on the cluster outskirts, revealing the onset of ram pressure stripping of cold galactic gas as the galaxy approaches the dense cluster environment.
          Lastly, in the middle-right panel we show the X-ray surface brightness in the $0.5-7$~keV energy band within a region $2~r_{500}$ on a side centred on the cluster.
        }
    \label{fig:map}
\end{figure*}

\subsection{Cluster visualization}
In Fig.~\ref{fig:map} we present a visualization of the gas properties of the most massive \fable\ cluster at $z=0$. The main panel shows the gas distribution in a $20 \times 20 \times 5$~Mpc slice centred on the gravitational potential minimum of the cluster. The gas surface density and temperature are encoded by the image brightness and hue/saturation, respectively, according to the colour map shown in the bottom-left of the figure.
The structure of the cluster gas is clearly laid out, revealing an array of interconnected filaments and infalling galaxies and groups, examples of which are shown in the panels on the right hand side of the figure.
During the formation of this cluster, a combination of adiabatic compression and shocks has heated the intracluster gas to high, X-ray emitting temperatures ($\gtrsim 10^7$~K). This is demonstrated in the middle right-hand panel, which shows the X-ray surface brightness within a region $2\,r_{500}$ on a side centred on the cluster.

\subsection{Star formation, stellar feedback and chemistry}
\label{subsec:SF}
Star formation and chemistry are treated with the same implementation as in Illustris, as described in detail in \cite{Vogelsberger2013} and \cite{Torrey2014}. In brief, the dense star-forming interstellar medium (ISM) is treated in a sub-resolution fashion using a slightly modified version of the \cite{Springel2003} sub-grid model. The ISM is modelled using an effective equation of state, where stars form stochastically from gas above a density threshold with a given star formation time scale.
Stellar mass loss and metal enrichment are treated by calculating the mass and chemical composition of ejected material for each active star particle at each time step and returning it to the nearby gas.

Stellar feedback in the form of galactic outflows is modelled via wind particles launched stochastically from star forming gas with a velocity based on the local dark matter velocity dispersion \citep{Vogelsberger2013}. The wind particles are briefly decoupled from hydrodynamic interactions until they have left the local ISM and deposit their mass, metals, momentum and thermal energy into the surrounding gas.
In Illustris, the energy given to the wind particles is purely kinetic. These cold winds push gas out of the galactic disc, preventing overcooling within the dense ISM and lowering the galactic star formation rate. However, in this model it is left fully up to hydrodynamical interactions to dissipate the kinetic energy to heat, which is often insufficient to prevent the ejected gas from quickly condensing back onto the galactic disc.
We reduce this possibility by imparting one-third of the wind energy as thermal rather than kinetic.
This slows the cooling of the ejected gas, causing it to remain outside of the dense ISM for a longer period of time, increasing the overall effectiveness of stellar feedback.
The same method has been used in \cite{Marinacci2014} (50 per cent thermal energy), the Auriga galaxy simulations (50 per cent; \citealt{Grand2017}) and IllustrisTNG (10 per cent; \citealt{Pillepich2018a}).

\subsection{Black hole accretion and feedback}\label{subsec:BH}
\subsubsection{Seeding and accretion}\label{subsubsec:accretion}
Black hole formation is modelled assuming that ``seed'' black holes (BHs) of mass $10^5 h^{-1} M_{\odot}$ form regularly enough that every halo above a mass threshold of $5 \times 10^{10} h^{-1} M_{\odot}$ contains a seed BH at its centre. Such haloes are identified by running a fast FoF algorithm on-the-fly. If a halo exceeds this mass threshold and does not already host a BH, the highest density gas cell in the halo is converted into a BH particle.

BHs are modelled as collisionless sink particles that are allowed to grow via mergers with other BHs and by accretion of ambient gas.
The prescription for BH accretion is described in detail in \cite{Vogelsberger2013}. Briefly, BHs accrete at an Eddington-limited Bondi-Hoyle-Lyttleton rate boosted by a constant factor $\alpha = 100$ \citep{Hoyle1939, Bondi1944, Springel2005}. A fraction $(1 - \epsilon_r)$ of the accreted mass is added to the mass of the BH, where $\epsilon_r = 0.1$ is the radiative efficiency. The remaining rest mass energy is made available as feedback energy.

\subsubsection{AGN Feedback}
\label{subsubsec:feedback}
For feedback from BHs we use an adapted form of the Illustris model.
This model distinguishes between two states: a high accretion rate quasar-mode \citep{DiMatteo2005, Springel2005} and a low accretion rate radio-mode \citep{Sijacki2007}. The quasar-mode is dominant at high redshift while the radio-mode becomes increasingly important at lower redshifts as the average accretion rate diminishes.

In the quasar-mode of feedback, a fraction $\epsilon_f = 0.1$ of the available feedback energy is coupled thermally and isotropically to the surrounding gas (see \citealt{Springel2005} for details).
In Illustris, BHs in the quasar-mode inject thermal energy into the surrounding gas \textit{continuously}. However, this approach can result in artificial overcooling whereby too little thermal energy is injected into too much mass and the energy is quickly radiated away.
For this reason, \cite{Booth2009} developed a model for thermal AGN feedback in smoothed particle hydrodynamics (SPH) simulations in which BHs store feedback energy until it is sufficient to raise the temperature of a given number of SPH particles by a given amount. This model was subsequently employed in the cosmological hydrodynamical simulations {\sc eagle} \citep{Schaye2014}, cosmo-OWLS \citep{LeBrun2014} and {\sc bahamas} \citep{McCarthy2016}.
We take a similar approach to reduce numerical overcooling in our simulations by introducing a duty cycle for the quasar-mode in which feedback energy is accumulated over a given time period, $\Delta t = 25$~Myr, before being released in a single event.

At low accretion rates below a fraction $\chi_{radio} = 0.01$ of the Eddington rate, we employ radio-mode feedback following \cite{Sijacki2007}.
In the radio-mode, hot buoyantly-rising bubbles are injected into the surrounding gas.
The duty cycle of the radio-mode is controlled by the mass growth of the BH such that a bubble is injected once the BH has increased in mass by a fraction $\delta_{\rm BH} \equiv \delta M_{\rm BH}/M_{\rm BH}$.
The energy content of the bubble is then $E_{\rm bub} = \epsilon_m \epsilon_r c^2 \delta M_{\rm BH}$ for which we choose a radio-mode coupling efficiency $\epsilon_m = 0.8$. The total radio-mode feedback efficiency as given by the product $\epsilon_m \epsilon_r$ is 8 per cent, similar to Illustris (7 per cent).
For the duty cycle we set the threshold for bubble triggering to $\delta_{\rm BH} = 0.01$, much smaller than $\delta_{\rm BH} = 0.15$ as used in Illustris. Since the bubble energy is proportional to the increase in mass, this corresponds to more frequent but less energetic bubbles.
For the bubble size and distance we assume the scaling relations defined in \cite{Sijacki2007}, with normalisation constants $D_{\textrm{bub},0} = 30 $~$h^{-1}$~kpc, $R_{{\rm bub},0} = 50 $~$h^{-1}$~kpc, $E_{{\rm bub},0} =$ $10^{60}$~erg and $\rho_{{\rm ICM},0} = 10^{4} h^2 M_{\odot}$~kpc$^{-3}$ for the bubble distance, radius, energy content and ambient density, respectively.

The remainder of the feedback energy that is not coupled to the surrounding gas by the quasar- or radio-mode goes into radiative electromagnetic feedback, which is approximated as an additional radiation field around the BH superposed with the redshift-dependent ultraviolet background. The full details of this model are given in \cite{Vogelsberger2013}.

In Appendix~\ref{A:models} we present several parametrizations of AGN feedback which were considered during the calibration process to demonstrate how changes to the duty cycle and energetics of the feedback modes impact the $z=0$ galaxy stellar mass function and the stellar and gas mass fractions of massive haloes.
In particular, we show that the introduction of a quasar-mode duty cycle can greatly suppress the growth of stellar mass in massive galaxies compared to continuous quasar-mode feedback.

\section{Comparing to Observations}\label{sec:comparison}
The physical properties of a simulation often do not correspond directly to those derived from real observations. This can result from a number of factors, including selection biases, projection effects, instrument systematics and methodology.
In this section we will discuss how we can mitigate some of these effects in our comparisons with observations.
We also discuss the derivation of quantities which are not intrinsic to the simulation.
For example, X-ray luminosity is a complicated combination of continuum and line emission that is not followed in the simulations. We therefore follow a procedure described in Section~\ref{subsec:xray} in which we create synthetic X-ray spectra in post-processing and analyse them similarly to observations.

\subsection{Galaxy stellar mass functions}\label{subsec:SMF_method}
In the field of numerical galaxy formation, the galaxy stellar mass function (GSMF) is one of the most important observationally constrained properties that simulations should match.
In constructing our GSMF we define galaxies as the self-bound subhaloes identified by \mbox{\sc subfind} and calculate their properties based on the corresponding subhalo catalogue.
With this definition, the total stellar mass of a galaxy is the total mass of star particles bound to the subhalo. However, this does not necessarily correspond to the stellar mass of the galaxy as would be measured by an observer. This is because a significant fraction of the stellar mass in massive systems can exist in the form of diffuse intracluster light (ICL), which is difficult to quantify in stellar mass measurements due to its low surface brightness (e.g. \citealt{Zibetti2005, Gonzalez2007, Morishita2016}).
Some studies choose to integrate a galaxy's light within a 2-D aperture of a given size (e.g. \citealt{Li2009}) while others integrate Sersic or other fits to the light profiles (e.g. \citealt{Baldry2012, Bernardi2013}). This choice can have a significant impact on the derived stellar mass function (see e.g. \citealt{Bernardi2013}).
We follow previous simulation studies in the literature and present our GSMF using multiple definitions of a galaxy's stellar mass: the total stellar mass bound to the corresponding subhalo and two aperture masses.
In one case we follow \cite{Genel2014}, who consider the stellar mass within a spherical aperture with radius equal to twice the stellar half-mass radius ($2 r_{\star, 1/2}$). In the other, we follow \cite{Schaye2014} who define the galaxy stellar mass as the mass within a fixed spherical aperture of radius 30 physical kiloparsec (pkpc). In each case only stellar mass bound to the subhalo is considered.

\subsection{X-ray properties}\label{subsec:xray}
Galaxy clusters are permeated with diffuse gas heated to temperatures on the order of $10^{7-8} \mathrm{K}$. This intracluster medium (ICM) emits strongly in X-rays, providing an invaluable means of detecting and studying the properties of clusters.
Previous studies have shown that the properties of hot gas derived from X-ray observations can be biased, for example due to complex thermal structure or clumping of the gas (e.g. \citealt{Mazzotta2004, Rasia2005, Nagai2007a, Khedekar2013}). A reliable comparison between simulations and observations therefore requires actual simulation of the spectral properties of the X-ray emission.
We derive X-ray luminosities and spectroscopic temperatures in the following manner.

For a given halo we consider gas within a projected aperture of radius $r_{500}$ centred on the minimum of the gravitational potential. This includes all gas along the line of sight (although for the zoom-in simulations we exclude gas outside the high-resolution region). This is intended to mimic X-ray observations, which measure projected luminosities and temperatures.
We exclude cold gas with a temperature less than $3 \times 10^4$~K as the lack of molecular cooling in our simulations implies that the temperature of such gas can be significantly overestimated and it should contribute negligibly to the X-ray emission.
We also exclude gas above the density threshold required for star formation, as the sub-grid multiphase model for star-forming gas does not reliably predict the thermal properties of the gas. In Appendix~\ref{A:cuts} we demonstrate that derived X-ray luminosities and spectroscopic temperatures are not particularly sensitive to the choice of temperature--density cut.

We project the emission measure of the gas onto temperature bins of width 0.02 keV between 0.01~keV and 24.0~keV, summing up the emission measure in each bin. Using the XSPEC package (\citealt{Arnaud1996}, version 12.8.0), we then produce a mock X-ray spectrum by summing APEC emission models \citep{Smith2001} generated for each temperature bin.
For simplicity we assume a constant metallicity of 0.3 times the solar value. This ensures that our X-ray analysis remains independent of the metal enrichment in the simulations, which is highly sensitive to the details of the feedback model. Even so, we find no significant difference in the derived X-ray luminosities and temperatures when using the metallicity of the gas in the simulations.
The mock spectrum is convolved with the response function of \textit{Chandra} and we adopt a large exposure time of $10^6$ seconds so as not to be limited by photon noise.
We then fit the spectrum with a single-temperature APEC model in the $0.5-10$ keV energy range with the temperature, metallicity and normalisation of the spectrum left as free parameters.
From the best-fitting spectrum we obtain the X-ray luminosity and spectroscopic temperature.

\subsection{Halo masses}\label{subsec:bias}
As previously discussed in Section~\ref{sec:intro}, the most common method to obtain halo masses for calibrating cluster scaling relations is the X-ray hydrostatic method, which some studies have shown can underestimate the true halo mass.
As there is no consensus as to the magnitude of the bias, any comparison between X-ray hydrostatic masses and true halo masses from the simulations must be carefully considered. We make use of the rising number of weak gravitational lensing studies to compare true masses from the simulations with both X-ray hydrostatic masses and halo masses measured with weak lensing. The latter are generally considered to be less biased due to the insensitivity of gravitational lensing to the equilibrium state of the gas or dark matter (for a review, see e.g. \citealt{Hoekstra2013, Mandelbaum2017}). On the other hand, lensing mass measurements can possess significantly more scatter due to the effects of cluster substructure, cluster triaxiality and projection effects (e.g. \citealt{Corless2007, Marian2010, Meneghetti2010, Becker2011}). Since there are also many more observed systems with available X-ray hydrostatic masses than weak lensing masses, especially toward lower halo masses, we choose to compare spherical overdensity masses measured from the simulations to both measures where possible.

\subsection{SZ properties}\label{subsec:SZ}
The thermal Sunyaev--Zel'dovich (tSZ) effect has long been recognised as a powerful tool for studying the physics of the ICM and the formation of large-scale structure \citep{Birkinshaw1999, Carlstrom2002}. The tSZ effect appears as a distortion in the cosmic microwave background (CMB) spectrum, arising from the inverse Compton scattering of CMB photons on energetic electrons in the ICM.
When integrated over the volume of a system, the tSZ flux provides a relatively clean measure of its total thermal energy.

The tSZ signal is characterised by $Y_{500}$, the Comptonisation parameter integrated over a sphere of radius $r_{500}$. Specifically,
\begin{equation}
  D_A^2(z) Y_{500} \equiv \frac{\sigma_T}{m_e c^2} \int^{r_{500}}_{0} P dV
\end{equation}
where $D_A(z)$ is the angular diameter distance, $\sigma_T$ the Thomson cross-section, $m_e$ the electron rest mass, $c$ the speed of light, and $P=n_ekT_e$ the electron pressure, equal to the product of the electron number density and electron temperature.

We compare the tSZ signal of our simulated systems with \cite{Planck2013XI}, who perform a stacking analysis on \textit{Planck} multi-frequency observations of a large sample of Locally Brightest Galaxies (LBGs) selected from the Sloan Digital Sky Survey (SDSS).
The selection criteria were designed to maximise the fraction of objects that are the central galaxies of their dark matter haloes. Correspondingly we measure the tSZ signal only for central galaxies and use $r_{500}$ calculated in spheres centred on each galaxy.
\cite{Planck2013XI} estimate the mean $Y_{500}$ in a series of stellar mass bins and convert this to a tSZ signal-halo mass relation by estimating an ``effective'' halo mass for each stellar mass bin using a mock sample of LBGs from the semi-analytic galaxy formation simulation of \cite{Guo2011a}.
We also compare to the weak lensing recalibration of the tSZ signal-halo mass relation given in \cite{Wang2016}, which reduces the dependency of the halo mass estimates on the galaxy formation model and explicitly accounts for uncertainties in both the modelling and the lensing measurements.

Due to the limited angular resolution of \textit{Planck}, the tSZ flux is actually measured within a (projected) aperture of radius $5\,r_{500}$. \cite{Planck2013XI} convert this measured flux, $Y_{5r_{500}}$, into the flux within a spherical aperture of radius $r_{500}$ by a conversion factor $Y_{500} = Y_{5r_{500}}/1.796$. This factor assumes the spatial template used in their matched filter, the universal pressure profile \citep{Arnaud2010}, and assumes no tSZ flux originates from beyond $5\,r_{500}$.
In our comparisons we remove the dependency on the assumed pressure distribution by converting the observationally inferred flux reported in \cite{Planck2013XI}, $Y_{500}$, back into the actual measured flux, $Y_{5r_{500}}$.
This is motivated by the work of \cite{LeBrun2015} who show that $Y_{500}$ is highly sensitive to the assumed spatial template. \cite{LeBrun2015} generate synthetic tSZ maps from a cosmological hydrodynamical simulation from the cosmo-OWLS project that reproduces a range of global SZ, X-ray and optical properties of local groups and clusters \citep{LeBrun2014}. Applying the same tools and assumptions as \textit{Planck}, \cite{LeBrun2015} show that the inferred flux, $Y_{500}$, is biased high by a factor $\sim 2$ at $M_{500} = 2.6 \times 10^{13} M_{\odot}$. This bias increases with decreasing halo mass, reaching nearly an order of magnitude overestimate below $\sim 10^{13} M_{\odot}$. The vast majority of the bias is due to the assumption of a fixed spatial template, which becomes an increasingly worse description of the hot gas in low mass haloes.

The tSZ flux as measured by \textit{Planck} for individual, low-mass systems may be biased due to source confusion (i.e. hot gas along the line-of-sight). Indeed, our tests have shown that the flux of individual low-mass systems can be significantly boosted by hot gas which overlaps with them in projection.
However, \cite{Planck2013XI} estimate the mean tSZ flux from stacking a large number of systems in mass bins, which can remain unbiased by source confusion. Indeed, \cite{LeBrun2015} demonstrate that the mean tSZ flux is not significantly biased by uncorrelated confusion across the whole mass range of the \cite{Planck2013XI} stacking analysis.
For this reason we integrate the tSZ flux of a galaxy within a spherical aperture of radius $5\,r_{500}$ and note that this may underestimate the observed flux if correlated or uncorrelated source confusion leads to a significant bias in the observations.

\subsection{ICM profiles}\label{subsec:prof}
Radial profiles of the intracluster medium characterize its distribution and thermodynamic history. The effects of non-gravitational processes such as AGN feedback cause ICM profiles to deviate from the self-similar relations predicted in the absence of such processes (e.g. \citealt{Voit2005}). A comparison of simulated profiles to observed ones is therefore a useful test of non-gravitational physics.

In X-ray observations, gas density profiles are derived from background-subtracted surface brightness profiles and gas temperature profiles are obtained from extracted spectra. Since these observations are sensitive only to hot X-ray emitting gas, in calculating the radial profiles of our simulated ICM we apply the same temperature--density cuts as used in our X-ray analysis described in Section~\ref{subsec:xray}. The gas is then divided into concentric spherical shells with logarithmically-spaced radii centred on the minimum of the gravitational potential of the halo. For each radial bin we calculate the volume-weighted electron number density, $n_e$, and the mass-weighted temperature, $T$. The former is defined as the total electron number divided by the volume of the bin. Since we may exclude some gas cells from the bin, we correct the bin volume by the ratio of the total volume of cells that are not excluded to the total volume of all cells in the bin.

\section{The galaxy population}\label{sec:galaxies}
\subsection{Galaxy stellar mass function at $z=0$}\label{subsec:GSMF}

\begin{figure}
	\includegraphics[width=\columnwidth]{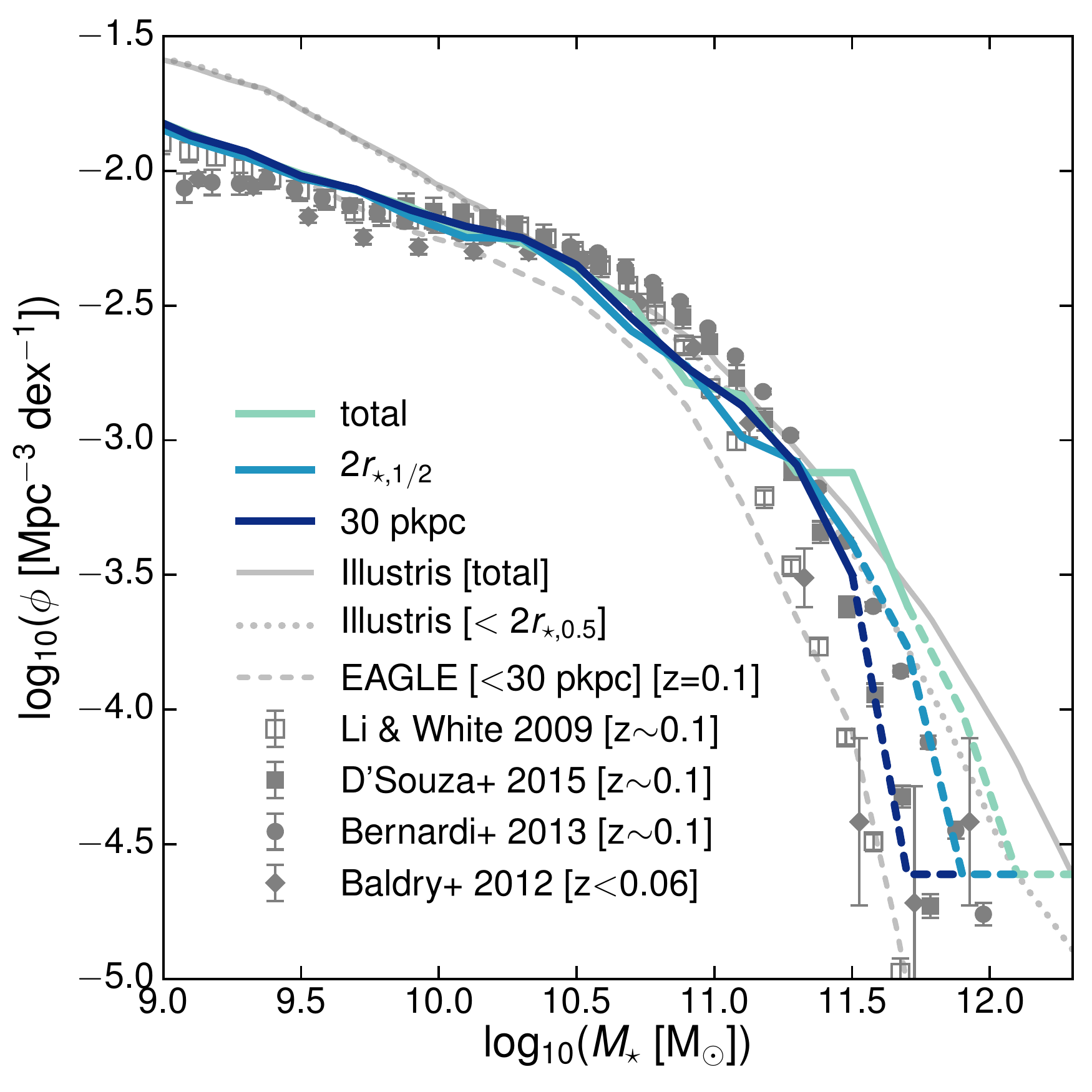}
        \caption{The \fable\ galaxy stellar mass function at $z=0$ for different definitions of galaxy stellar mass (lines) in comparison to observations (symbols with error bars). We consider three definitions of a galaxy's stellar mass: the total mass of all star particles bound to the subhalo, those within twice the stellar half-mass radius and those within 30 pkpc.
          Lines become dashed at the high-mass end when there are fewer than 10 objects per 0.2 dex stellar mass bin.
          Data points show measurements with 1-sigma error bars from \protect\cite{Li2009}, \protect\cite{DSouza2015}, \protect\cite{Bernardi2013} and \protect\cite{Baldry2012}. The grey solid and dotted lines are the $z=0$ GSMFs of Illustris \protect\citep{Genel2014} using the total bound stellar mass of each galaxy and the mass within twice the stellar half-mass radius, respectively. The grey dashed line is the $z=0.1$ GSMF of {\sc eagle} \protect\citep{Schaye2014} using the stellar mass within a spherical 30 pkpc aperture.
          All mass functions assume a \protect\cite{Chabrier2003} IMF.
        }
    \label{fig:SMF_best}
\end{figure}

Fig.~\ref{fig:SMF_best} shows the $z=0$ galaxy stellar mass function (GSMF) for all galaxies in the (40 $h^{-1}$ Mpc)$^3$ simulation volume. We plot the GSMF for three different definitions of a galaxy's stellar mass, as discussed in Section~\ref{subsec:SMF_method}: these are the total bound stellar mass, the mass within twice the stellar half-mass radius ($2\,r_{\star, 1/2}$) and the mass within a radius of 30~pkpc.
We compare with the Illustris and {\sc eagle} simulations (grey lines) and observational estimates of the GSMF from \cite{Li2009}, \cite{Baldry2012}, \cite{Bernardi2013} and \cite{DSouza2015} (symbols with error bars).
We plot the $z=0$ GSMF from Illustris for both the total bound mass and the mass within $2\,r_{\star, 1/2}$ \citep{Genel2014} and the $z=0.1$ GSMF of {\sc eagle}, which uses the mass within a 30~pkpc aperture \citep{Schaye2014}.

At the low mass end of the GSMF, \fable\ is in excellent agreement with the observational data. This contrasts with Illustris, which significantly overestimates the abundance of galaxies below the knee ($M_{\star} \lesssim 10^{10} M_{\odot}$).
While some of this improvement is owed to changes to the stellar feedback model, we have determined that the dominant cause is a difference in the gravitational softening compared to Illustris. We have chosen somewhat larger softening lengths compared to Illustris and find that this results in fewer low-mass galaxies. Somewhat surprisingly, the difference is not only present near the resolution limit but extends almost two decades in stellar mass.
We have tested this explicitly by performing a simulation identical to the \fable\ periodic box described in Section~\ref{subsec:sims} but with a gravitational softening length that is approximately 2.5 times smaller.
Quantitatively, the effect is a systematic decrease of $\sim 0.2$~dex in the mass function for stellar masses ranging from near the resolution limit ($\sim 10^{8} M_{\odot}$) to just below the knee ($\sim 10^{10} M_{\odot}$). This offset is already present at $z=8$ and persists until $z=0$. As there is no general consensus regarding the choice of gravitational softening lengths, our improved agreement at the low mass end of the GSMF is largely serendipitous. We also note that, like Illustris, we are unlikely to be fully converged with respect to resolution.

The knee of the GSMF is in agreement with the data, although it is slightly lower than in Illustris. The difference is largely the result of changes to the stellar feedback model. As discussed in Section~\ref{subsec:SF}, we have modified the stellar feedback model of Illustris by assigning one-third of the galactic wind energy as thermal.
We find that this reduces the abundance of galaxies mostly around the knee of the $z=0$ GSMF, leaving lower masses ($M_{\star} \lesssim 10^{10} M_{\odot}$) relatively unaffected.
Physically, star formation is suppressed because warmer winds increase the buoyancy of the gas outflow thereby reducing the rate at which gas is recycled back onto the galaxy.

The high mass end of the total and $2\,r_{\star, 1/2}$ GSMFs are very similar to those of Illustris, with a slight reduction in the abundance of the most massive galaxies. This is encouraging given that we have greatly reduced the burstiness of the radio-mode of AGN feedback compared to Illustris and therefore reduced its ability to suppress star formation in massive haloes. Instead we have managed to efficiently suppress star formation in massive galaxies through the use of a duty cycle for the quasar-mode of feedback. In Appendix~\ref{A:models} we show how this affects the high mass end of the $z=0$ GSMF and the gas fractions of massive haloes compared with continuous quasar-mode feedback.

The GSMF of {\sc eagle} was calibrated to reproduce the mass function derived by \cite{Li2009} for a complete spectroscopic sample from the Sloan Digital Sky Survey (SDSS) (open symbols in Fig.~\ref{fig:SMF_best}). \cite{Li2009} measure galaxy flux within a projected Petrosian aperture, which \cite{Schaye2014} show yields a GSMF similar to a spherical aperture of radius 30~pkpc for galaxies in {\sc eagle}.
At low masses, $M_{\star} \lesssim 10^{10} M_{\odot}$, our $30$~pkpc aperture mass function is almost identical to that of {\sc eagle}. Around the knee of the GSMF ($M_{\star} \lesssim 10^{11} M_{\odot}$) we are actually in better agreement with \cite{Li2009} compared with {\sc eagle}, which somewhat underestimates the GSMF there.
At larger masses we overestimate \cite{Li2009} by $\sim 0.3$~dex, although the difference may be exaggerated as \cite{Bernardi2017} argue that \cite{Li2009} use mass-to-light ratios that are biased low for massive galaxies.

At the high mass end of the GSMF there is significant variation between different observational studies.
Some of the dominant causes for this variation are discussed in \cite{Bernardi2017}. In particular, the observed GSMF depends on how the total light associated with a galaxy is determined. The potential impact of this effect is apparent from the simulated total and aperture mass functions shown in Fig.~\ref{fig:SMF_best}, which differ by $\sim 0.3$~dex on the vertical axis at the high mass end. In addition, there is some freedom in the stellar population modelling used to estimate the mass associated with the total stellar light (i.e. the mass-to-light ratio).

The GSMF derived from the total bound stellar mass of galaxies slightly exceeds observational constraints at the high mass end, suggesting that the \fable\ simulations produce slightly too many massive galaxies. The degree to which this is true depends on what fraction of the total mass in massive galaxies is accounted for in observations. In particular, a significant fraction ($\sim 30$ per cent) of the total stellar mass in our massive galaxies is contained in the ICL, the diffuse nature of which makes it difficult to quantify from observations.
Studies which measure the galaxy flux within a particular aperture will typically exclude the majority of the ICL associated with massive galaxies. For example, the Petrosian aperture used by \cite{Li2009} is known to significantly underestimate the flux of galaxies with extended surface brightness profiles \citep{Blanton2001, Graham2005, Bernardi2010, Bernardi2013}.
More recent studies such as \cite{Baldry2012}, \cite{Bernardi2013} and \cite{DSouza2015} attempt to measure a better estimate of the total flux of galaxies by integrating models fit to their surface brightness distributions.
\cite{Baldry2012} fit Sersic profiles to $z < 0.06$ galaxies from the Galaxy And Mass Assembly (GAMA) survey while \cite{Bernardi2013} fit Sersic-exponential models to a magnitude-limited sample of SDSS galaxies.
\cite{DSouza2015} use the same sample as \cite{Li2009} but integrate the galaxy flux from exponential or de Vaucouleurs profile fits and derive flux corrections from stacked SDSS images, which provide a more accurate measurement of the total amount of light owing to the increased signal-to-noise ratio. The extra light returned by this method compared to \cite{Li2009} results in a larger abundance of massive galaxies, as evident in Fig.~\ref{fig:SMF_best}.
These model profiles can only be fit to the central, high signal-to-noise regions of galaxies and must make assumptions about the outer regions of the galaxy profile. Which profile is the most appropriate at the high mass end is still debated and can lead to a significant bias in the total estimated flux and resultant stellar mass estimate (see e.g. discussion in \citealt{Bernardi2013}).

Using an aperture of radius $2\,r_{\star, 1/2}$, the \fable\ GSMF is in very good agreement with \cite{Bernardi2013} at the high mass end but is overestimated compared to \cite{DSouza2015} and \cite{Baldry2012}.
\cite{Bernardi2017} show that the Sersic-exponential fits used by \cite{Bernardi2013} return galaxy fluxes similar to the corrected fluxes of \cite{DSouza2015}. Similarly, \cite{Bernardi2013} show that their luminosity function is in good agreement with that used in \cite{Baldry2012}. This implies that the difference between these three studies at the massive end of the GSMF is due to differences in their assumed mass-to-light ratios rather than their methods for estimating galaxy flux.
\cite{Bernardi2017} state that the \cite{Bernardi2013} model is oversimplified and may overestimate the mass-to-light ratio. On the other hand, the stellar population modelling used by \cite{DSouza2015} results in a mass function that is $\sim 0.3$ dex lower on the vertical axes above $10^{11.5} M_{\odot}$ compared to more recent estimates of the mass-to-light ratio based on the same IMF \citep{Bernardi2017}.
Given that there is no consensus as to the best approach to stellar population modelling, in addition to the uncertainty in how to fit the light profiles of massive galaxies, there is arguably little point in tuning the simulated galaxy stellar mass function to a specific dataset.
Overall, we are satisifed that the difference between the simulated and observed GSMFs is similar to the variation between different observational studies.

\subsection{Galaxy stellar mass function at $z \leq 3$}
\begin{figure}
	\includegraphics[width=\columnwidth]{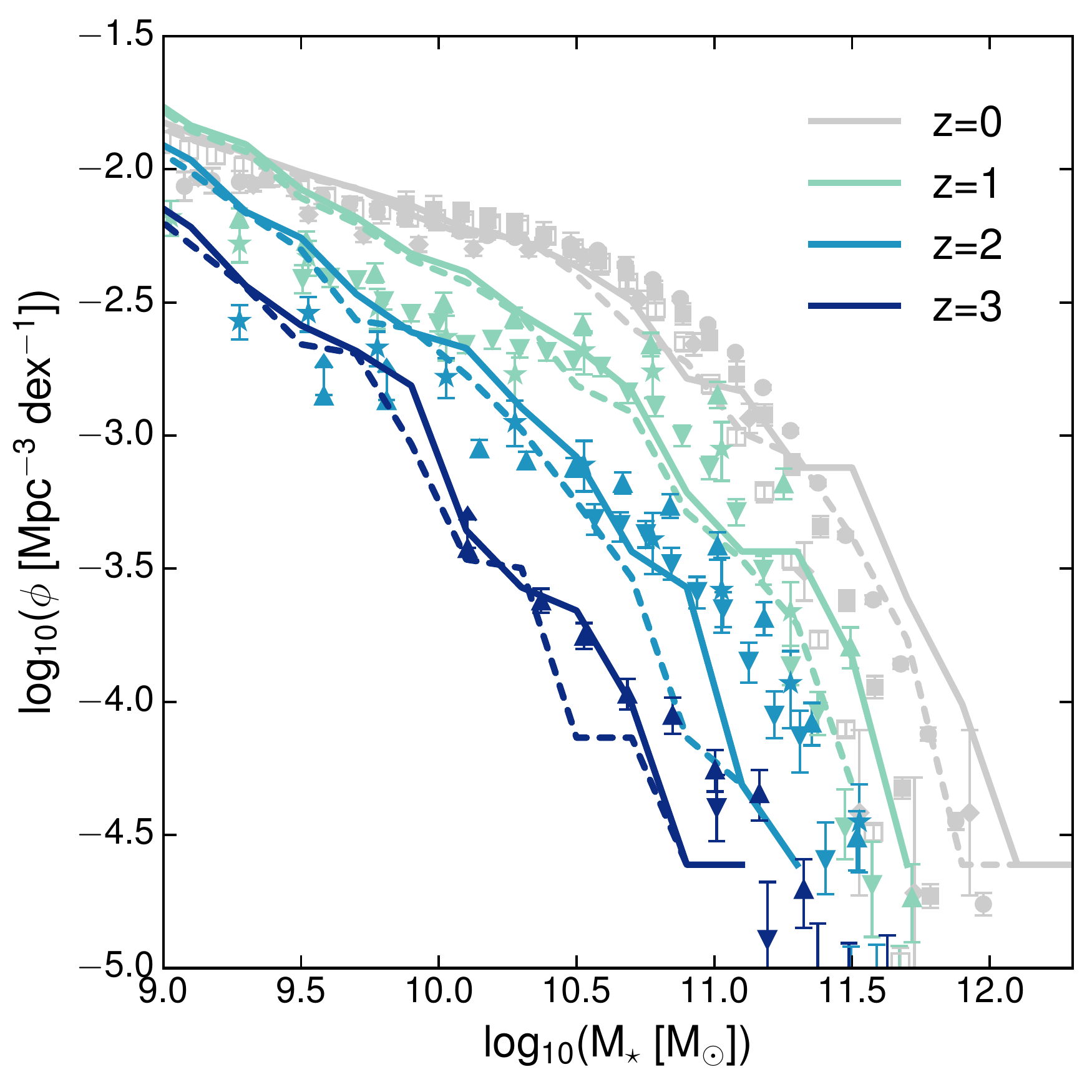}
        \caption{The galaxy stellar mass function at redshifts $0 \leq z \leq 3$ (lines) compared to observations (symbols with error bars). Two definitions are used for a simulated galaxy's stellar mass: all stellar mass bound to the subhalo (solid lines) and bound stellar mass within twice the stellar half-mass radius (dashed lines).
          The $z=0$ data are the same as in Fig.~\ref{fig:SMF_best}.
          At $z=1$ we compare to observed GSMFs in the redshift ranges $1.0 \leq z < 1.5$, $0.8 < z < 1.1$ and $1.0 < z < 1.25$ for \protect\cite{Muzzin2013} (downward triangles), \protect\cite{Ilbert2013} (upward triangles) and \protect\cite{Tomczak2013} (stars), respectively. At $z=2$ and $z=3$ we compare with the GSMFs for redshift ranges $2.0 < z < 2.5$ and $3.0 < z < 3.5$, respectively.
          Only stellar mass bins above the mass completeness limit are plotted in each case. Stellar masses have been converted to a \protect\cite{Chabrier2003} IMF where necessary by subtracting $0.25$ dex or $0.05$ dex for a \protect\cite{Salpeter1955} or \protect\cite{Kroupa2001} IMF, respectively.
        }
    \label{fig:SMF_z}
\end{figure}

In Fig.~\ref{fig:SMF_z} we plot the GSMF at $z \leq 3$ in comparison to observational data from \cite{Muzzin2013} and \cite{Ilbert2013}, two independent estimates both based on UltraVISTA DR1, and \cite{Tomczak2013}, for the FourStar Galaxy Evolution Survey (ZFOURGE).

We continue to have good agreement with the data beyond $z=0$.
Although the \fable\ model for AGN feedback has been calibrated to match the $z=0$ GSMF, the agreement is not guaranteed at higher redshifts.
The high mass end of the GSMF is in good agreement with the data at each of the redshifts shown, except for a slight underestimate at $z=2$. This may be due to small number statistics imposed by our finite box size, as the two highest occupied mass bins at $z=2$ contain only one or two galaxies.

At $z \geq 1$ the low mass end of the GSMF is somewhat overestimated, although this is not entirely unexpected given that this was also the case in Illustris. We have slightly altered the stellar feedback model of Illustris by implementing thermal winds, however this largely affects galaxies around the knee of the GSMF rather than low mass galaxies and any significant changes are limited to $z < 2$.

\section{Global Group and Cluster Properties}\label{sec:global}

\subsection{Stellar mass fractions}\label{subsec:star_fracs}

\begin{figure}
	\includegraphics[width=\columnwidth]{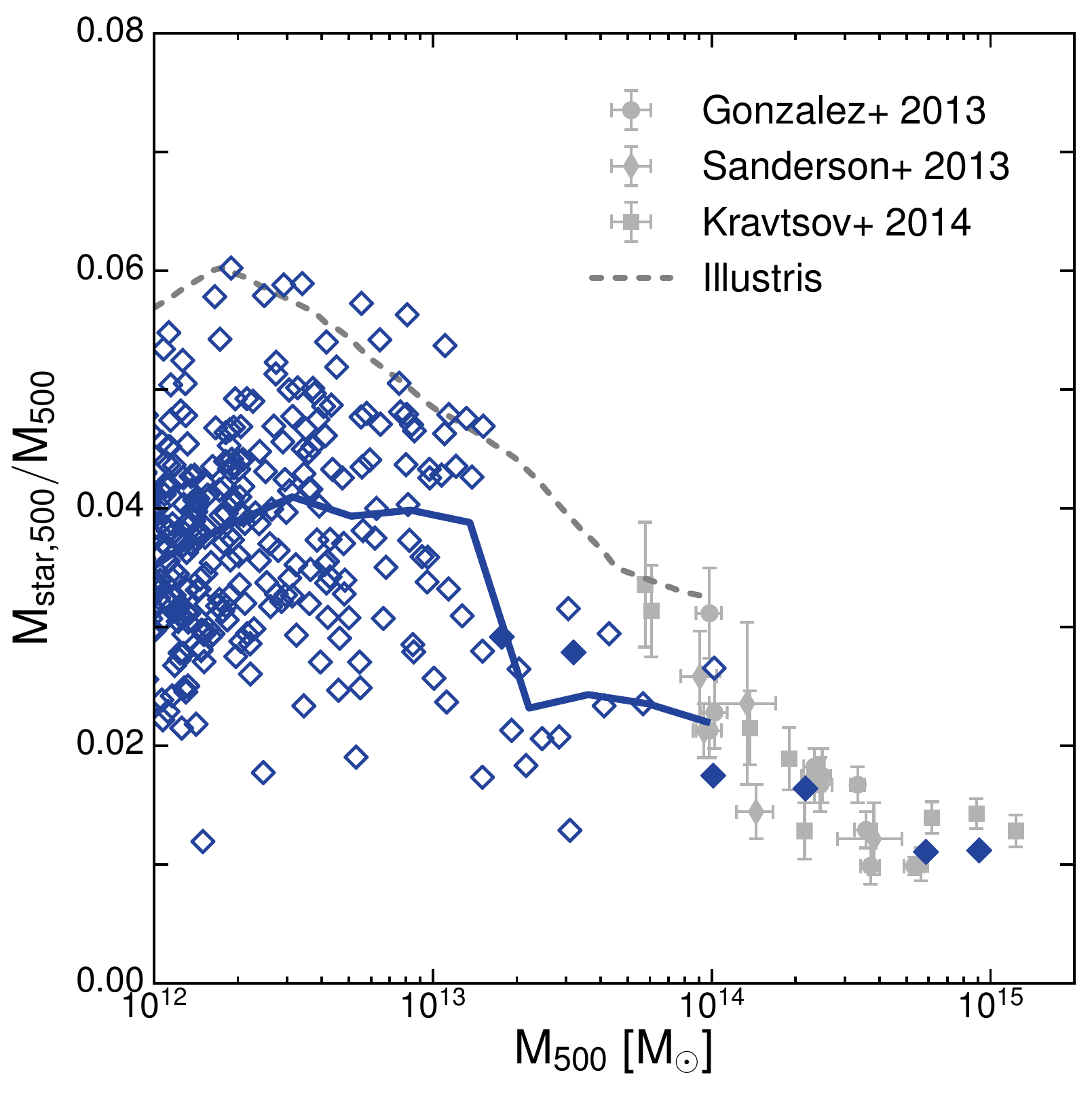}
        \caption{Stellar mass fraction within $r_{500}$ as a function of halo mass at $z=0$.
          Open diamonds show haloes from the $40\,h^{-1}$~Mpc periodic box while filled diamonds show the main halo of each zoom-in simulation.
          The solid line shows the mean relation in bins of halo mass for comparison with the mean relation from Illustris at $z=0$ (grey dashed line).
          Symbols with error bars show $z\simeq0$ observations from \protect\cite{Gonzalez2013}, \protect\cite{Sanderson2013} and \protect\cite{Kravtsov2014}. Following \protect\cite{Chiu2015}, the stellar mass measurements of \protect\cite{Gonzalez2013} and \protect\cite{Sanderson2013} have been reduced by 24 per cent to ensure all stellar masses are appropriate for a Chabrier IMF.
        }
    \label{fig:star_fracs}
\end{figure}

In Fig.~\ref{fig:star_fracs} we plot the stellar mass fraction within $r_{500}$ as a function of halo mass at $z=0$.
We consider all FoF haloes in the periodic volume (open diamonds), plus the main FoF halo in each of the zoom-in simulations (filled diamonds).
We do not discriminate between stars in satellite galaxies, the brightest central galaxy (BCG) or the ICL because the distinction between BCG and ICL is not well defined. We therefore compare to studies which take into account contributions from all three components \citep{Gonzalez2013, Sanderson2013, Kravtsov2014}.
The grey dashed line is the mean relation from the Illustris simulation \citep{Genel2014}.

The \fable\ simulations are a very good match to observed stellar mass fractions across a wide range of halo masses from galaxy groups (a few times $10^{13} M_{\odot}$) to high mass clusters ($\sim 10^{15} M_{\odot}$).
Although our AGN feedback model was tuned to reproduce the $z=0$ GSMF, the cosmological volume used in the tuning process contains only one halo with $M_{500} \sim 10^{14} M_{\odot}$. It is therefore reassuring that our model yields a realistic buildup of stellar mass even in dense cluster environments.
We reproduce the observed trend with halo mass, with an increase in the stellar fraction toward lower mass systems.
The relationship between stellar fraction and halo mass is similar between \fable\ and Illustris, with an offset in the normalisation. The difference in normalisation at $M_{500} \gtrsim 10^{13} M_{\odot}$ is due to more efficient suppression of star formation by AGN feedback in \fable, consistent with the offset in the GSMFs.
The offset at $M_{500} \lesssim 10^{13} M_{\odot}$ is partly due to our choice of larger gravitational softening lengths compared with Illustris and partly our change to the stellar feedback model.

We caution that the observations plotted in Fig.~\ref{fig:star_fracs} derive halo masses from X-ray data, which could potentially underestimate the true mass (see Section~\ref{subsec:bias}). Correcting for such a bias (if present) would shift the observational data to lower stellar fractions, away from our relation.
On the other hand, observations may underestimate the stellar mass of the BCG and associated ICL, which can contribute $\gtrsim 50$ per cent of the total stellar mass both observationally (e.g. \citealt{Gonzalez2013}) and in our simulations.
The characteristically diffuse emission of the ICL makes it particularly difficult to quantify and thus a non-negligible fraction of a cluster's total stellar content may be missed.

\subsection{Gas mass fractions}\label{subsec:gas_fracs}

\begin{figure}
	\includegraphics[width=\columnwidth]{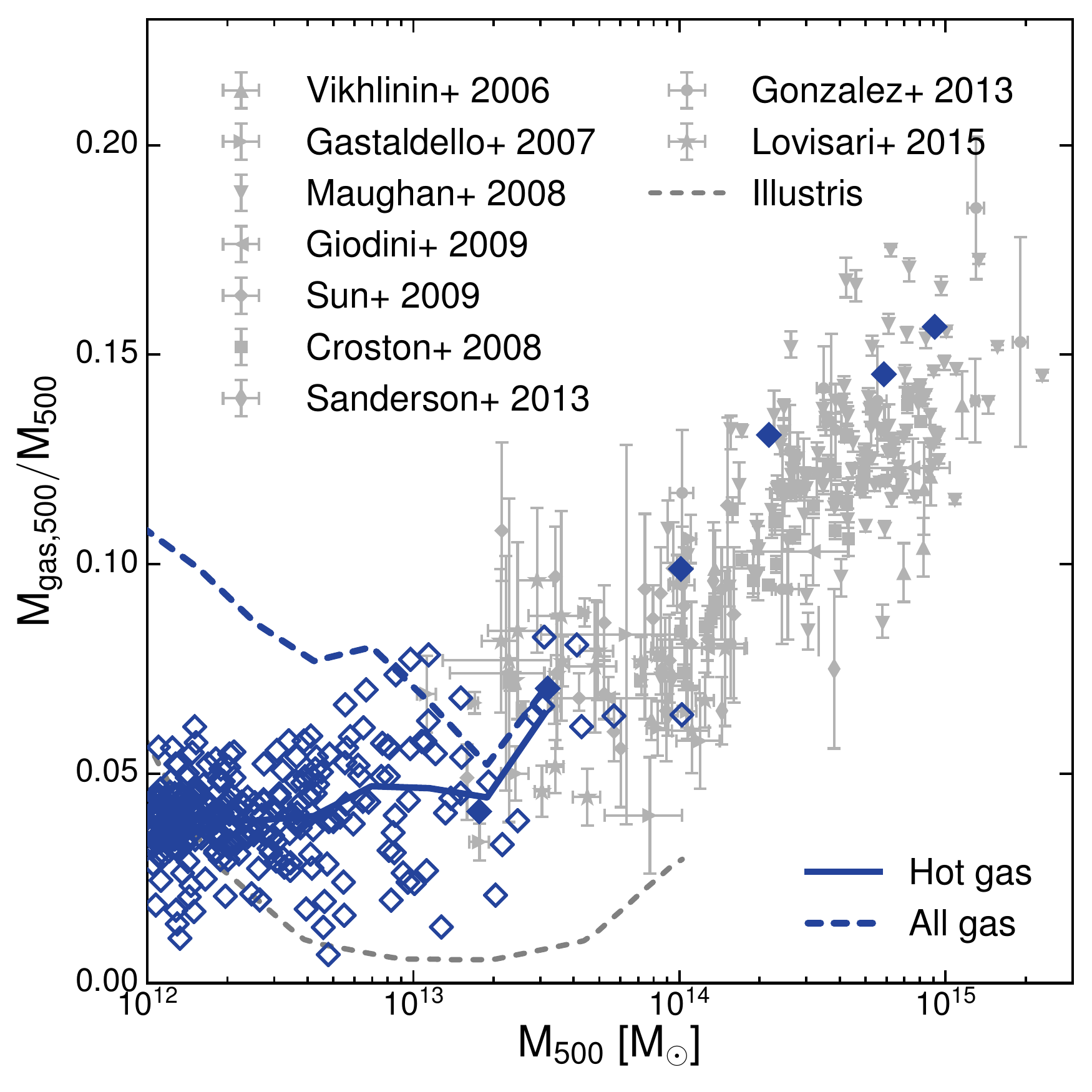}
        \caption{Gas mass fraction within $r_{500}$ as a function of halo mass at $z=0$. Marker styles are the same as in Fig.~\ref{fig:star_fracs}. The solid line shows the mean relation of these points in halo mass bins. Cold and multiphase gas has been excluded as described in Section~\ref{subsec:xray}. The dashed line shows the mean relation when all gas is included. The grey dashed line is the mean relation from Illustris at $z=0$ \citep{Genel2014}. Symbols with error bars show $z \simeq 0$ observations from \protect\cite{Vikhlinin2006}, \protect\cite{Gastaldello2007}, \protect\cite{Maughan2008}, \protect\cite{Giodini2009}, \protect\cite{Sun2009}, \protect\cite{Croston2008}, \protect\cite{Sanderson2013}, \protect\cite{Gonzalez2013} and \protect\cite{Lovisari2015}.
        }
    \label{fig:gas_fracs}
\end{figure}

In Fig.~\ref{fig:gas_fracs} we plot the total gas mass fraction within $r_{500}$ as a function of halo mass at $z=0$. The solid line shows the mean relation in bins of halo mass. We compare to a range of observational data (symbols with error bars) as well as the mean relation in Illustris (grey dashed line).
Since observed gas fractions are obtained from X-ray observations, we exclude cold gas and star-forming gas that is followed only with a simplified multiphase model, in accordance with our simulated X-ray spectra (see Section~\ref{subsec:xray}). For completeness, we also show the mean gas fraction--halo mass relation when all gas is included (dashed line). The difference becomes significant below $\sim 10^{13} M_{\odot}$, however there is very little available X-ray data at these masses.

We have calibrated the \fable\ AGN feedback model to observed gas fractions only for haloes in the $40\,h^{-1}$~Mpc periodic box, which are represented by the open diamonds in Fig.~\ref{fig:gas_fracs}. Although the model was not calibrated to cluster scales ($\gtrsim 10^{14} M_{\odot}$), the zoom-in simulations (filled diamonds) are in good agreement with the observations even in massive clusters ($\approx 10^{15} M_{\odot}$). This was certainly not guaranteed, as the much deeper potentials of clusters make it more difficult for AGN feedback to eject gas beyond $r_{500}$.
Indeed, the mean gas fractions of our simulated clusters do lie slightly above the mean observed relation, which suggests that the AGN feedback may need to be slightly more efficient at these high masses.
We also note that this offset would be exacerbated in the case of a significant X-ray mass bias, which would shift the observed data away from our relation.

At $M_{500} \approx 10^{13}-10^{14} M_{\odot}$ the mean gas mass fraction in \fable\ is significantly higher than in Illustris.
In the Illustris model, radio-mode AGN feedback ejected large gas masses out of massive haloes, resulting in significantly underestimated gas fractions.
In our updated model, radio-mode feedback events are more frequent but less energetic than in Illustris, and are therefore less able to eject gas from massive haloes. In Appendix~\ref{A:models} we show that radio-mode feedback in \fable\ is able to lower gas fractions in massive haloes but that it is also assisted by the quasar-mode, whose periodic rather than continuous feedback reduces gas fractions by suppressing the accumulation of gas at early times.

\subsection{X-ray Luminosity-Mass Relation}

\begin{figure*}
  \includegraphics[width=0.497\textwidth]{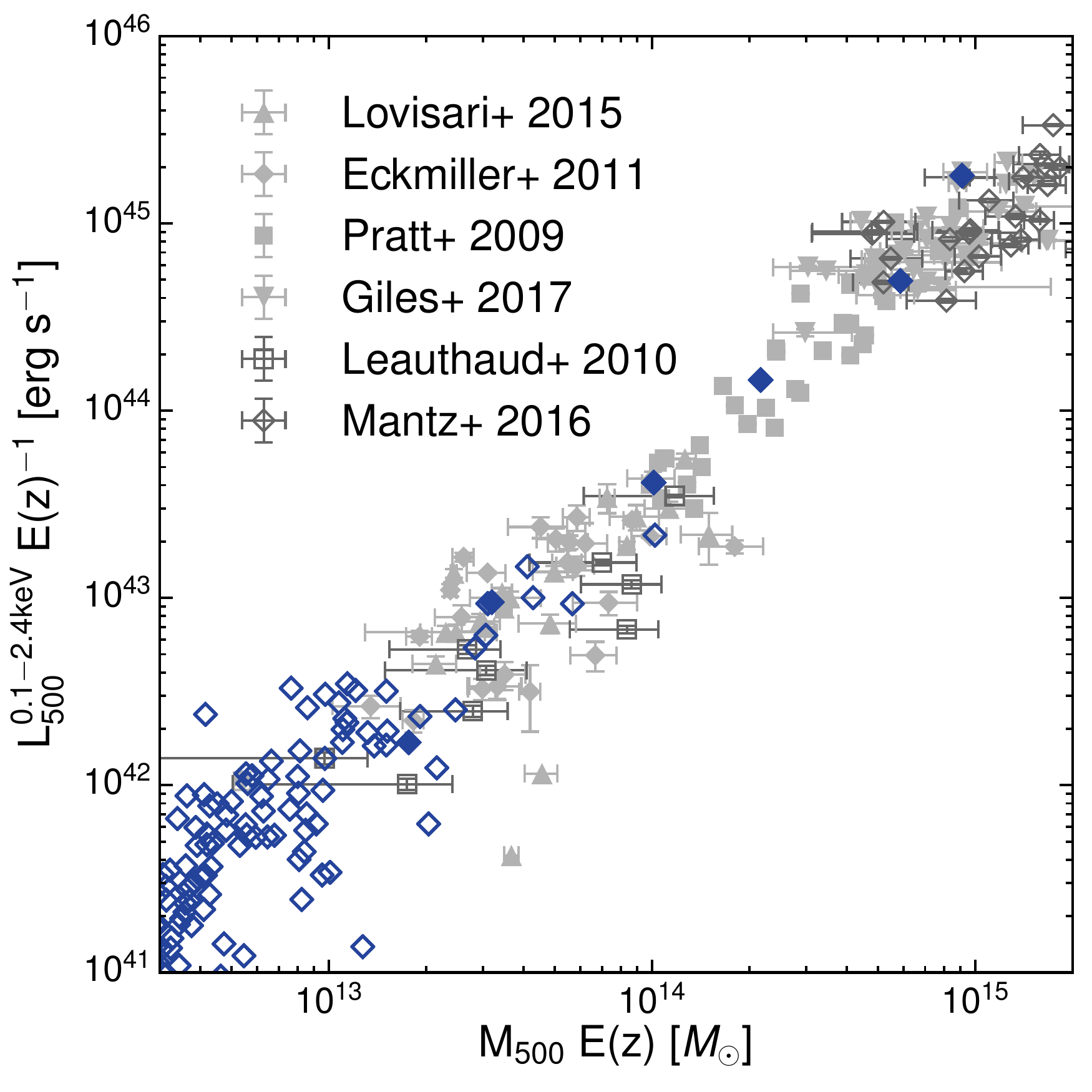}
  \includegraphics[width=0.497\textwidth]{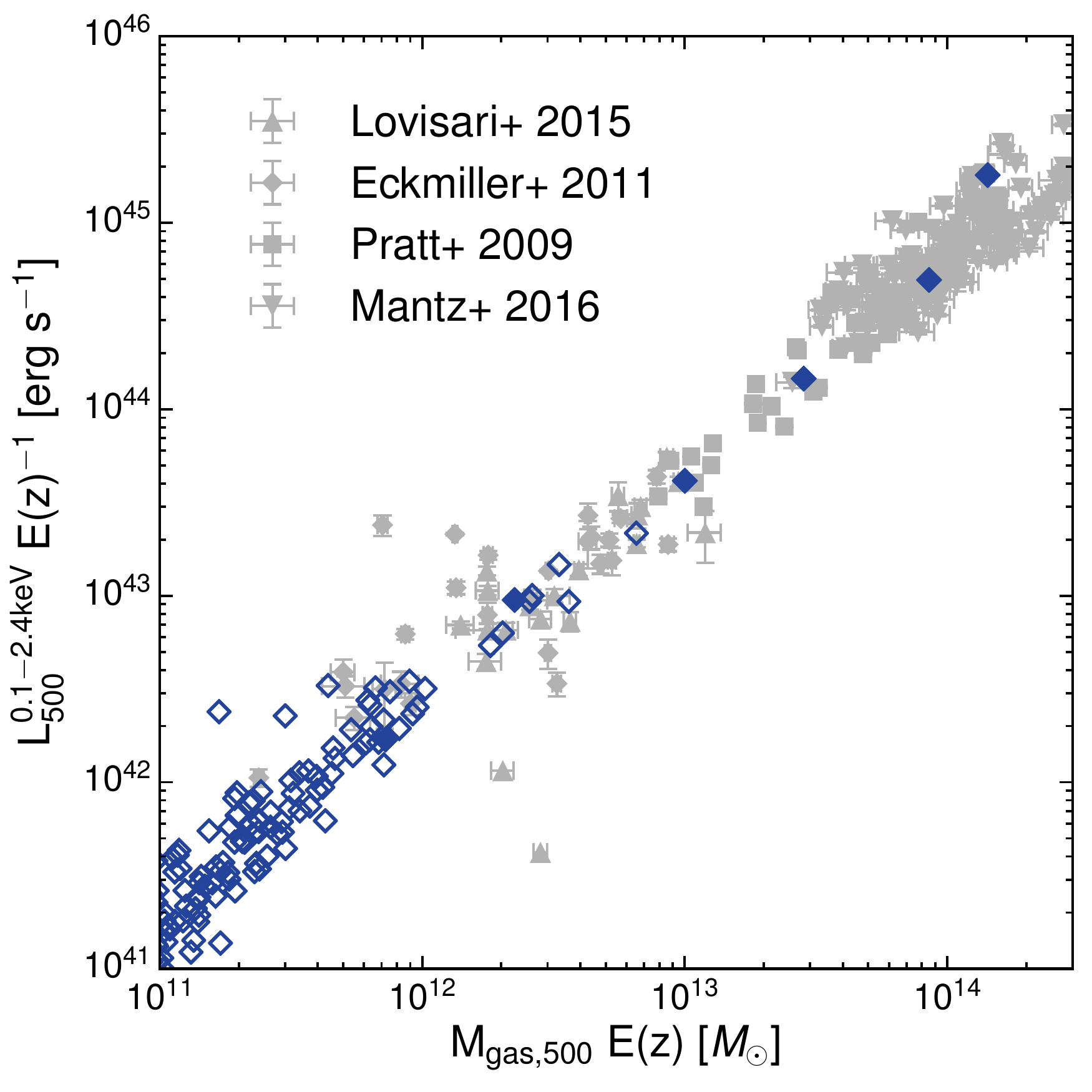}
  \caption{X-ray Luminosity in the $0.1-2.4$~keV band as a function of total mass (left) and gas mass (right) at $z=0$. Marker styles are the same as in Fig.~\ref{fig:star_fracs}.
    In the left hand panel, light grey symbols with error bars represent observational data for which total masses were estimated from X-ray observations assuming hydrostatic equilibrium \protect\citep{Lovisari2015, Eckmiller2011, Pratt2009, Giles2017}. Dark grey symbols with error bars show observational data for which total masses were estimated from weak gravitational lensing \protect\citep{Leauthaud2010, Mantz2016}.
    In the right hand panel we compare to X-ray luminosities and gas masses from \protect\cite{Lovisari2015}, \protect\cite{Eckmiller2011}, \protect\cite{Pratt2009} and \protect\cite{Mantz2016}.
        }
    \label{fig:Lsoft-M}
\end{figure*}

Here we present X-ray luminosity as a function of halo mass and gas mass in the \fable\ simulations in comparison with observations.
X-ray luminosities are calculated as described in Section~\ref{subsec:xray} in one of two bands: a soft X-ray band in the range $0.1-2.4$~keV and a bolometric band in the range $0.01-100$ keV.
We scale luminosity by $E(z)^{-1}$ and mass by $E(z)$, where $E(z) = \sqrt{ \Omega_{\rm M} (1+z)^3 + (1 - \Omega_{\rm M} - \Omega_{\rm \Lambda})(1+z)^2 + \Omega_{\rm \Lambda}}$ describes the redshift evolution of the Hubble parameter.
Halo masses are those measured directly from the simulation. The gas mass is the total mass of gas included in the creation of the synthetic X-ray spectrum, i.e., after excluding cold and multiphase gas (see Section~\ref{subsec:xray}).

In Fig.~\ref{fig:Lsoft-M} we plot the $L_{500}-M_{500}$ and $L_{500}-M_{\mathrm{gas}}$ relations in the soft X-ray band. At group-scales we compare to \cite{Lovisari2015}, a complete X-ray selected sample of 20 galaxy groups observed with \textit{XMM-Newton}, and \cite{Eckmiller2011}, a sample of 26 groups observed with \textit{Chandra}.
At the high mass end we compare to \cite{Pratt2009} who investigate the luminosity scaling relations of 31 local ($z<0.2$) clusters from the Representative XMM-Newton Cluster Structure Survey (REXCESS; \citealt{Bohringer2007}). REXCESS halo masses were estimated iteratively from the $M_{500}-Y_X$ relation of \cite{Arnaud2007} and gas masses are taken from \cite{Croston2008}.
For the $L_{500}-M_{500}$ relation we supplement the high mass end with data from \cite{Giles2017}, a complete sample of 34 galaxy clusters at $0.15 \leq z \leq 0.3$ observed with Chandra.
Each of these studies uses X-ray hydrostatic mass estimates.

We also compare to the weak lensing calibrated $L_{500}-M_{500}$ relations of \cite{Leauthaud2010} and \cite{Mantz2016}.
\cite{Leauthaud2010} measure mean halo masses via stacked weak gravitational lensing for a sample of 206 X-ray selected groups in the COSMOS field. We have rescaled their mean halo mass, $\left\langle M_{200} \right\rangle$, to the mass within $r_{500}$ using the best-fit NFW profile of each mass bin.
\cite{Mantz2016} measure X-ray luminosities from \textit{Chandra} and \textit{ROSAT} data for 27 clusters with weak lensing mass estimates as part of the \textit{Weighing The Giants} project \citep{Applegate2014, Kelly2014, VonderLinden2014a}.
We also plot the $L_{500}-M_{\mathrm{gas}}$ measurements of \cite{Mantz2016} for their full sample of 139 clusters.

Compared with observed $L_{500}-M_{500}$ relations based on X-ray hydrostatic masses, the \fable\ simulations are in excellent agreement over the full mass range from groups to massive clusters. However, there is tentative evidence that the predicted relation is overestimated compared with observations based on weak lensing masses.
This may point to the existence of an X-ray mass bias, which would manifest itself as a systematic difference between the $L_{500}-M_{500}$ relations derived from weak lensing masses and those derived from X-ray hydrostatic masses.
If a significant X-ray mass bias is indeed present, the observed halo gas mass fractions shown in Fig.~\ref{fig:gas_fracs} would be shifted to lower values, away from our relation. In this case, the \fable\ clusters would be too gas-rich and as a result their X-ray luminosities would be too high relative to the weak-lensing calibrated $L_{500}-M_{500}$ (assuming weak lensing masses are less biased).
Given that X-ray luminosity is very sensitive to the density distribution of the X-ray emitting gas, the luminosities of the simulated systems could be biased by high-density clumps with high X-ray emission. However, the excellent agreement with the observed $L_{500}-M_{\mathrm{gas}}$ relation in the right hand panel of Fig.~\ref{fig:Lsoft-M} suggests that this is not the case, i.e., that the gas content of our simulated haloes has a realistic clumping factor.

\begin{figure*}
  \includegraphics[width=0.497\textwidth]{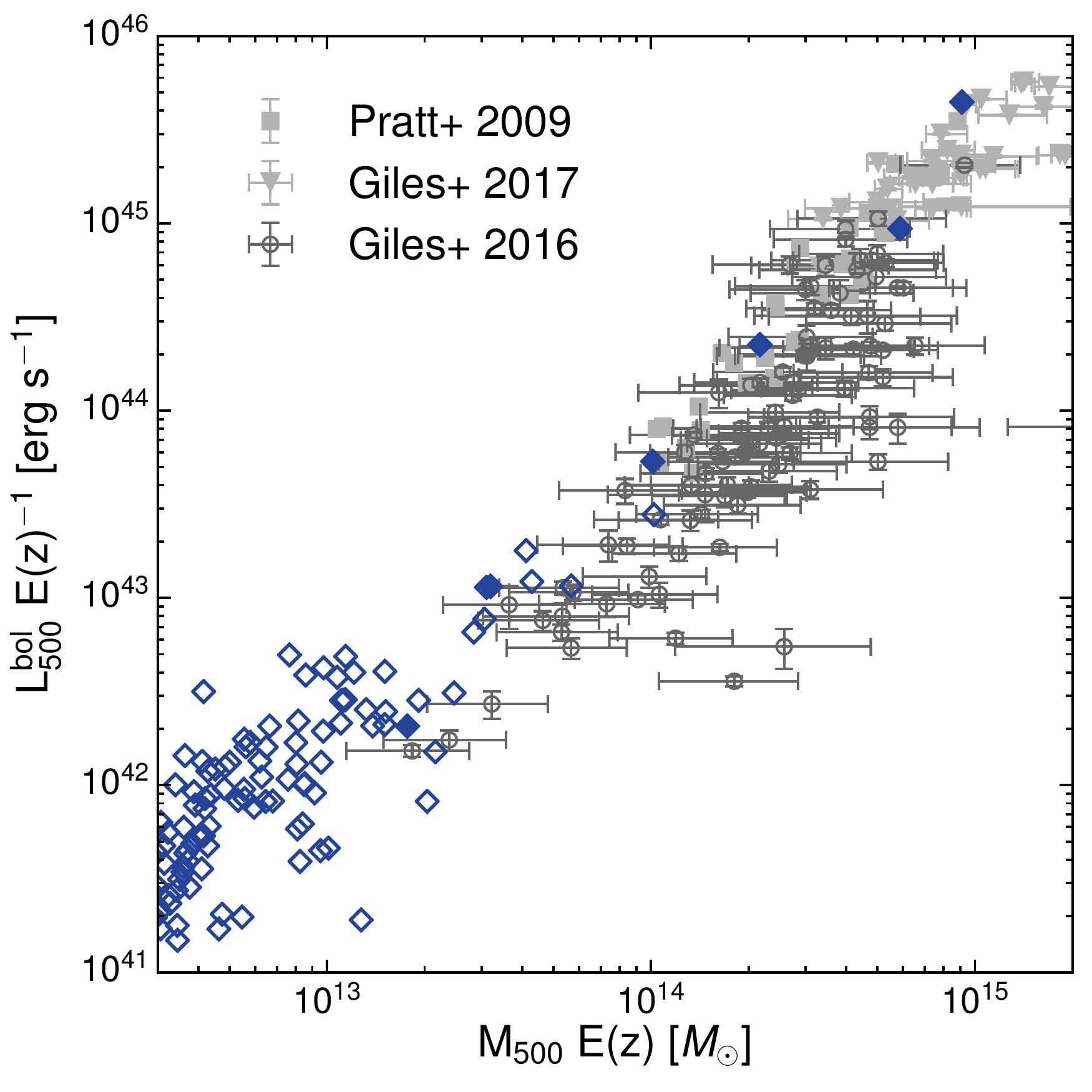}
  \includegraphics[width=0.497\textwidth]{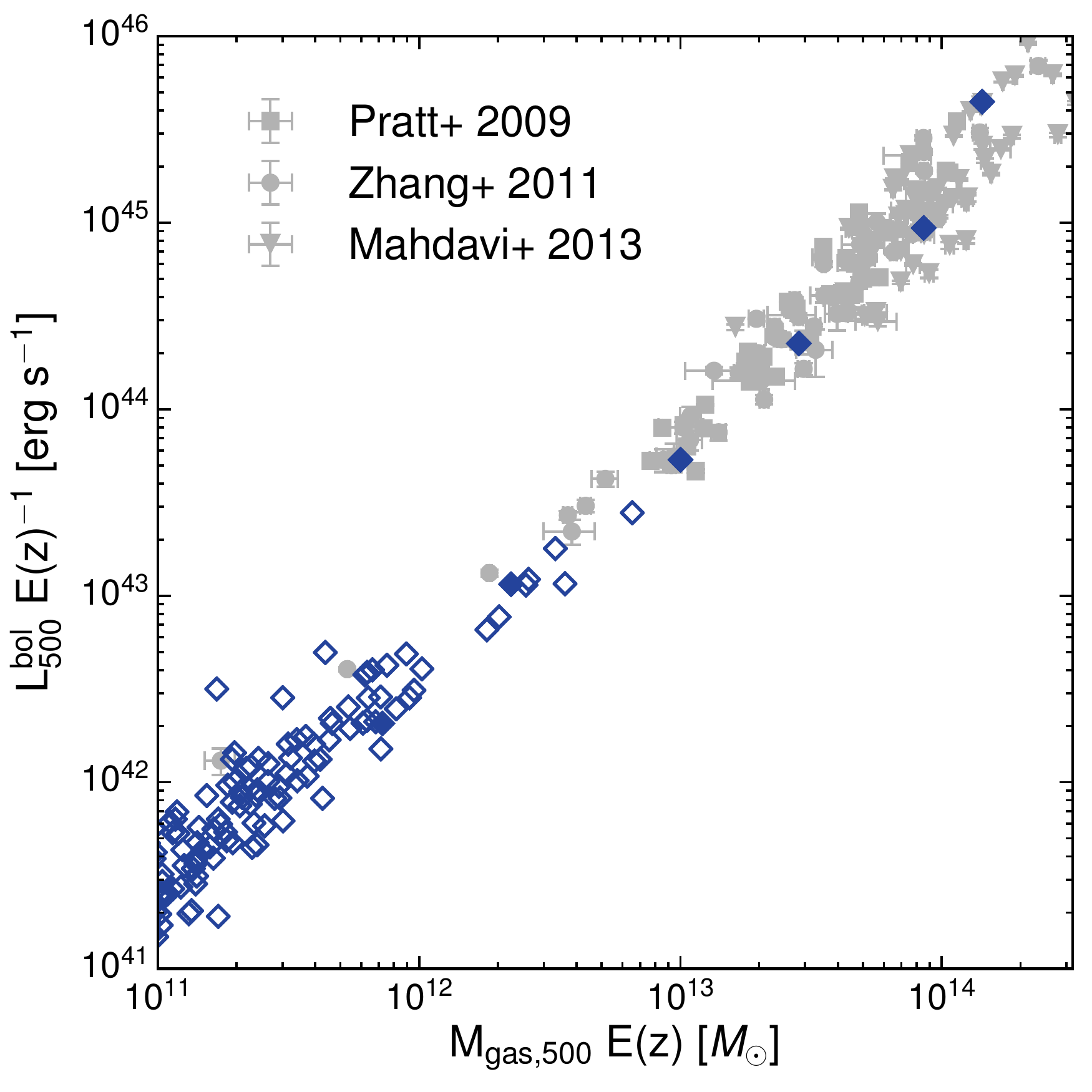}
  \caption{Bolometric ($0.01-100$ keV) X-ray luminosity as a function of total mass (left) and gas mass (right) at $z=0$.
    In the left hand panel, light grey symbols with error bars represent observational data based on X-ray hydrostatic masses \protect\citep{Pratt2009, Giles2017}. Dark grey symbols with error bars are the XXL-100-GC clusters \protect\citep{Giles2016} for which total masses were estimated from the internally calibrated weak lensing mass--temperature relation presented in \protect\cite{Lieu2016}.
    In the right hand panel we compare with data from \protect\cite{Pratt2009}, \protect\cite{Zhang2011a} and \protect\cite{Mahdavi2013}.
  }
    \label{fig:Lbol-M}
\end{figure*}

In Fig.~\ref{fig:Lbol-M} we show the $L_{500}-M_{500}$ and $L_{500}-M_{\mathrm{gas}}$ relations using the bolometric X-ray luminosity.
For the $L_{500}-M_{500}$ relation we compare to REXCESS \citep{Pratt2009} and \cite{Giles2017}, both of whom use X-ray hydrostatic masses. We also compare with weak lensing calibrated data from \cite{Giles2016} for the XXL-100-GC sample, which consists of the 100 brightest clusters in the XXL survey \citep{Pacaud2016}. Halo masses for the XXL-100-GC are estimated from the weak lensing mass--temperature relation presented in \cite{Lieu2016}.
For $L_{500}-M_{\mathrm{gas}}$ we complement the low mass end with data from \cite{Zhang2011a}, who analyse 62 galaxy clusters in the HIFLUGCS sample with \textit{XMM-Newton} and \textit{ROSAT} data.
At the high mass end of $L_{500}-M_{\mathrm{gas}}$ we compare with \cite{Mahdavi2013}, a sample of 50 clusters with weak lensing mass estimates and X-ray data from \textit{Chandra} or \textit{XMM-Newton}. Note that the luminosity and gas mass for this sample are measured within $r_{500}$ as derived from weak lensing.

Again we achieve excellent agreement with the relation based on X-ray hydrostatic masses.
However, as was hinted at in the soft band $L_{500}-M_{500}$ relation, there appears to be a systematic offset between observed bolometric $L_{500}-M_{500}$ relations based on weak lensing masses and those based on X-ray hydrostatic masses.
Under the assumption that weak lensing masses are less biased, this suggests that the X-ray luminosities of the \fable\ groups and clusters may be slightly too high for a given halo mass.
This conclusion is, however, complicated by the relatively large measurement uncertainties and overall scatter in weak lensing mass estimates compared to X-ray masses (see e.g. \citealt{Meneghetti2010}) and the relatively poor sampling of the $L_{500}-M_{500}$ plane by weak lensing data.

\subsection{X-ray Luminosity-Temperature relation}\label{subsec:L-T}

\begin{figure*}
  \includegraphics[width=0.497\textwidth]{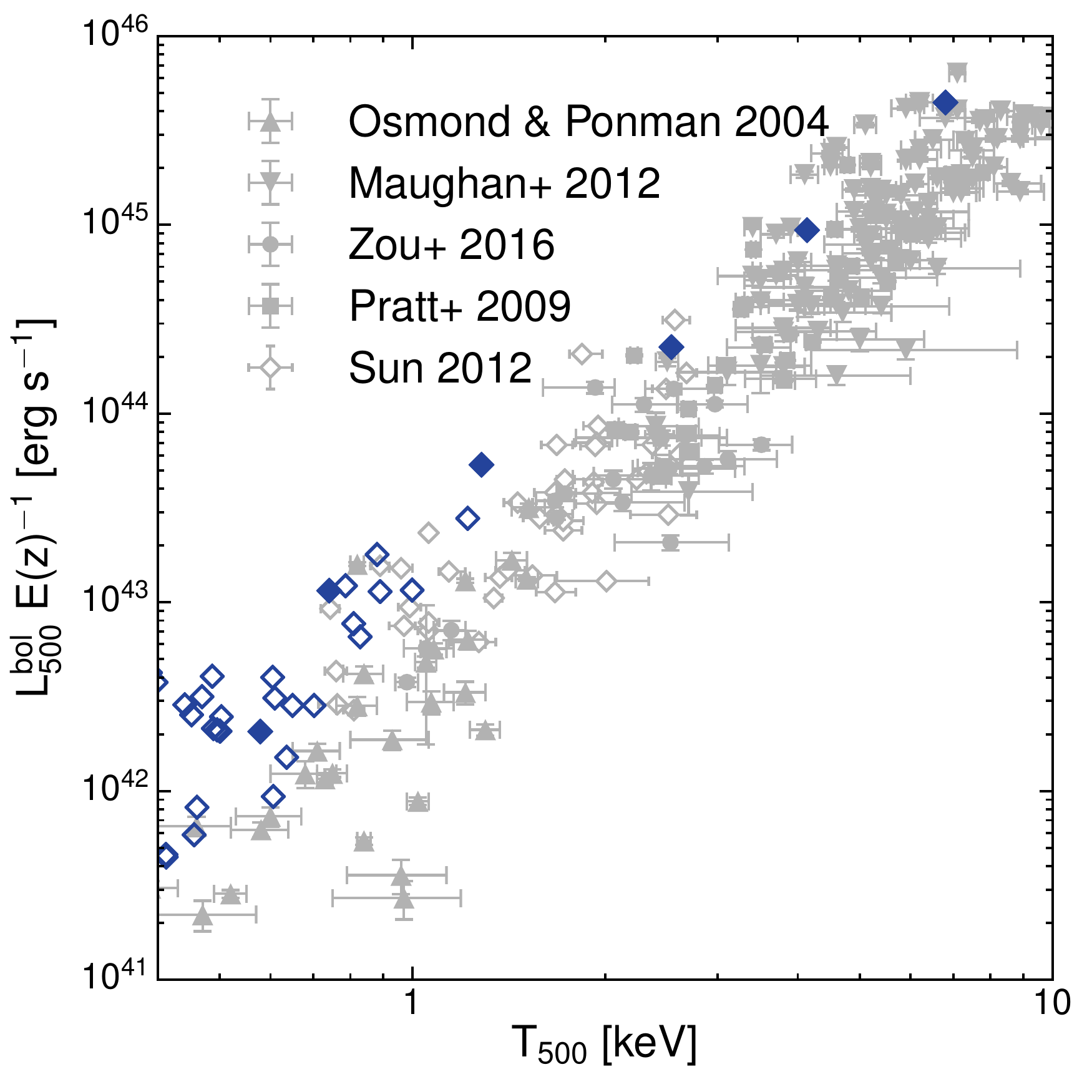}
  \includegraphics[width=0.497\textwidth]{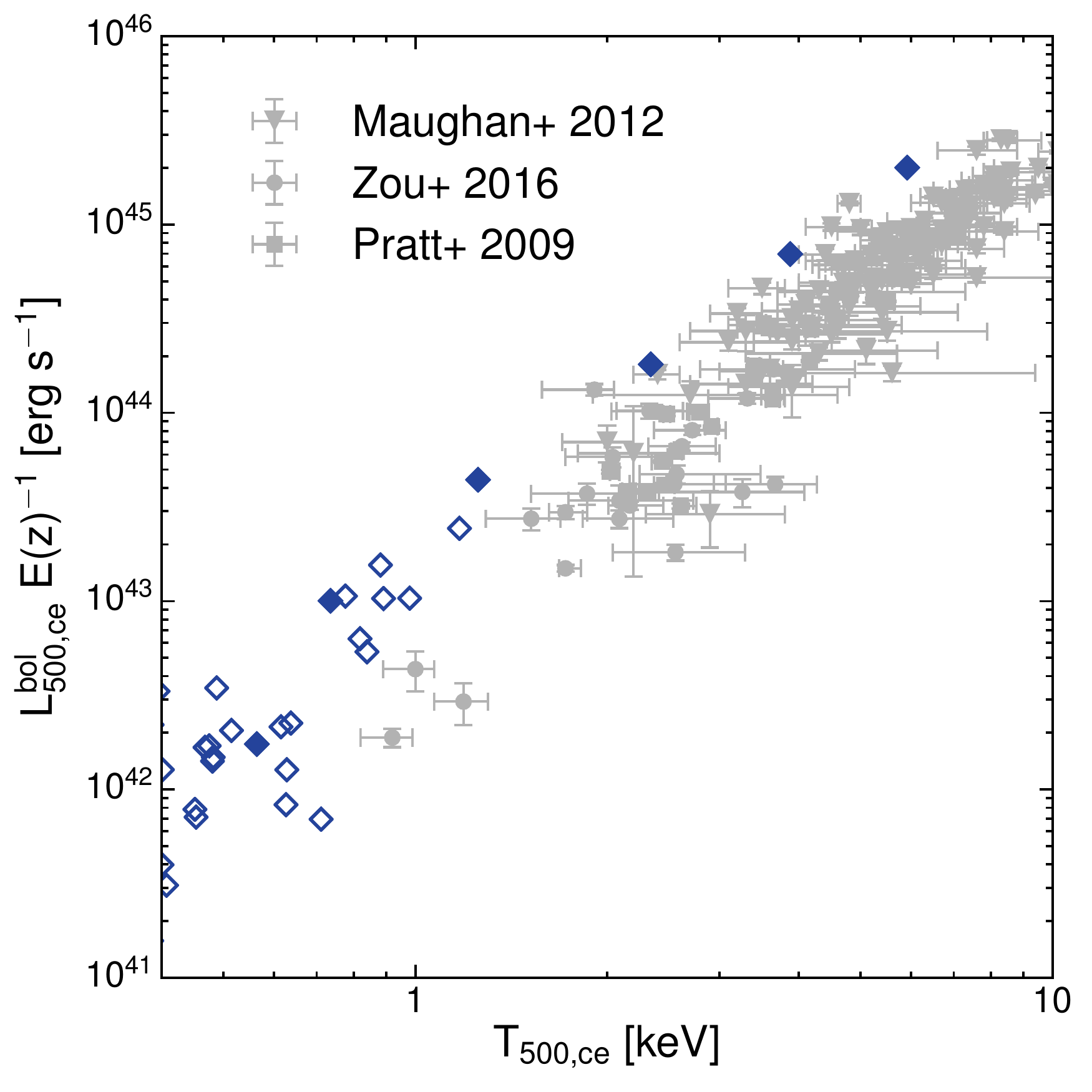}
  \caption{Bolometric ($0.01-100$ keV) X-ray luminosity as a function of spectroscopic temperature measured in the $(0-1)$~$r_{500}$ aperture (left) and the $(0.15-1)$~$r_{500}$ aperture (right) at $z=0$. Data from several X-ray studies are shown for comparison \protect\citep{Osmond2004, Maughan2012, Zou2016, Pratt2009, Sun2012}.
  }
  \label{fig:L-T}
\end{figure*}

In Fig.~\ref{fig:L-T} we plot bolometric X-ray luminosity as a function of X-ray spectroscopic temperature. In the left and right hand panels we consider luminosities and temperatures measured with and without the cluster core, respectively. We define a core radius of $0.15$~$r_{500}$ as used in the comparison studies.
We use the spectroscopic temperature estimates of \fable\ groups and clusters for consistency with the X-ray luminosity determinations, although we show in Appendix~\ref{A:temp_diff} that the difference between spectroscopic and mass-weighted temperature is small ($\lesssim 0.1$~dex) for temperatures above $\approx 0.5$~keV.

We compare against data from \cite{Osmond2004}, \cite{Maughan2012}, \cite{Zou2016}, REXCESS \citep{Pratt2009} and \cite{Sun2012}.
The \cite{Osmond2004} sample contains 35 systems with group-scale X-ray emission observed with \textit{ROSAT}. Although not statistically representative, the sample corresponds to some of the lowest temperature systems that have been observed.
The data from \cite{Maughan2012}, \cite{Zou2016} and \cite{Sun2012} each consist of groups and clusters observed with \textit{Chandra}.
The \cite{Maughan2012} sample consists of 114 clusters originally described in \cite{Maughan2008}, the \cite{Zou2016} sample is statistically complete and contains 23 groups and low-mass clusters from the 400d survey \citep{Burenin2007}, and
\cite{Sun2012} presents the X-ray luminosity--temperature relation of the \cite{Sun2009} sample of 43 galaxy groups with which we compare radial profiles of the ICM in Section~\ref{sec:profiles}.

We find that the \fable\ groups and clusters lie on the upper end of the scatter in the observations.
This is true whether or not the cluster core is excised, which implies that our X-ray luminosities or temperatures are not biased by, for example, an overabundance of dense cool cores.
At $T_{500} \sim 1$~keV we are actually in good agreement with the \cite{Sun2012} data, however, we expect that their sample is biased towards high X-ray luminosities due to their selection criteria, which require that the group emission can be traced to at least $r_{2500}$ ($\approx 0.47$ $r_{500}$; \citealt{Sun2009}).
Indeed, their average luminosity at $T_{500} \sim 1$~keV is notably higher than the \cite{Osmond2004} and \cite{Zou2016} samples.
Similarly, \cite{LeBrun2014} find that the \cite{Sun2009} sample has a significantly higher mean X-ray luminosity for groups with masses $M_{500} \sim 10^{13-13.5} M_{\odot}$ compared with the galaxy group studies of \cite{Osmond2004}, \cite{Rozo2009} and \cite{Leauthaud2010}.

It is not clear whether the small discrepancy between the simulation prediction and the observed relations is a result of overestimated total luminosities, underestimated global temperatures, or a combination of the two. The former explanation may be consistent with our comparison to weak lensing studies of the $L_{500}-M_{500}$ relation (Figs.~\ref{fig:Lsoft-M} and \ref{fig:Lbol-M}), for which there is some evidence that our luminosities are too high for a given halo mass.
However, the poor sampling and relatively large scatter of the weak lensing measurements means that we cannot rule out the possibility that the gas in our groups and clusters possess too-low average temperatures.
In the following section we therefore investigate the role of temperature using the total mass--temperature relation.

\subsection{Mass-Temperature relation}

\begin{figure*}
  \includegraphics[width=0.497\textwidth]{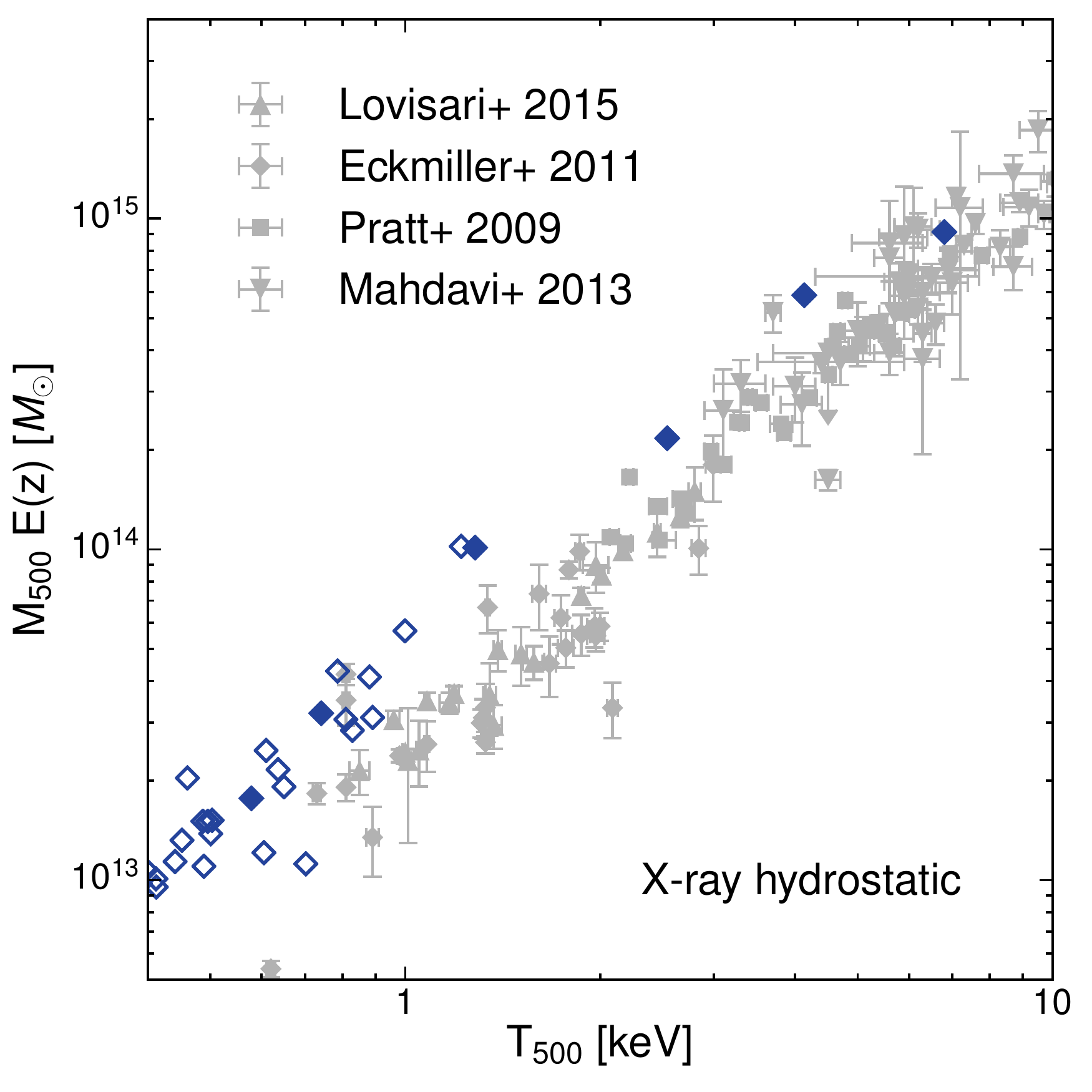}
  \includegraphics[width=0.497\textwidth]{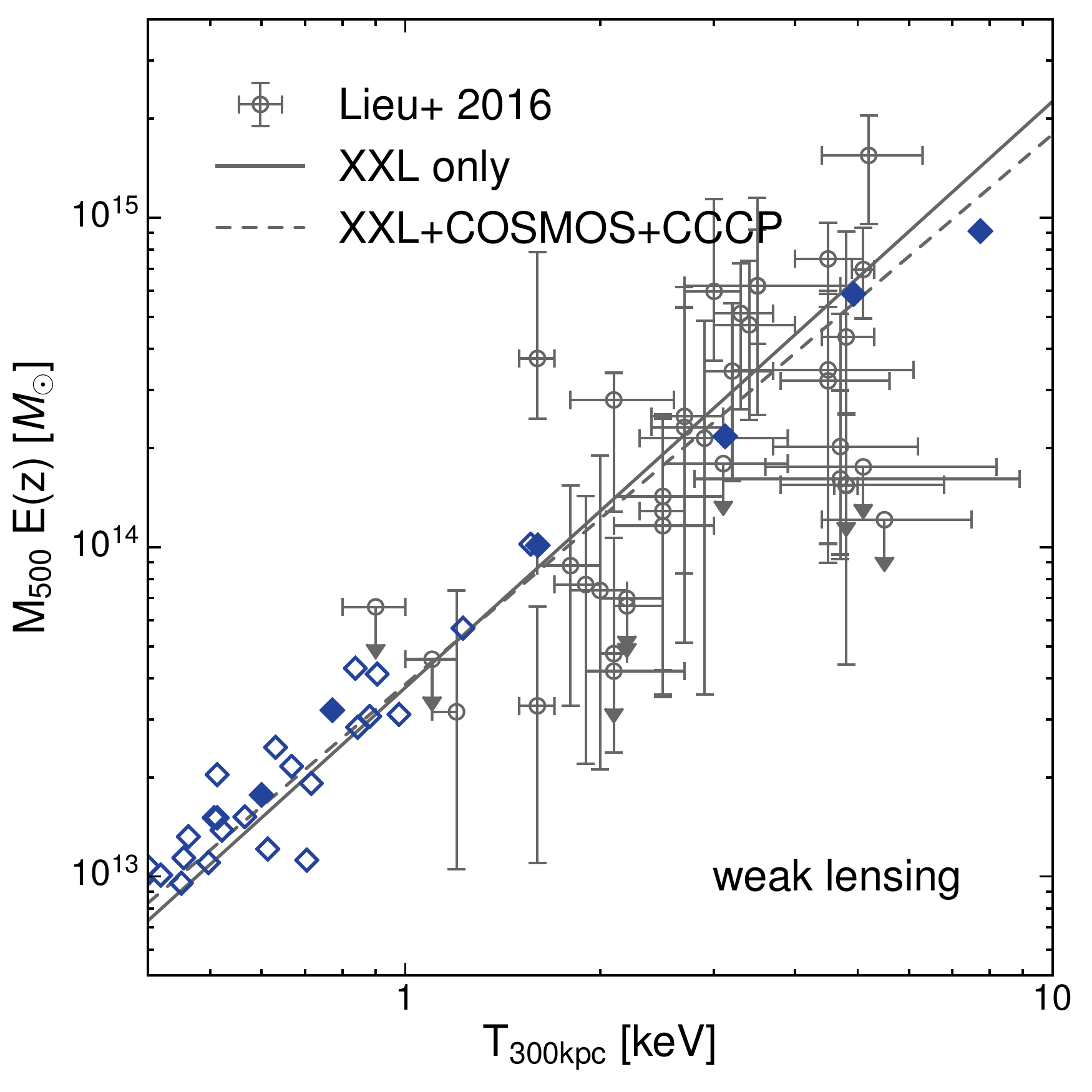}
  \caption{
    Total mass as a function of spectroscopic temperature compared to observations based on X-ray hydrostatic masses (left) and weak lensing masses (right).
    In the left hand panel temperatures are measured within a projected aperture of radius $r_{500}$. Symbols with error bars show data from \protect\cite{Lovisari2015}, \protect\cite{Eckmiller2011}, \protect\cite{Pratt2009} and \protect\cite{Mahdavi2013}.
    In the right hand panel we compare to \protect\cite{Lieu2016} whom measure weak-lensing masses for a sample of the brightest clusters in the XXL survey (symbols with error bars). We mimic \protect\cite{Lieu2016} by measuring temperatures within a projected aperture of radius $300$~pkpc. The solid line is the best-fit to the XXL data and the dashed line is the best-fit to an extended sample including additional groups and clusters from COSMOS \protect\citep{Kettula2013a} and CCCP \protect\citep{Hoekstra2015}.
  }
    \label{fig:M-T}
\end{figure*}

In Fig.~\ref{fig:M-T} we show the total mass as a function of spectroscopic temperature in \fable\ compared to observational samples that use either X-ray hydrostatic masses (left panel) or weak lensing masses (right panel).

For the X-ray mass comparison we show data from \cite{Lovisari2015}, \cite{Eckmiller2011}, REXCESS \citep{Pratt2009} and \cite{Mahdavi2013}.
We calculate the spectroscopic temperature within $r_{500}$, appropriate for REXCESS and \cite{Mahdavi2013}, but caution that the data of \cite{Lovisari2015} and \cite{Eckmiller2011}, which populate group-scale masses, cover only a fraction of $r_{500}$ and this fraction is not consistent between different systems.

For the weak lensing mass comparison we show the mass--temperature relation derived by \cite{Lieu2016} for 38 clusters from the XXL-100-GC sample that lie within the footprint of the Canada-France-Hawaii Telescope Lensing Survey (CFHTLenS), using the CFHTLenS shear catalogue for the mass measurements \citep{Heymans2012, Erben2013}. As in \cite{Lieu2016}, we measure spectroscopic temperatures within a projected aperture of radius 300~pkpc. We also show the best-fitting $M_{500}-T_{\mathrm{300kpc}}$ relation from \cite{Lieu2016} for an extended sample including 10 galaxy groups from COSMOS \citep{Kettula2013a} and 48 massive clusters from the Canadian Cluster Comparison Project (CCCP; \citealt{Hoekstra2015}).

We find that the \fable\ groups and clusters lie systematically above the $M_{500}-T_{500}$ relation derived from X-ray masses (left hand panel of Fig.~\ref{fig:M-T}). As the size of the offset is similar to the mean offset between the simulated and observed $L_{500}-T_{500}$ relations, both discrepancies could be the result of systematically low average temperatures.
On the other hand, the \fable\ systems are a good match to the weak lensing calibrated $M_{500}-T_{\mathrm{300kpc}}$ relation from \cite{Lieu2016} (right hand panel of Fig.~\ref{fig:M-T}), which would otherwise suggest that their average temperatures are not significantly biased.
Although the statistical uncertainties are large, the \cite{Lieu2016} $M_{500}-T_{\mathrm{300kpc}}$ relation has a significantly higher normalisation than the $L_{500}-T_{500}$ relations based on X-ray hydrostatic masses.
The difference in aperture may have a small effect at high temperatures, since a 300~pkpc aperture is smaller than $r_{500}$ for the majority of $T>1$~keV systems and could yield a slightly higher temperature measurement due to the declining temperature profiles of such systems.
However, this would lower rather than boost the normalisation of the $M_{500}-T$ relation.
Furthermore, the difference between these measures is small in our simulations ($\lesssim 0.1$ dex) and \cite{Giles2016} find no systematic differences between the two temperatures for the XXL-100-GC sample.
This implies that the offset between the X-ray and weak lensing calibrated $M_{500}-T$ relations is due to an offset in halo mass.
Under the assumption that weak lensing masses are unbiased, the \cite{Lieu2016} results suggest that X-ray hydrostatic masses are biased low and our agreement with their results suggests that \fable\ systems possess realistic global temperatures.

We find that the simulation predictions of the $M_{500}-T$ relation are rather robust with respect to variations of the feedback modelling. The better agreement of the simulations with the weak lensing calibrated relation may hence provide further circumstantial evidence that weak lensing mass measurements are less biased. We caution, however, that changes in the feedback modelling beyond those considered here may results in larger variations in the predicted normalisation.

\subsection{SZ-Mass relation}
\begin{figure}
	\includegraphics[width=\columnwidth]{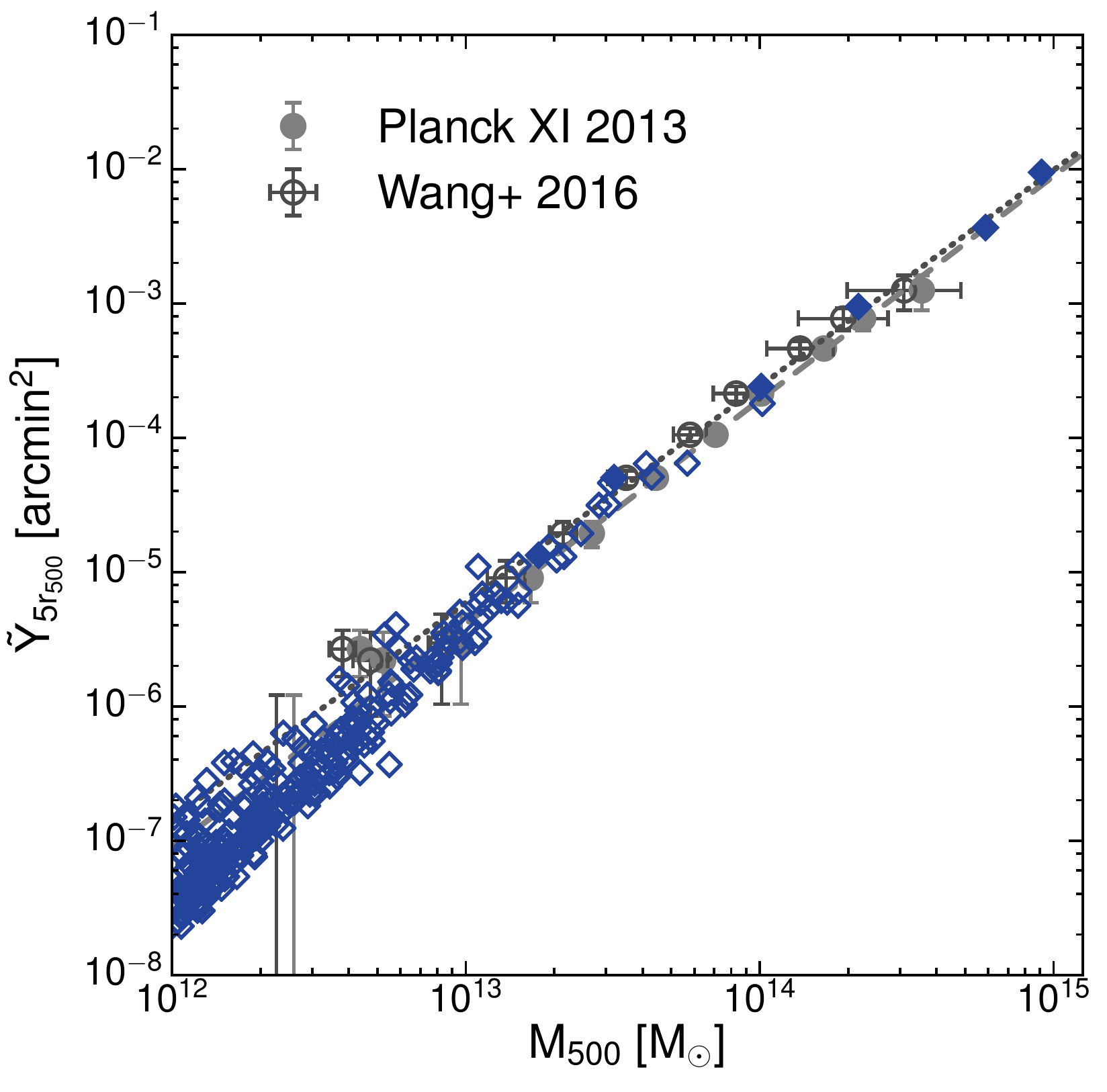}
        \caption{tSZ flux as a function of halo mass at $z=0$.
          The tSZ flux is calculated as described in Section~\ref{subsec:SZ} and has been scaled to $z=0$ and a fixed angular diameter distance of 500~Mpc.
	Symbols with error bars show the mean tSZ signal in bins of halo mass for a sample of SDSS locally brightest galaxies presented in \protect\cite{Planck2013XI}.
	Filled circles correspond to halo masses derived by \protect\cite{Planck2013XI} from a semi-analytic galaxy formation model and open circles show the recalibrated halo masses and associated uncertainties determined by \protect\cite{Wang2016} using stacked weak lensing analyses.
          The dashed line shows the best-fitting relation from \protect\cite{Planck2013XI} and the dotted line shows the best-fitting relation for the \protect\cite{Wang2016} recalibration.
        }
    \label{fig:SZ}
\end{figure}

In Fig.~\ref{fig:SZ} we plot the mean tSZ flux--halo mass relation calculated as described in Section~\ref{subsec:SZ}.
As in \cite{Planck2013XI}, we self-similarly scale the tSZ flux to redshift $z=0$ and to a fixed angular diameter distance of 500~Mpc, yielding the tSZ signal, $\mathrm{\tilde{Y}_{5r_{500}}}$, in units of square arcminutes.

The \fable\ simulations produce a power-law relation extending from massive galaxies to clusters in good agreement with both the original \cite{Planck2013XI} relation, which is based on halo masses derived from a semi-analytic galaxy formation model, and the weak lensing calibrated relation from \cite{Wang2016}.
At $\sim 5 \times 10^{12} M_{\odot}$ there is a slight upturn in the observed relation not seen in the simulations, however, these two mass bins correspond to detections of less than two sigma. Furthermore, \cite{Planck2013XI} state that the three lowest mass bins are noticeably affected by dust contamination and may be more uncertain than the statistical measures indicate.
Indeed, \cite{Greco2014a} explicitly model the dust emission from each LBG in the sample and find that, for the low-mass LBGs with $M_{500} \lesssim 10^{13.3} M_{\odot}$, the stacked signal from dust emission is comparable to or larger than the stacked tSZ signal.

\section{ICM profiles}\label{sec:profiles}

\begin{figure*}
  \includegraphics[width=0.497\textwidth]{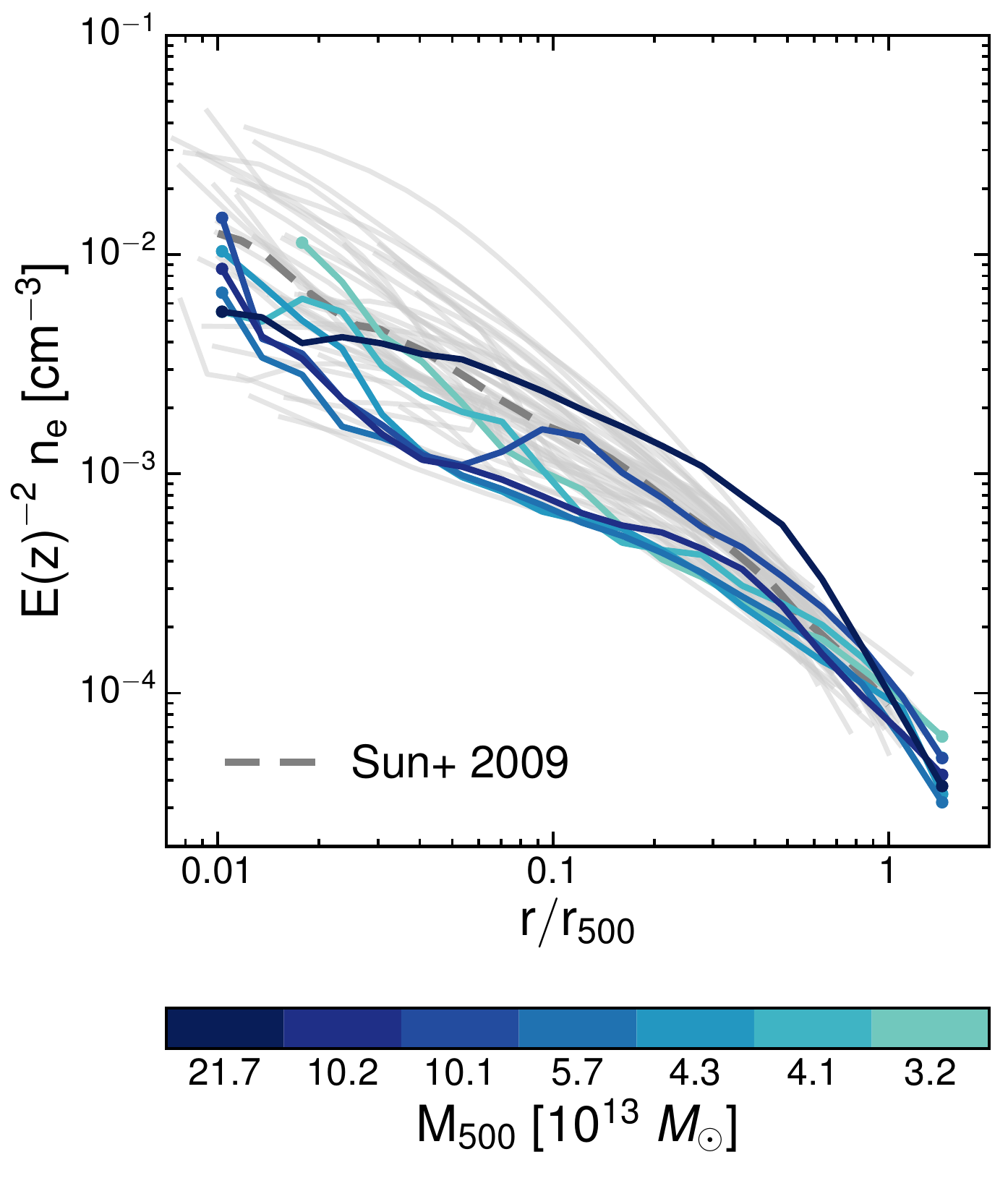}
  \includegraphics[width=0.497\textwidth]{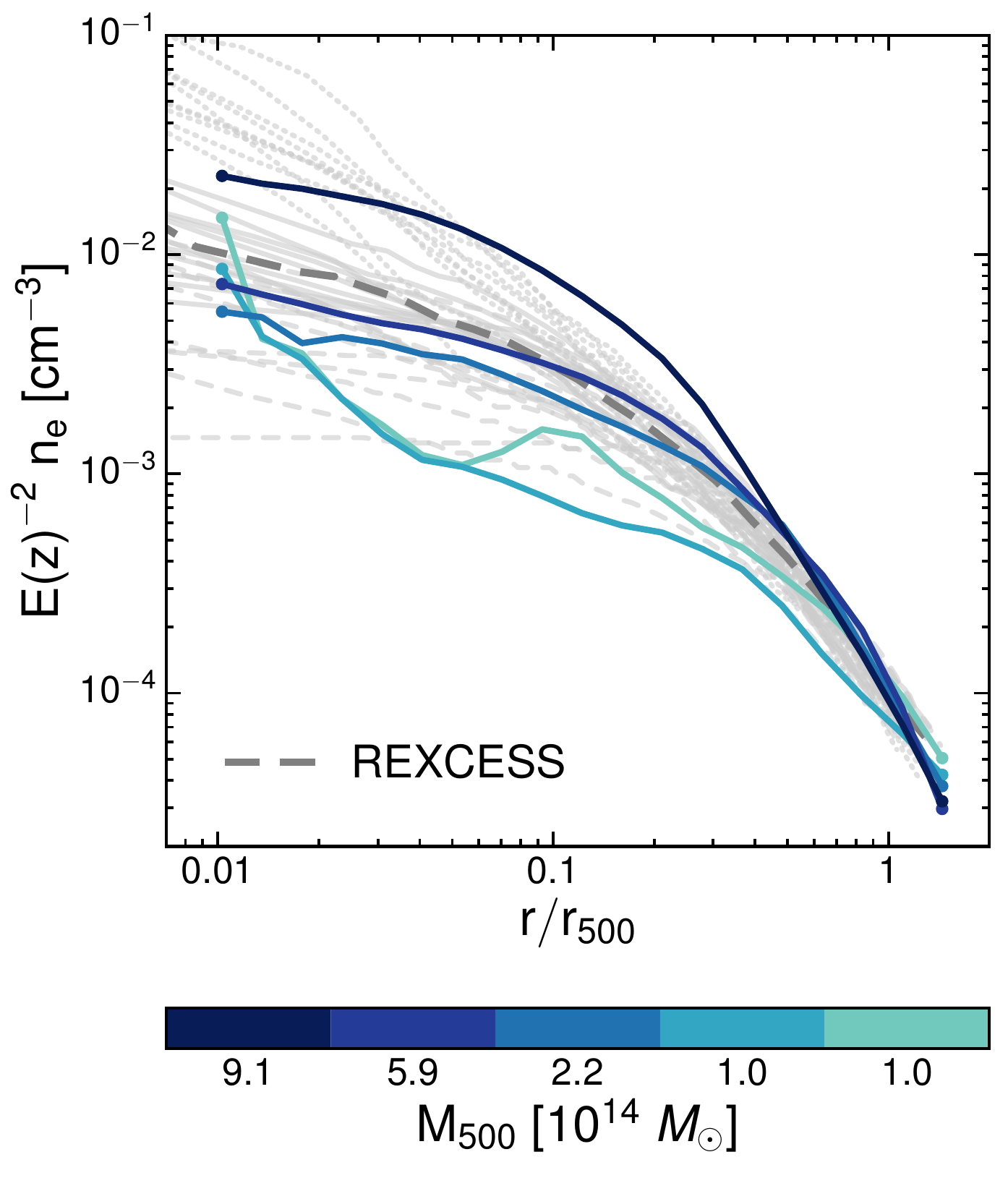}
  \caption{Density profiles of the ICM in \fable\ groups and clusters at $z=0$ in comparison to observed profiles. Lines show individual profiles of simulated systems colour coded by halo mass. All profiles have been self-similarly scaled in redshift. In the left hand panel we plot the density profiles of the \protect\cite{Sun2009} galaxy groups (grey lines) and the profiles of a mass-selected sample of simulated group-scale systems with a similar median halo mass (see main text).
          The thick dashed line shows the median of the \protect\cite{Sun2009} sample.
          In the right hand panel we show density profiles for all cluster-scale systems with $\mathrm{M}_{500} \geq 10^{14} M_{\odot}$ in comparison to those of the REXCESS cluster sample (grey lines; \protect\citealt{Croston2008}). For the observed sample, solid lines correspond to relaxed clusters, dashed lines to disturbed clusters and dotted lines to cool-core clusters according to the definitions of \protect\cite{Pratt2009}. The thick dashed line shows the median REXCESS profile.
        }
    \label{fig:density}
\end{figure*}

\begin{figure*}
	\includegraphics[width=0.497\textwidth]{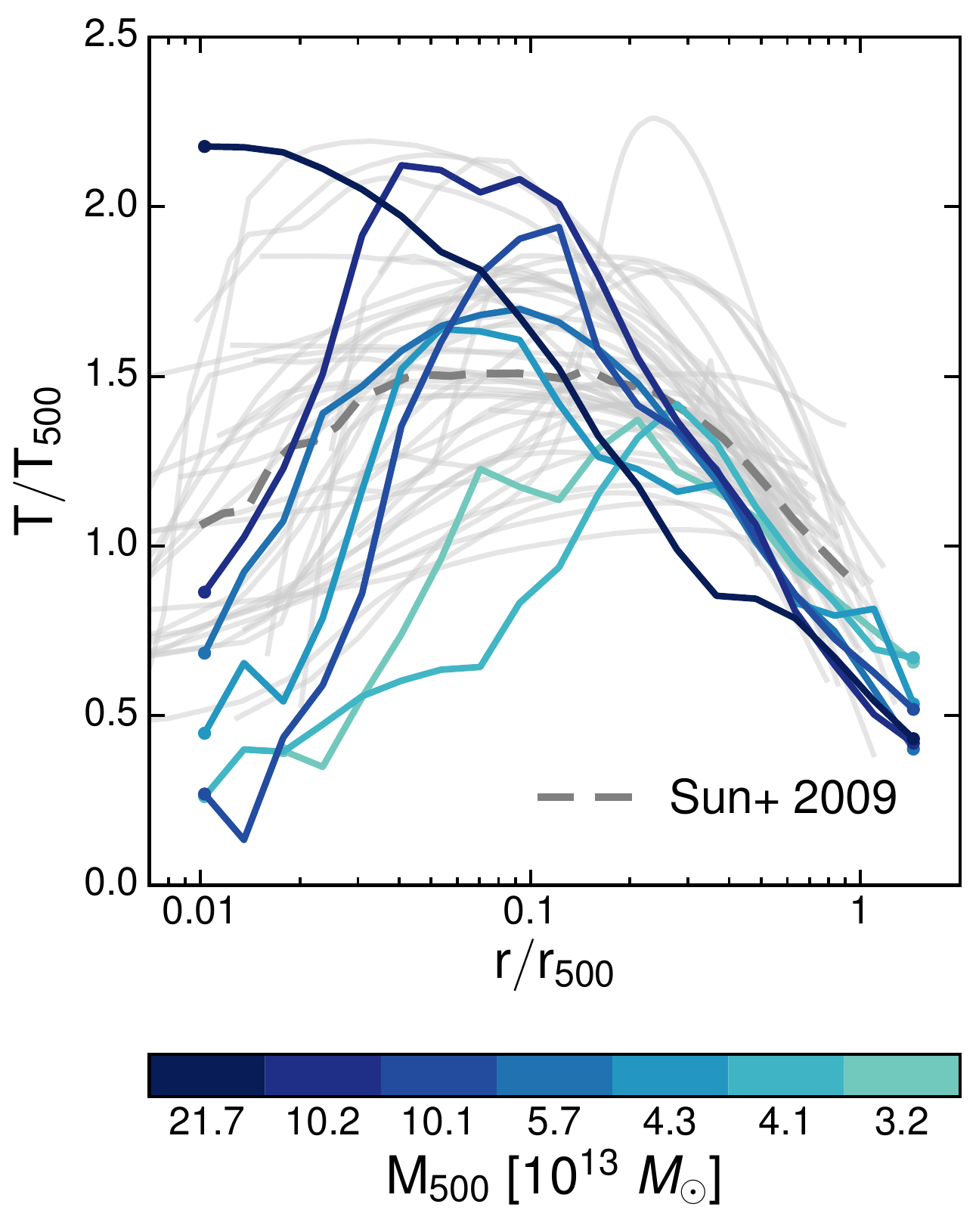}
	\includegraphics[width=0.497\textwidth]{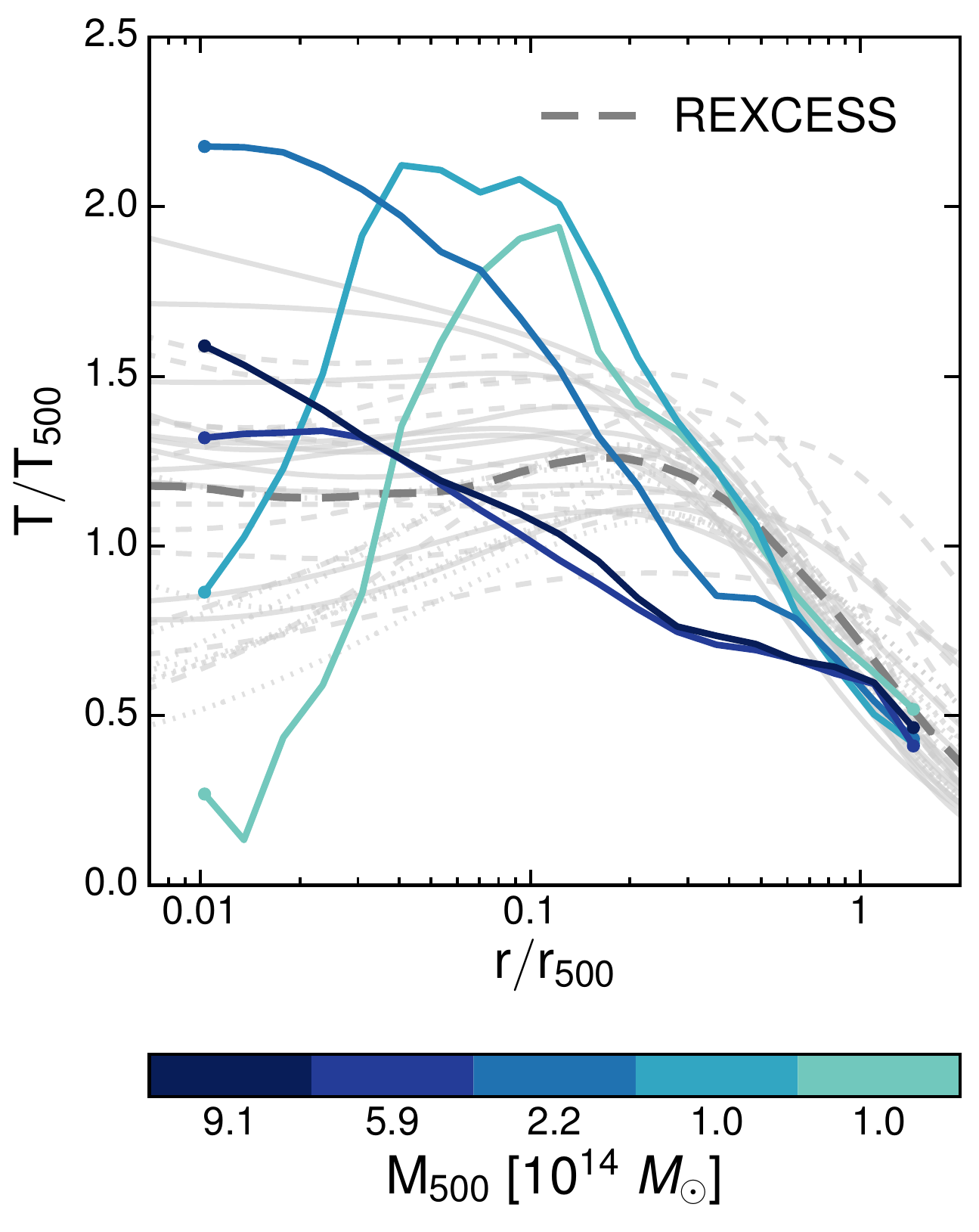}
        \caption{Dimensionless temperature profiles at $z=0$ for the same simulated and observed samples as shown in Fig.~\ref{fig:density}. Profiles are normalised by the characteristic temperature defined in equation~\ref{eq:T500}.
          In the left hand panel, solid grey lines show the individual deprojected temperature profiles of the \protect\cite{Sun2009} sample. The thick dashed line shows the median profile.
          In the right hand panel, grey lines show individual temperature profiles for REXCESS clusters derived from the best-fitting pressure \protect\citep{Arnaud2010} and entropy \protect\citep{Pratt2010} profiles. Line styles are the same as shown in Fig.~\ref{fig:density}.
        }
    \label{fig:temp}
\end{figure*}

\begin{figure*}
	\includegraphics[width=0.497\textwidth]{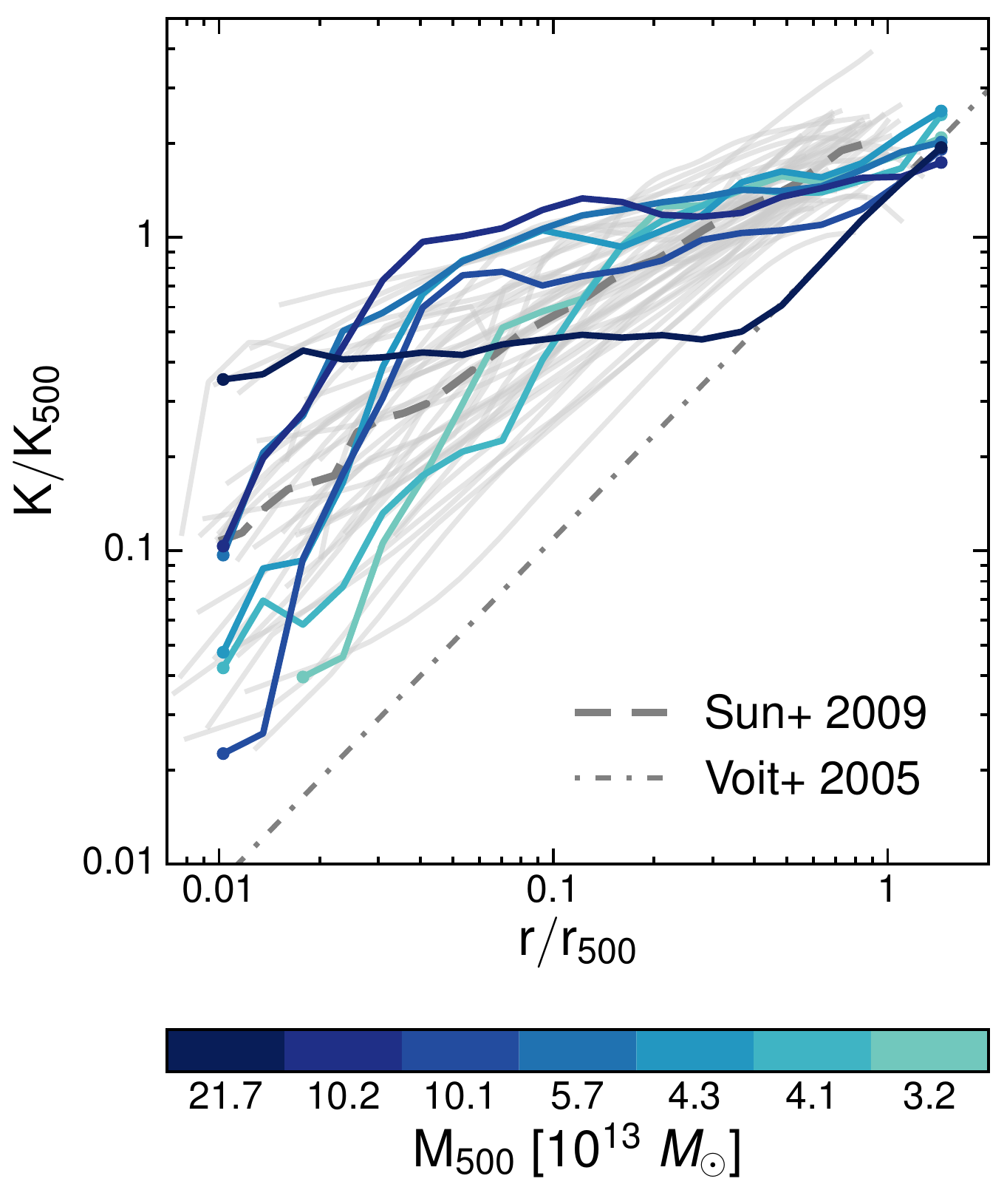}
	\includegraphics[width=0.497\textwidth]{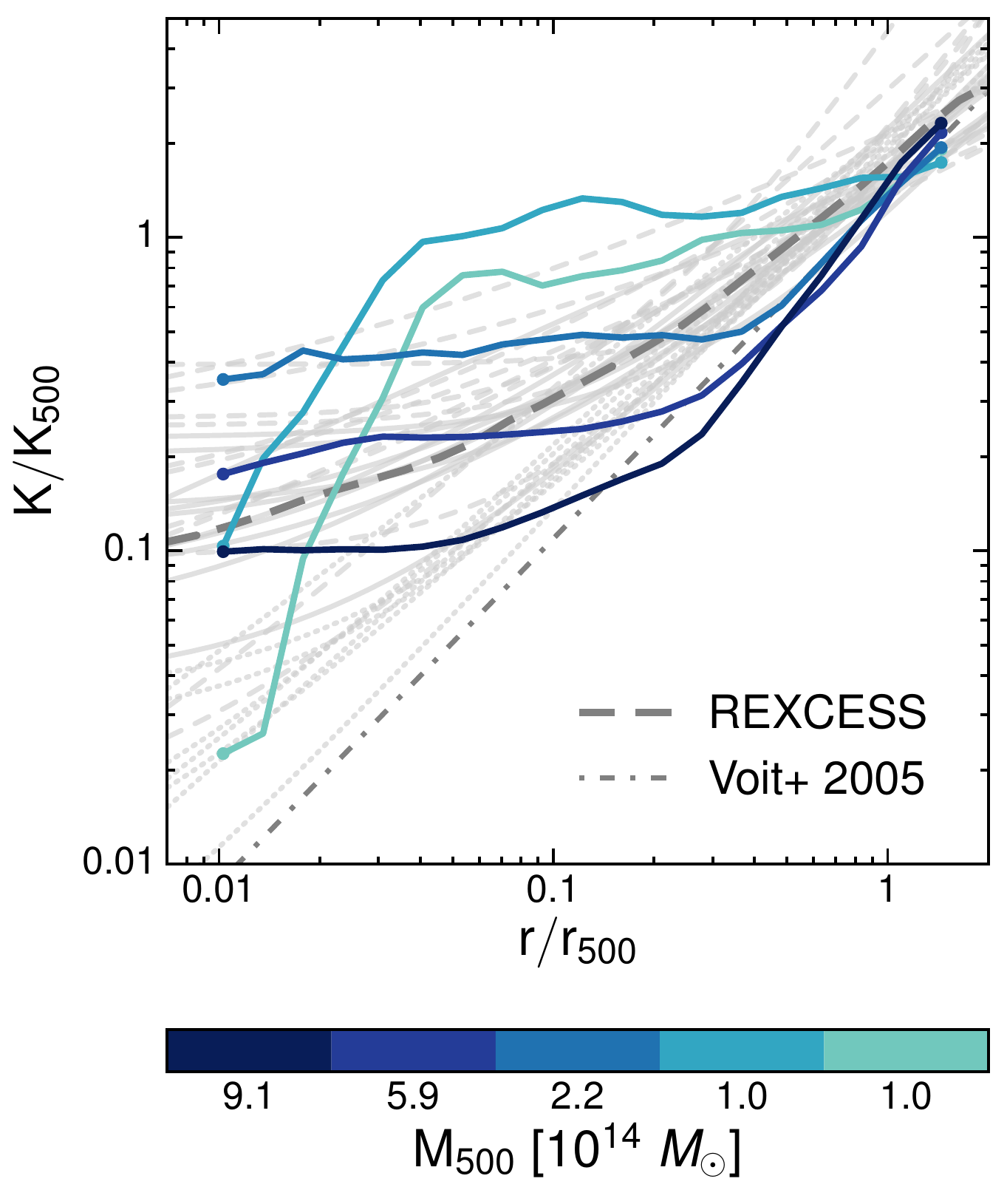}
        \caption{Dimensionless entropy profiles at $z=0$ for the same simulated and observed samples as shown in Fig.~\ref{fig:density}.
          Profiles are scaled by the characteristic entropy scale $K_{500}$ as defined in the text. The dash-dotted line shows the baseline ICM entropy profile derived by \protect\cite{Voit2005a} from non-radiative simulations.
          In the left hand panel, grey lines show the entropy profiles of the \protect\cite{Sun2009} groups derived from the density and temperature profiles shown in Fig.~\ref{fig:density} and \ref{fig:temp}.
          In the right hand panel, grey lines show the best-fitting entropy profiles of the REXCESS clusters \protect\citep{Pratt2010}. Line styles are the same as shown in Fig.~\ref{fig:density}.
        }
    \label{fig:entropy}
\end{figure*}

\begin{figure*}
  \includegraphics[width=0.497\textwidth]{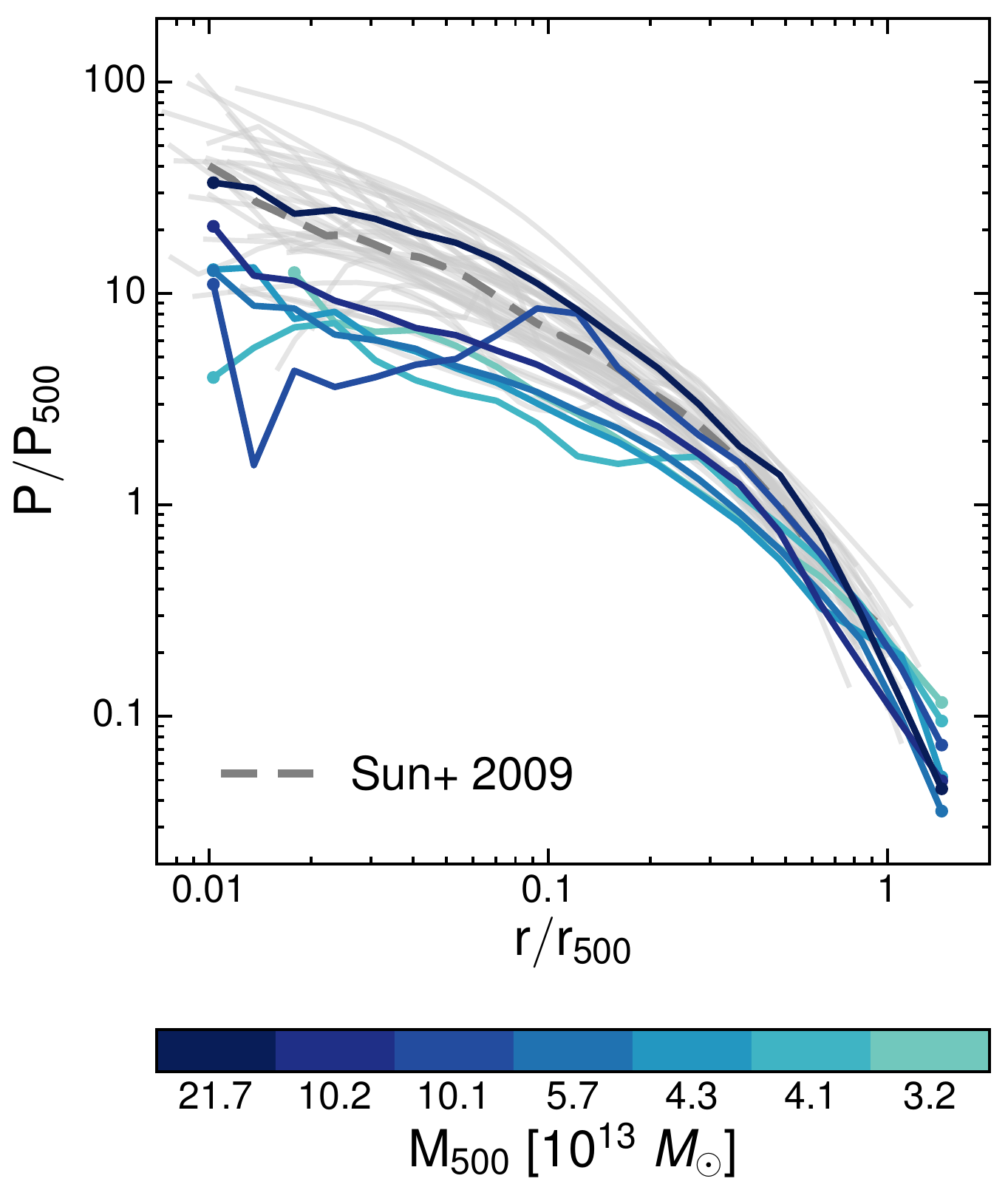}
  \includegraphics[width=0.497\textwidth]{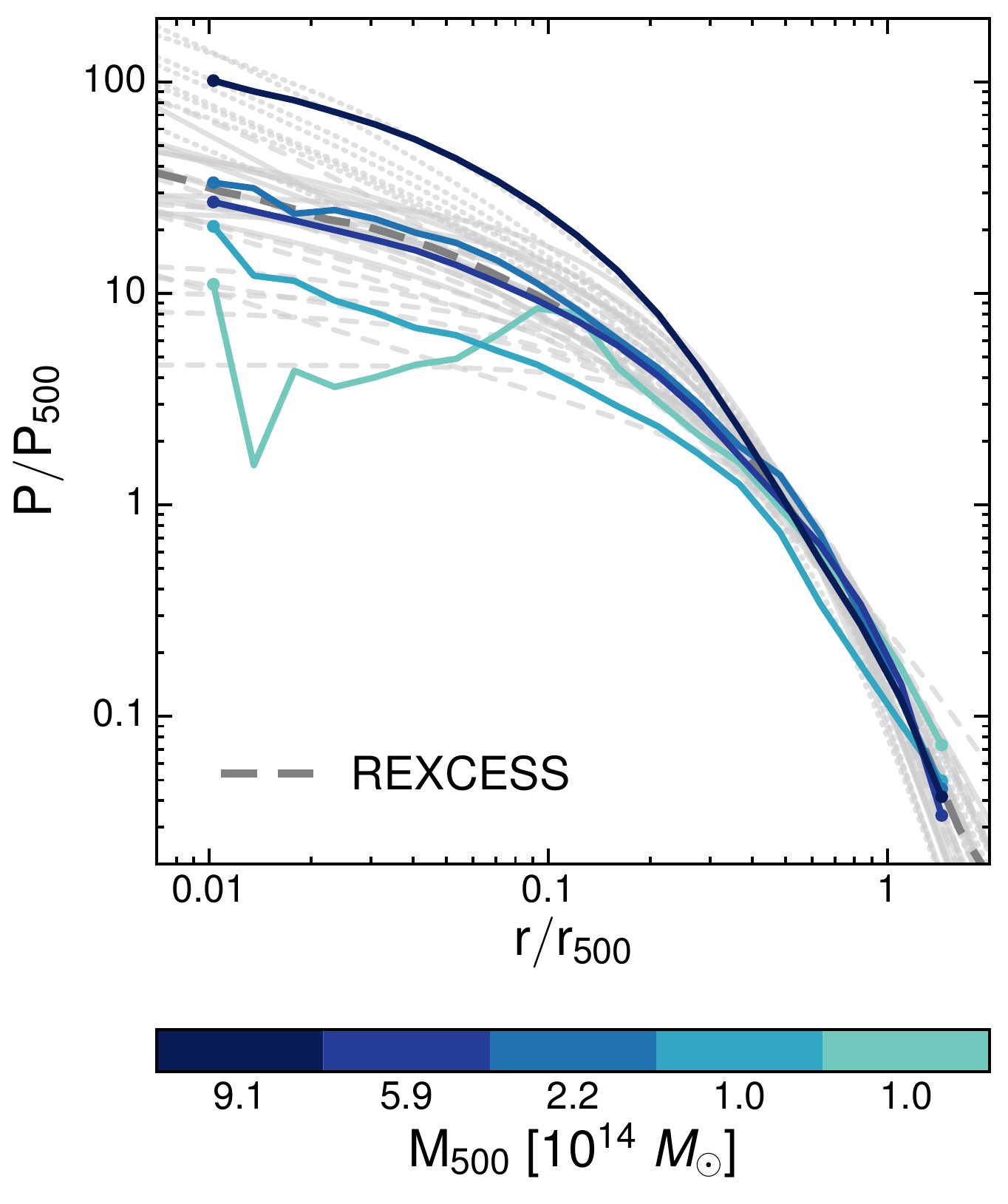}
  \caption{Dimensionless pressure profiles at $z=0$ for the same simulated and observed samples as shown in Fig.~\ref{fig:density}. Profiles are scaled by the characteristic pressure $P_{500}$ as defined in the text.
    In the left hand panel, grey lines show the pressure profiles of the \protect\cite{Sun2009} groups derived from the density and temperature profiles shown in Fig.~\ref{fig:density} and \ref{fig:temp}.
    In the right hand panel, grey lines show the best-fitting pressure profiles of the REXCESS clusters \protect\citep{Pratt2010}. Line styles are the same as shown in Fig.~\ref{fig:density}.
  }
    \label{fig:pressure}
\end{figure*}

\subsection{Density profiles}\label{subsec:density}

In Fig.~\ref{fig:density} we show spherically-averaged radial density profiles of the ICM in \fable\ groups and clusters at $z=0$. In the left hand panel we compare group-scale systems to the density profiles of the \cite{Sun2009} sample of groups observed with \textit{Chandra}.
The \cite{Sun2009} sample consists of 43 groups with X-ray hydrostatic mass estimates in the range $1.6 \times 10^{13}~M_{\odot} \leq M_{500} \leq 1.8 \times 10^{14}~M_{\odot}$ with a median mass of $7.3 \times 10^{13} M_{\odot}$. The sample was drawn from \textit{Chandra} archival data with the requirement that gas properties could be derived out to at least $r_{2500}$ ($\approx 0.47$ $r_{500}$).
We compare to a halo mass-selected sample of simulated groups with $3.2 \times 10^{13}~M_{\odot} \leq M_{500} \leq 2.2 \times 10^{14}~M_{\odot}$ and a median mass of $5.7 \times 10^{13} M_{\odot}$.
The simulated density profiles are in good agreement with the \cite{Sun2009} profiles beyond $\sim 0.3\,r_{500}$ but slightly underestimate the median density at smaller radii.
At $\gtrsim 0.3\,r_{500}$ the agreement is consistent with our match to observed gas mass fractions (see Fig.~\ref{fig:gas_fracs}), as this region contains $\sim 90$ per cent of the total gas mass.

In Section~\ref{sec:global} we argued that if there is a significant X-ray mass bias then the \fable\ systems may be over-luminous in X-rays and that this may be the result of an overabundance of gas (the observed gas fractions in Fig.~\ref{fig:gas_fracs} being biased high).
In this case, we would expect to overestimate the gas density in the outer regions compared to the data. The fact that this is not seen does not, however, rule out the possibility of a significant X-ray mass bias.
One explanation is the difference between the halo mass distributions of the observed and simulated samples. Although we have tried to match the median halo masses of the observed and simulated samples as closely as possible given the small sample size, the median mass of the simulated sample is 22 per cent lower than that of the \cite{Sun2009} sample and would decrease further in the case of an X-ray mass bias (a 45 per cent difference if X-ray masses are biased low by 30 per cent). As the gas content of massive haloes is a relatively strong function of their halo mass (see e.g. Fig.~\ref{fig:gas_fracs}), this could result in a mismatch between the two samples.
Another explanation is that the average X-ray luminosity of the \cite{Sun2009} sample is biased high due to their requirement that group emission be traced out to a significant fraction of $r_{500}$.
Indeed, as we discussed in Section~\ref{subsec:L-T}, the \cite{Sun2009} groups with $T_{500} \sim 1$~keV possess slightly higher X-ray luminosities compared to the \cite{Osmond2004} and \cite{Zou2016} samples, such that the $L^{bol}_{500}-T_{500}$ relation of our simulated groups is actually in good agreement with the \cite{Sun2009} data. This could explain why our simulated groups match the density profile of the \cite{Sun2009} groups yet seem to overestimate the X-ray luminosity relative to other studies.

At $r \lesssim 0.3\,r_{500}$ the simulated density profiles lie within the observed scatter but largely fall below the median observed profile.
This suggests that our AGN feedback model may displace slightly too much gas from the central regions.
On the other hand, the observed groups are detected via their X-ray flux, which may preferentially select systems with high central densities compared to our halo mass-selected sample, particularly if the \cite{Sun2009} sample is biased toward high luminosities relative to other X-ray selected samples.

In the right hand panel of Fig.~\ref{fig:density} we compare our simulated clusters to the density profiles of the REXCESS clusters, a representative sample of 31 clusters observed with \textit{XMM-Newton} \citep{Bohringer2007, Croston2008}.
The REXCESS clusters were chosen such that $r_{500}$ lies well within the field of view of \textit{XMM-Newton}, allowing detailed local background modelling and increased measurement precision at large radii \citep{Croston2008}.
The sample is unbiased with respect to cluster morphology or dynamical state, containing a representative distribution of relaxed, cool-core and morphologically disturbed clusters (as defined in \citealt{Pratt2009}), which correspond to the solid, dotted and dashed lines in Fig.~\ref{fig:density}, respectively.
The REXCESS sample covers the mass range $1.0 \times 10^{14} M_{\odot} \leq M_{500} \leq 7.8 \times 10^{14} M_{\odot}$ with a median mass of $2.6 \times 10^{14} M_{\odot}$. Our comparison sample consists of all five \fable\ clusters with $M_{500} \geq 1.0 \times 10^{14} M_{\odot}$ and has a comparable median mass of $2.1 \times 10^{14} M_{\odot}$.

Overall the density profiles of our simulated clusters are a good match to the REXCESS clusters. At $r \gtrsim 0.3\,r_{500}$ the densities of the three most massive \fable\ clusters are somewhat high compared to the median observed profile, which is consistent with the excess gas we might expect in our simulated clusters in the case of an X-ray mass bias. Indeed, the cumulative gas fraction (not shown) rises more steeply between $\sim 0.1-0.6~r_{500}$ than the REXCESS profiles \citep{Pratt2010}.
This suggests that AGN feedback may act too violently, pushing gas from the cluster centre and causing it to pile up at larger radii. A similar trend was found for clusters in IllustrisTNG \citep{Barnes2017}.

The simulated profiles are similar in shape to the relaxed or disturbed REXCESS clusters. Of the five simulated clusters only one has a central density comparable to observed (weak) cool-core clusters. Potentially, heating of the cluster centre by AGN feedback might be preventing cool-cores from forming in the same proportion as observed at $z \sim 0$ (10 of 31 REXCESS clusters and $\sim 30-40$ per cent in SZ surveys, e.g. \citealt{Planck2011XI, Andrade-Santos2017, Rossetti2017}). A larger sample will be needed to assess this in detail.
Reproducing the observed fraction of cool-core galaxy clusters in cosmological simulations with feedback is a notoriously difficult problem (e.g. \citealt{Borgani2011, Kravtsov2012}).
There has been some recent success in this area (e.g. \citealt{Rasia2015, Barnes2017, Hahn2017}), however, these simulations tend to underestimate the observed cool-core fraction when compared to low-redshift SZ-selected samples.
It is clear that an AGN feedback model that is capable of reproducing the global properties of clusters does not necessarily provide an effective description of the physical processes responsible for the creation of cool cores: additional processes such as AGN-driven turbulence, cosmic rays, stabilisation from magnetic fields or anisotropic thermal conduction may be required.

\subsection{Temperature profiles}\label{subsec:temp}

In Fig.~\ref{fig:temp} we plot the dimensionless temperature profiles at $z=0$ for the same group- and cluster-scale samples described in the previous section.
We facilitate a comparison between different halo masses by normalising the profiles by the characteristic temperature,
\begin{equation}
  kT_{500} = \mu m_p G M_{500}/2 r_{500},
  \label{eq:T500}
\end{equation}
the temperature of an isothermal sphere of mass $M_{500}$ and radius $r_{500}$. Here $\mu$ is the mean molecular weight, which we take as $\mu = 0.59$, and $m_p$ is the proton mass.

In the left hand panel of Fig.~\ref{fig:temp} we compare to the deprojected temperature profiles of the \cite{Sun2009} groups.
We find that beyond the core ($\gtrsim 0.2\,r_{500}$) the temperature profiles of the \fable\ groups have a similar slope to the median observed profile. The normalisation is somewhat lower than observed, however, this may be the result of an X-ray hydrostatic mass bias in the \cite{Sun2009} mass estimates: since the halo mass, $M_{500}$, is related to the characteristic temperature scaling as $T_{500} \propto M_{500}^{2/3}$, a bias toward lower masses would shift the observed dimensionless temperature profiles to higher values.
Within the core, the simulated temperature profiles show similar scatter to observed and in general follow the same shape as the observed profiles.

For the most massive simulated system in the group sample, the temperature rises steadily towards the centre rather than dropping within the core. This may be a side-effect of our relatively simple model for radio-mode AGN feedback, which injects bubbles of thermal energy at irregular intervals. In reality, such bubbles are expected to be supported by non-thermal pressure from, e.g., cosmic rays and should only contribute to the observed temperature profile once the injected energy has thermalised.

In the right hand panel of Fig.~\ref{fig:temp} we compare to the dimensionless temperature profiles of the REXCESS clusters.
The REXCESS temperature profiles rise with roughly constant slope from $r_{500}$ down to $\sim 0.3\,r_{500}$ before dropping slowly or levelling out within the cluster core.
Close to $r_{500}$ there is a slight offset between the predicted and observed temperatures. As for the \cite{Sun2009} group comparison, this could be explained by a bias in the X-ray hydrostatic mass estimates.
Between $\sim 0.3-1\,r_{500}$, three of the five simulated temperature profiles show a similar slope to observed but for the two most massive clusters in the sample the slope is somewhat shallower.

The REXCESS temperature profiles are not as strongly peaked as the \cite{Sun2009} profiles. Hence the two \fable\ systems with $M_{500} \sim 1 \times 10^{14} M_{\odot}$, which are included in both the group- and cluster-scale samples, are a reasonable match to one or more of the \cite{Sun2009} systems but reach higher peak temperatures compared to the REXCESS clusters. This is unsurprising given that the halo masses of these systems are on the boundary between groups and clusters and correspond only to the very lowest masses in the REXCESS sample.
The three most massive \fable\ clusters possess temperature profiles that continue to rise within the cluster core ($\lesssim 0.1\,r_{500}$), unlike the observed temperature profiles, which tend to level out.
As noted above, this could be because AGN bubble feedback is modelled via injection of thermal energy rather than of non-thermal components. This may result in an overheating of the ICM in the central regions.

\subsection{Entropy profiles}\label{subsec:entropy}

In Fig.~\ref{fig:entropy} we plot the dimensionless entropy profiles at $z=0$ for the same group- and cluster-scale samples described in Section~\ref{subsec:density}.
We use the widely adopted definition of ICM ``entropy'', $K = kT/n_e^{2/3}$, and normalise by the characteristic entropy scale, $K_{500} = kT_{500}/n_{e,500}^{2/3}$, which reflects the mass variation expected in a self-similar model. The characteristic temperature $kT_{500}$ is defined in equation~\ref{eq:T500}. $n_{e,500}$ is the mean electron density within $r_{500}$ and is defined as
\begin{equation}
  n_{e,500} = 500 f_b \rho_c(z) / (\mu_e m_p),
  \label{eq:ne500}
\end{equation}
assuming the global baryon fraction $f_b = \Omega_b/\Omega_{\rm M}$ in a universe with critical density $\rho_c(z)$ at redshift $z$ and a mean molecular weight per free electron $\mu_e$. We adopt a value of $\mu_e = 1.14$ and use $f_b$ and $\rho_c(z)$ corresponding to our assumed cosmology.

In the left hand panel of Fig.~\ref{fig:entropy} we plot the entropy profiles of the group-scale systems. Close to $r_{500}$ the entropy profiles tend to the power-law prediction from \cite{Voit2005a}, which was derived from hydrodynamic simulations in the absence of non-gravitational processes. At smaller radii the entropy profiles deviate from this relation due to non-gravitational processes such as AGN feedback, which ejects and heats gas in the group centre, and star formation, which removes low entropy gas.
The simulated entropy profiles are in good agreement with the \cite{Sun2009} profiles over the full range of radii. The exception is the most massive system, which follows the non-radiative relation before quickly flattening into an isentropic core at $\sim 0.4\,r_{500}$.
This is consistent with the temperature profile of this system, which rises steadily towards the cluster centre instead of dropping slowly within the core.

In the right hand panel of Fig.~\ref{fig:entropy} we see that the observed entropy profiles of the REXCESS clusters run approximately parallel to the \cite{Voit2005a} relation at large radii and slowly flatten toward the cluster centre. This change in slope occurs less rapidly than in the less massive \cite{Sun2009} groups, since the deeper potentials of more massive haloes make non-gravitational processes such as AGN feedback less effective. The cool-core REXCESS clusters have mostly power-law-like entropy profiles while the disturbed clusters tend to have cored entropy profiles.

The two \fable\ clusters with $M_{500} \sim 1 \times 10^{14} M_{\odot}$ have fairly flat entropy profiles down to $\sim 0.05\,r_{500}$ before dropping rapidly within the core. These profiles lie within the intrinsic scatter of the \cite{Sun2009} sample but are unlike the entropy profiles of the REXCESS clusters. This suggests that these systems are more similar to high-mass groups than low-mass clusters.
For the three most massive \fable\ clusters, the entropy at $\gtrsim 0.3\,r_{500}$ closely follows the baseline relation of \cite{Voit2005a}. This suggests that AGN feedback has little effect on the thermodynamics of the gas at large radii.
The observed profiles lie somewhat higher than the baseline and the simulations, however, the observed profiles may be slightly overestimated in the case of an X-ray mass bias.
At smaller radii the three most massive clusters show flat central entropy distributions.
These cored entropy profiles are characteristic of many of the relaxed and disturbed REXCESS clusters, although the cores extend to somewhat larger radii than observed, which is likely related to the slight overprediction of the gas density in these systems at $\sim 0.3\,r_{500}$ (see Sec.~\ref{subsec:density}).
This is consistent with a picture in which AGN feedback in \fable\ clusters is overly effective at heating and expelling gas in the central regions but is relatively ineffective at large radii.

\subsection{Pressure profiles}\label{subsec:pressure}

In Fig.~\ref{fig:pressure} we plot the pressure profiles of the ICM at $z=0$ for the group- and cluster-scale samples described in Section~\ref{subsec:density}.
We normalise the pressure profiles, $P(r) = kT(r) n_e(r)$, by the characteristic pressure, $P_{500} = kT_{500} n_{e,500}$, where $kT_{500}$ and $n_{e,500}$ are defined in equations~\ref{eq:T500} and \ref{eq:ne500}.

In the left hand panel of Fig.~\ref{fig:pressure} we compare to the \cite{Sun2009} pressure profiles. At $r \gtrsim 0.3\,r_{500}$ the simulated profiles are a good match to the data. Similar to the dimensionless temperature and entropy profiles, there is a slight offset in normalisation with respect to the median observed profile, possibly due to an X-ray mass bias.
Inside $\sim 0.3\,r_{500}$ the most massive system remains in good agreement with the data, however the central pressure in the less massive simulated groups is slightly underestimated. This is consistent with the density profiles shown in Fig.~\ref{fig:density}, which are slightly underestimated at $r \lesssim 0.3\,r_{500}$ compared to the median observed profile. As we discussed in Section~\ref{subsec:density}, this suggests that AGN feedback may be ejecting too much gas from the central regions of galaxy groups, although selection effects in the \cite{Sun2009} sample may also play a role.

In the right hand panel of Fig.~\ref{fig:pressure} we find excellent agreement with the REXCESS pressure profiles across the full range of radii.
In the outskirts of the \fable\ clusters ($r \geq 0.5\,r_{500}$) the dimensionless pressure profiles coincide over a wide range of halo masses. The same is true for the observed clusters, which suggests that both the simulated and observational samples represent fairly self-similar populations.
At small radii there is a departure from the self-similar scaling due to the effects of non-gravitational processes. The dispersion increases toward the cluster centre in both the observed and simulated samples to a similar degree.

\section{Discussion}\label{sec:discussion}
The \fable\ simulations employ an updated version of the Illustris galaxy formation model. Specifically we have updated the sub-grid models for feedback from stars and AGN in order to reproduce the $z=0$ galaxy stellar mass function and the present-day gas mass fractions of massive haloes.
The latter were not considered during the Illustris calibration and were severely underestimated with respect to observations.
By adopting a model that reproduces observed gas fractions, we have obtained significantly more realistic galaxy groups and clusters while, at the same time, maintaining a good match to the observed galaxy stellar mass function in the field.

The \fable\ model produces a very similar galaxy stellar mass function to Illustris (see Fig.~\ref{fig:SMF_best}), despite significantly changing the way in which AGN feedback regulates star formation in massive galaxies.
In Illustris, quasar-mode feedback was continuous such that a relatively small amount of accreted feedback energy was injected into the surrounding gas at every timestep. This often resulted in too little thermal energy being input into too much gas, allowing the energy to be efficiently radiated away.
The quasar-mode was therefore inefficient at suppressing star formation and strong radio-mode feedback was needed to suppress stellar mass buildup in massive haloes. This was achieved with a long duty cycle for the radio-mode such that feedback events were infrequent but highly energetic. The major side-effect of this model was that the radio-mode acted too violently on the gas, resulting in severely underestimated gas fractions in Illustris at $z=0$ (see Fig.~\ref{fig:gas_fracs}).
In \fable\ we have introduced a modification to the quasar-mode in the form of a duty cycle. Rather than a continuous injection of thermal energy as in Illustris, the available feedback energy is stored over a 25~Myr time period before being injected into the surrounding gas in a single event. This heats the gas to much higher temperatures, resulting in a longer cooling time and overall more efficient feedback.
The updated quasar-mode feedback is more effective at suppressing the stellar mass buildup of massive galaxies, thereby allowing the radio-mode to operate on a much shorter duty cycle.
This gentler form of radio-mode feedback has a smaller impact on the gas content of massive haloes and has enabled us to produce groups and clusters with realistic gas fractions (see Fig.~\ref{fig:gas_fracs}).
We point out that the agreement with observed gas fractions at cluster scales ($M_{500} \gtrsim 10^{14} M_{\odot}$) was not guaranteed, since the feedback model was not calibrated at such scales.
By applying our calibrated model to cluster-scale objects simulated using the zoom-in technique, we are able to compare the predictions of the \fable\ model to a variety of observational constraints across a wide range of halo masses. We demonstrate very good agreement with observations for a number of group and cluster properties, including stellar mass fractions, X-ray luminosity--mass relations, integrated Sunyaev--Zel'dovich flux and radial profiles of the intracluster medium.

Yet there remain some discrepancies with our model compared to observations. In particular we find that the X-ray luminosity--temperature ($L_{500}-T$) relation lies on the upper end of the observed scatter (see Fig.~\ref{fig:L-T}).
From the $L_{500}-T$ relation alone it is unclear whether the cause of this offset is dominated by overestimated X-ray luminosities or underestimated spectroscopic temperatures.
We aim to gain some insight on the discrepancy by comparing to other observed scaling relations such as the halo mass--temperature ($M_{500}-T$) relation and the X-ray luminosity--halo mass ($L_{500}-M_{500}$) relation.
However, the difference between relations based on X-ray hydrostatic mass estimates versus weak lensing mass estimates means that the conclusion is dependent upon which is used for the comparison.

If we compare the $M_{500}-T_{500}$ relation of the \fable\ simulations to observational data based on X-ray derived halo masses (left hand panel of Fig.~\ref{fig:M-T}), we would conclude that the average X-ray temperatures of our systems are systematically underestimated by $\sim 0.2$~dex. This is large enough to explain the discrepancy in $L_{500}-T_{500}$ without affecting the $L_{500}-M_{500}$ relations, which are in excellent agreement with observed $L_{500}-M_{500}$ relations based on X-ray derived halo masses (Fig.~\ref{fig:Lsoft-M} and \ref{fig:Lbol-M}).

On the other hand, we have very good agreement with the $M_{500}-T_{\mathrm{300 kpc}}$ relation of \cite{Lieu2016}, which uses halo masses measured via weak lensing (right hand panel of Fig.~\ref{fig:M-T}). This would suggest that our simulated systems possess realistic global temperatures.
Similarly, our agreement with the weak lensing calibrated $Y-M_{500}$ relation (Fig.~\ref{fig:SZ}) implies that the mass-weighted temperatures of our simulated systems are realistic (the integrated tSZ flux being proportional to the total thermal energy content of the gas). Given that our spectroscopic temperature estimates are not systematically lower than the mass-weighted temperature (see Appendix~\ref{A:temp_diff}), this provides further evidence that the discrepancy in $L_{500}-T_{500}$ relation is unlikely to be due to underestimated temperatures.

There is a significant offset in normalisation between the weak lensing $M_{500}-T$ relation of \cite{Lieu2016} and the $M_{500}-T$ relations based on X-ray masses, albeit with large scatter in the weak lensing data. In fact, \cite{Lieu2016} perform a comparison of the normalisation of different $M_{500}-T$ relations from the literature and find that relations based on weak lensing masses favour $\sim 40$ per cent higher normalisations than those based on X-ray hydrostatic masses. This implies a systematic difference between halo masses measured from weak lensing and masses measured from X-rays.
Indeed, a number of observational studies have found that, within $r_{500}$, X-ray hydrostatic masses are biased low compared to weak lensing masses by $\sim 25-30$ per cent (e.g. \citealt{Donahue2014, VonderLinden2014, Hoekstra2015, Simet2017}). A slightly larger X-ray mass bias of $\sim 40$ per cent, as seemingly preferred by the \cite{Lieu2016} sample, is large enough to reconcile the results of cluster abundance studies with cosmological constraints from \textit{Planck} measurements of the primary CMB \citep{PlanckXXIV2015}.

Under the assumption that weak lensing masses are less biased than X-ray hydrostatic masses, we would deduce that the spectroscopic temperatures of \fable\ groups and clusters are realistic and that the discrepancy in $L_{500}-T_{500}$ is largely the result of overestimated X-ray luminosities. This is consistent with the $L_{500}-M_{500}$ relations shown in Fig.~\ref{fig:Lsoft-M} and \ref{fig:Lbol-M}, which suggest that the predicted X-ray luminosities may be overestimated as a function of halo mass compared to observations based on weak lensing (rather than X-ray) mass measurements.
Furthermore, if X-ray hydrostatic masses are biased low, then the observational constraints on stellar and gas mass fractions plotted in Fig.~\ref{fig:star_fracs} and \ref{fig:gas_fracs} would be biased high.
For example, \cite{Eckert2016} find that the weak lensing calibrated gas fraction of XXL-100-GC clusters is significantly lower than independent results based on X-ray hydrostatic masses.
Since we have calibrated our feedback model to reproduce observed gas mass fractions assuming a negligible X-ray mass bias, this would imply that \fable\ groups and clusters are too gas rich. The excess gas could then explain our overpredicted X-ray luminosities at fixed temperature.

  On the other hand, a large X-ray hydrostatic mass bias ($\gtrsim 30$ per cent) implies a baryon depletion factor significantly exceeding that predicted by numerical simulations (e.g. \citealt{Eckert2016}). Moreover, a number of studies find results consistent with little to no bias (e.g. \citealt{Gruen2014, Israel2014, Smith2015, Applegate2016, Maughan2015, Andreon2017}). In a future paper we will investigate the level of X-ray and hydrostatic mass bias in the \fable\ simulations as a function of cosmic time (Henden et al., in preparation). With this knowledge in hand we hope to explain the origin of the discrepancies between the predicted and observed scaling relations.

Few cosmological hydrodynamical simulations manage to convincingly reproduce the observed X-ray scaling relations.
The cosmo-OWLS and {\sc bahamas} projects obtain a good match to the observed X-ray luminosity--halo mass relation with relatively low-resolution simulations and when modelling a signficant X-ray mass bias.
At much higher resolution, the {\sc c-eagle} clusters, which employ the {\sc eagle} galaxy formation model, are slightly over-luminous for a given halo mass due to the clusters being too gas rich.
Similarly, the hydrodynamical cluster simulations presented in \cite{Truong2018} possess lower than observed temperatures and approximately 30 per cent higher X-ray luminosities than observed, although the latter discrepancy they suggest is at least partly due to sample selection.
Mesh-based cosmological hydrodynamical simulations encounter similar issues.
For example, the Rhapsody-G \citep{Hahn2017} suite of cluster zoom-in simulations, which use an Eulerian adapative mesh refinement (AMR) method, show X-ray luminosities as a function of halo mass consistently higher than observed (by $\sim 20$ per cent at $M_{500} \approx 10^{15} M_{\odot}$ and about a factor of 2 at $M_{500} \approx 10^{14} M_{\odot}$). Interestingly, \cite{Hahn2017} find that the normalisation of the X-ray luminosity--halo mass relation is insensitive to the AGN feedback parameters, including drastic changes to the length of the duty cycle, contrary to the results of SPH simulations with a similar AGN feedback model (e.g. \citealt{LeBrun2014}).

In our simulations, which are run with the {\sc arepo} moving-mesh code, we find that changes to the duty cycle and energetics of AGN feedback can have a large impact on the gas mass fractions of massive haloes (see Appendix~\ref{A:models}).
We expect that the \fable\ model could likely be adjusted to produce somewhat lower gas mass fractions and thus lower X-ray luminosities in massive haloes by lengthening the duty cycle of radio-mode feedback. This would make individual events more energetic and therefore more effective at ejecting gas beyond the virial radius.
In Appendix~\ref{A:models} we show that increasing the burstiness of the radio-mode in this way can significantly lower halo gas fractions without drastically altering the $z=0$ galaxy stellar mass function, which is already in good agreement with observations in our fiducial model.

On the other hand, further tuning of our relatively simplistic model for thermal bubble feedback is unlikely to significantly improve the ICM profiles of our simulated clusters, which show indications of over-heating and removal of too much gas in the central regions, while gas at $\gtrsim 0.5\,r_{500}$ is relatively unaffected (see e.g. entropy profiles in Section~\ref{subsec:entropy}).
A similar though more extreme predicament applies to the {\sc c-eagle} clusters \citep{CEAGLE}, which show lower than observed gas density in the core and entropy profiles with significantly larger cores and higher central entropies than observed.
\cite{CEAGLE} suggest that AGN feedback in {\sc c-eagle} is too active at late times, increasing the central entropy of clusters and preventing the formation of cool-core systems with steep central density and entropy profiles.
The \fable\ clusters also do not contain an obvious strong cool-core system, although a larger sample will be needed to assess this issue in detail.
Other numerical works such as \cite{Rasia2015} and \cite{Hahn2017} have reproduced the observed dichotomy between cool-core and non-cool-core clusters, however, this does not necessarily imply good agreement with observational constraints on the global properties of clusters, since both models tend to produce clusters with somewhat higher than observed X-ray luminosities \citep{Hahn2017, Truong2018}.
IllustrisTNG also produce a fraction of cool-core clusters that is in agreement with observations between $0.25 < z < 1.0$, however, the cool-core fraction is underpredicted at $z < 0.25$, with more clusters showing close to isentropic cores at $z=0$ than observed \citep{Barnes2017}.

Given that essentially all numerical simulations are unable to convincingly reproduce the observed thermodynamic profiles of cluster core regions, it seems vital that simulation models for AGN feedback should continue to be improved, for example, by inflating bubbles self-consistently with AGN jet feedback \citep{Bourne2017, Weinberger2017b}, while also incorporating additional physical processes that have previously been neglected, such as cosmic-rays \citep{Jacob2016, Pfrommer2016}, outflows driven by radiation pressure from AGN \citep{Costa2017a, Costa2017, Ishibashi2018} and anisotropic thermal conduction \citep{Kannan2016, Kannan2016a}.
In addition, further improvement to the realism of simulated groups and clusters will rely on a better understanding of the issues of mass bias and selection effects so that reliable comparisons to unbiased observational data sets can be performed.

\section{Conclusions}\label{sec:conclusion}
We have presented the new \fable\ suite of cosmological hydrodynamical simulations, which is based on the framework of the Illustris project. The simulations consist of a (40~$h^{-1}$~Mpc)$^3$ cosmological volume and a number of zoom-in simulations of individual galaxy groups and clusters. We have employed the moving mesh code {\sc arepo} and an updated version of the Illustris galaxy formation model. We have adapted the sub-grid models for stellar and AGN feedback in order to reproduce galaxy groups and clusters with more realistic gas fractions compared to Illustris whilst maintaining a similarly high level of agreement with the observed present-day galaxy stellar mass function.
In this paper we have presented various other comparisons with observations, including X-ray and SZ scaling relations and radial profiles of the ICM over a wide range of halo masses. Our main conclusions are as follows:

\begin{itemize}
\item We obtain very good agreement with observed galaxy stellar mass functions. The high mass end of the $z=0$ mass function is similar to that of Illustris, despite significant changes to the way in which AGN feedback suppresses the buildup of stellar mass. While the \fable\ model was calibrated to reproduce the $z \approx 0$ mass function, the fact that the agreement with observations continues to higher redshift is a success of the model.
\item The stellar mass fractions of \fable\ galaxy groups and clusters are also an excellent match to low-redshift observations, including in massive clusters that were not present in the calibration volume.
\item The $z=0$ halo gas mass fractions represent a major improvement over Illustris and are now in good agreement with observations. This can be attributed to much less energetic but more frequent thermal energy injections in the radio-mode of AGN feedback, which remove less gas from haloes compared to the Illustris model.
\item The predicted X-ray luminosity--total mass ($L_{500}-M_{500}$) relations are in excellent agreement with observed relations based on X-ray hydrostatic mass estimates but seem to overestimate the X-ray luminosity for a given halo mass when compared with weak lensing mass estimates. The difference between observed relations is consistent with a significant X-ray mass bias. Similarly, a comparison of observed total mass--spectroscopic temperature ($M_{500}-T$) relations reveals a systematic difference between those based on X-ray hydrostatic masses and weak lensing masses. The \fable\ simulations are in good agreement with $M_{500}-T$ data based on weak lensing masses but have significantly lower global temperatures/higher masses compared to relations using only X-ray data.
\item The slope of the predicted X-ray luminosity--spectroscopic temperature ($L_{500}-T_{500}$) relation is in excellent agreement with observations. The normalisation of the relation lies, however, on the upper end of the scatter in the data.
The size of this offset is similar to the offset with weak lensing-based $L_{500}-M_{500}$ relations and X-ray-only $M_{500}-T_{500}$ relations. This implies that the discrepancy in $L_{500}-T_{500}$ could be due to either overestimated X-ray luminosities or underestimated global temperatures. We lean towards the former explanation, as this is consistent with the general expectation that weak lensing masses are less biased than X-ray hydrostatic masses. An improved understanding of mass bias will be important for making further progress here.
\item The simulations are also in excellent agreement with the mean Sunyaev--Zel'dovich flux--total mass ($Y_{5r500}-M_{500}$) relation derived from \textit{Planck} observations of locally bright galaxies. This implies that the global temperatures of our simulated systems are not significantly underestimated, consistent with our match to the weak lensing $M_{500}-T$ relation.
\item In general, the radial profiles of the ICM are a good match to observations outside $\sim 0.3\,r_{500}$, where the majority of the ICM is located. Density and pressure profiles of the ICM are in good agreement with observations of both group- and cluster-scale systems. The group-scale profiles have slightly lower-than-observed density/pressure within $\sim 0.3\,r_{500}$, however, this may be (partly) due to selection effects in the observed sample. The temperature and entropy profiles of $\lesssim 10^{14} M_{\odot}$ haloes are also in good agreement with observations, while for more massive systems the scatter in the simulated profiles somewhat exceeds that of the observed samples.
\end{itemize}

The \fable\ simulations represent a major improvement over Illustris in the galaxy group and cluster regime. The baryonic content and global X-ray and tSZ properties of the ICM are a good match to observations across a wide range of scales. The ICM is realistically distributed with residual deviations arising in the thermodynamic properties only toward the cluster centre.
Our results are consistent with numerous other simulation studies and suggest that a subtle interplay between AGN feedback and a number of supplementary physical phenomena may be needed to explain the observational properties of galaxy clusters and groups in the core and to the outskirts.
In future papers we aim to study the formation and evolution of cluster galaxies and make predictions for various quantities relevant to cluster cosmology, such as the scatter and redshift evolution of cluster scaling relations.

\section*{Acknowledgements}
We would like to thank Ming Sun for providing us with their data. We also thank Helen Russell, Stephen Walker and Elena Rasia for helpful discussions about X-ray data analysis.
We are grateful to Volker Springel for making the {\sc arepo} moving-mesh code available to us and to the Illustris collaboration for their development of the Illustris galaxy formation model on which this work is based.
NAH is supported by the Science and Technology Facilities Council (STFC).
EP acknowledges support by the Kavli Foundation.
DS and SS acknowledge support by the STFC and the ERC Starting Grant 638707 ``Black holes and their host galaxies: co-evolution across cosmic time''.
This work made use of the following DiRAC facilities (\href{www.dirac.ac.uk}{www.dirac.ac.uk}):
the Data Analytic system at the University of Cambridge [funded by a BIS National E-infrastructure capital grant (ST/K001590/1), STFC capital grants ST/H008861/1 and ST/H00887X/1, and STFC DiRAC Operations grant ST/K00333X/1] and the COSMA Data Centric system at Durham University [funded by a BIS National E-infrastructure capital grant ST/K00042X/1, STFC capital grant ST/K00087X/1, DiRAC Operations grant ST/K003267/1 and Durham University]. DiRAC is part of the National E-Infrastructure.

\bibliographystyle{mnras}
\bibliography{Paper1}

\appendix
\section{Testing AGN feedback models}\label{A:models}
The \fable\ simulations implement a series of physical models based on those of the Illustris galaxy formation simulation.
With the Illustris model as our starting point, we have updated the sub-grid model for AGN feedback to reproduce the massive end of the present-day galaxy stellar mass function (GSMF) and the hot gas content of massive haloes with $M_{500} \approx 10^{13}-10^{14} M_{\odot}$.
Multiple variations of AGN feedback were tested, with our preferred model corresponding to the fiducial model presented in this paper.

Here we present three additional variations of our AGN feedback model and their corresponding impact on the $z=0$ GSMF and the stellar and gas fractions of massive haloes.
Each model has been implemented in a periodic box of length 40 $h^{-1}$ (comoving) Mpc on a side with initial conditions as described in Section~\ref{subsec:sims}. These simulations differ only by the AGN feedback parameters listed in Table~\ref{tab:simtable}. The meaning of each parameter is explained in Section~\ref{subsubsec:feedback} and qualitative descriptions of the different models are given below.
For reference, we also list the parameters used in Illustris. We point out that our choice for the radiative efficiency, $\epsilon_r$, is half that of Illustris. However, the coupling efficiency of the quasar-mode, $\epsilon_f$, has been doubled such that the fraction of accreted rest mass energy used for quasar-mode feedback (the product of $\epsilon_r$ and $\epsilon_f$) is kept the same.

\begin{table*}
  \centering
  \caption{Parameter values for the AGN feedback models studied in this paper in comparison to Illustris. The parameter $\chi_{\rm radio}$ is the fraction of the Eddington accretion rate below which BHs operate in the radio-mode and above which they operate in the quasar-mode; $\epsilon_r$ is the radiative efficiency of BH accretion; $\epsilon_f$ is the thermal coupling efficiency of the quasar-mode; $\Delta t$ is the accumulation time period between quasar-mode feedback events in Myr; $\epsilon_m$ is the coupling efficiency of the radio-mode and $\delta_{\rm BH}$ is the fractional increase in BH mass required to trigger a radio-mode feedback event.}
  \vspace{8pt}
  \begin{tabular}{@{}lccccccccccccc@{}}
  \label{tab:simtable}
 & $\chi_{radio}$  & $\epsilon_r$  & $\epsilon_f$  & $\Delta t$ (Myr) & $\epsilon_m$  & $\delta_{BH}$  \\
  \hline
\textsc{Illustris} & 0.05 & 0.2 & 0.05 & -- & 0.35 & 0.15 \\
\textsc{weak radio} & 0.05 & 0.1 & 0.1 & -- & 0.4 & 0.001  \\
\textsc{stronger radio} & 0.05 & 0.1 & 0.1 & -- & 0.8 & 0.01  \\
\textsc{quasar duty cycle} & 0.05 & 0.1 & 0.1 & 25 & 0.4 & 0.001  \\
\textsc{fiducial} & 0.01 & 0.1 & 0.1 & 25 & 0.8 & 0.01  \\
\hline
\end{tabular}
\end{table*}

The `weak radio' AGN feedback model is similar to the Illustris model but has a significantly shorter radio-mode duty cycle. Specifically, we have greatly reduced the fractional increase in black hole mass required to trigger a radio-mode event, $\delta_{BH}$, such that bubbles are created more frequently but with less energy. In addition, although the radio-mode coupling efficiency, $\epsilon_m$, is approximately the same as Illustris, our lower radiative efficiency means that the overall efficiency with which AGN convert their accreted rest mass energy into bubbles (the product of $\epsilon_r$ and $\epsilon_m$) is approximately half that of Illustris.
In the model `stronger radio', we increase the length of the radio-mode duty cycle such that bubbles are an order of magnitude more energetic compared to the `weak radio' model, although correspondingly less frequent.
We also increase the radio-mode coupling efficiency by a factor of two relative to the `weak radio' model so that the overall efficiency is approximately the same as Illustris.
The third model, `quasar duty cycle', implements a duty cycle for the quasar-mode of feedback as described in Section~\ref{subsubsec:feedback}. In this model, AGN accumulate feedback energy over a period $\Delta t = 25$~Myr before releasing the energy in a single event.
The fourth model is our fiducial model and combines the stronger radio-mode feedback of the second model with the quasar-mode duty cycle of the third model. In addition, we have lowered the accretion rate threshold for switching between quasar- and radio-mode feedback, $\chi_{radio}$, such that black holes spend overall more time in the quasar-mode.
We point out that our fiducial model and the Illustris model convert a very similar fraction of accreted mass into feedback energy ($\epsilon_r \epsilon_f$ and $\epsilon_r \epsilon_m$ for the quasar-mode and radio-mode, respectively).
The major difference compared to the Illustris model is in the duty cycles of the two modes.

\begin{figure}
	\includegraphics[width=\columnwidth]{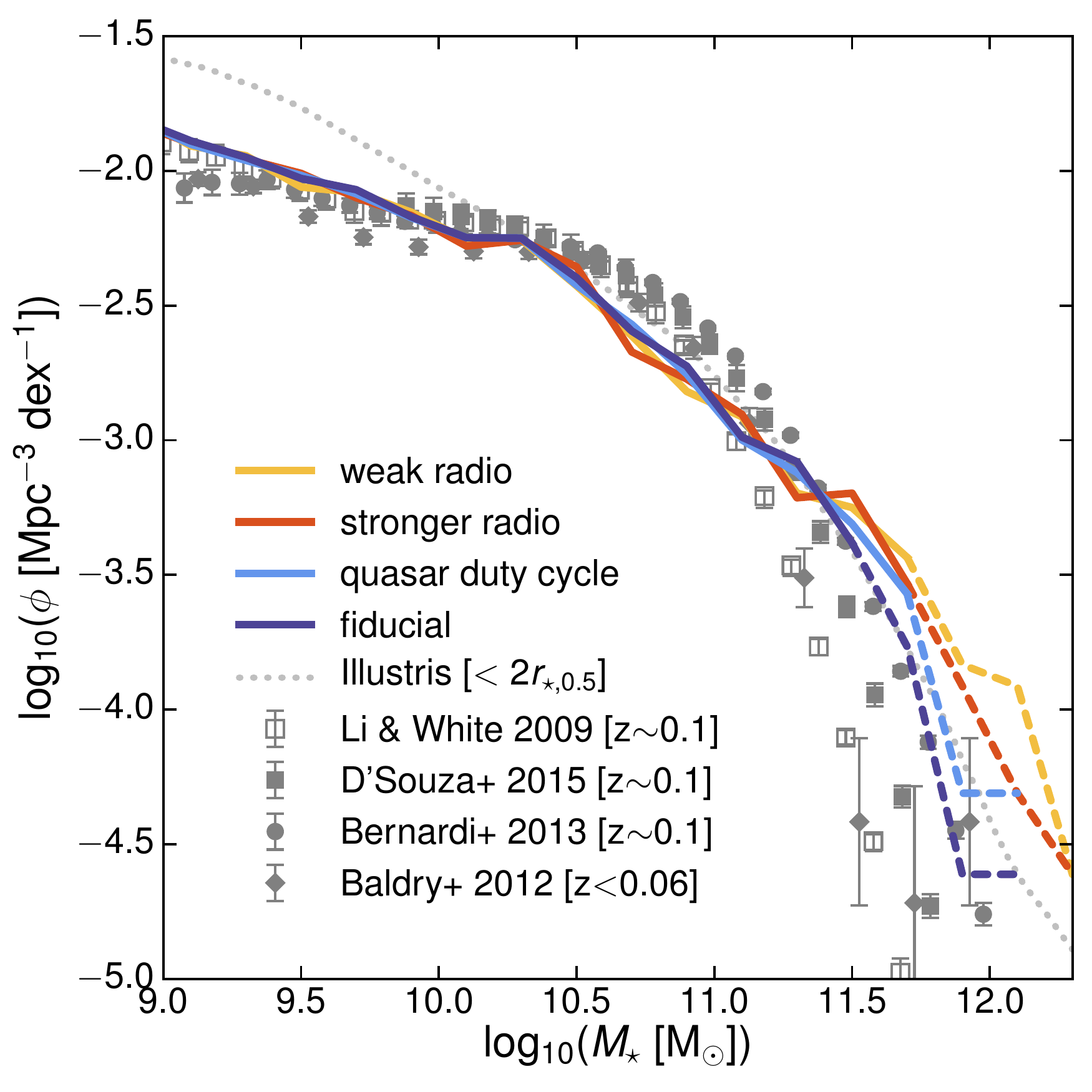}
        \caption{The galaxy stellar mass function at $z=0$ for different AGN feedback models (lines) compared to observations (symbols with error bars). Lines become dashed at the high-mass end when there are fewer than 10 objects per 0.2 dex stellar mass bin. Here the stellar mass of a galaxy is defined as the mass of stars bound to the subhalo within twice the stellar half-mass radius. The grey dotted line shows the equivalent stellar mass function in Illustris. The observational data are as shown in Fig.~\ref{fig:SMF_best}.
        }
    \label{fig:SMF_comparison}
\end{figure}

In Fig.~\ref{fig:SMF_comparison} we compare the $z=0$ GSMFs of the four models.
The `stronger radio' model produces a slightly lower abundance of galaxies at the high mass end compared with the `weak radio' model. This implies that less frequent but more energetic bubble injections are slightly more efficient at suppressing star formation in massive haloes, partly by displacing gas from the dense central regions and partly by heating the surrounding gas to higher temperatures. The change is remarkably small given that there is an order-of-magnitude difference between the duty cycle parameter, $\delta_{BH}$, of the two models.
This explains the need for strong radio-mode feedback in the Illustris model to reduce stellar mass buildup to the degree seen in observations of the high mass end of the GSMF.

\begin{figure*}
  \includegraphics[width=0.497\textwidth]{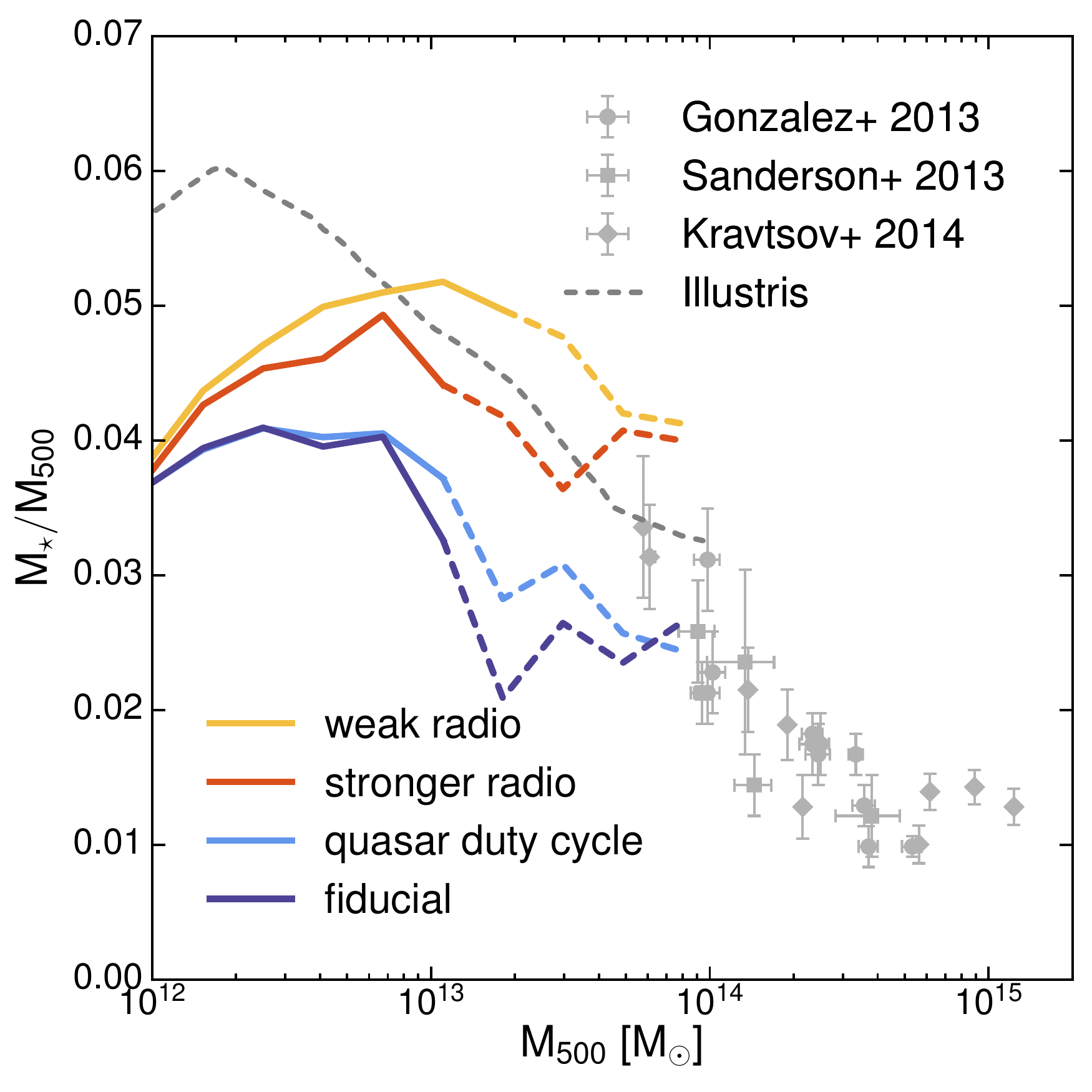}
  \includegraphics[width=0.497\textwidth]{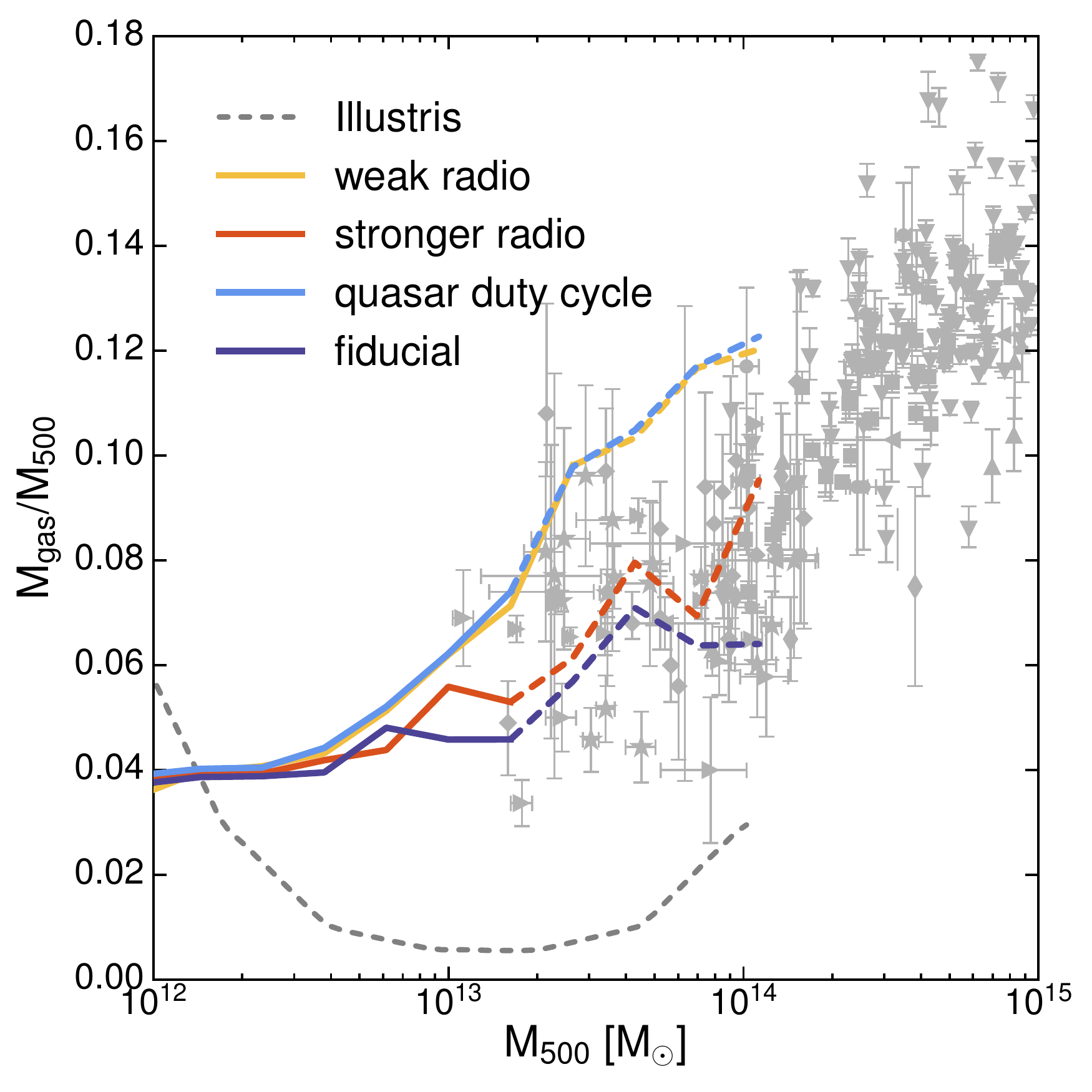}
  \caption{Stellar mass (left) and gas mass (right) fractions inside $r_{500}$ as a function of halo mass at $z=0$ for different AGN feedback models. Lines show the mean relation in halo mass bins of width $0.2$ dex. Lines become dashed when there are fewer than 10 haloes per bin.
    The grey dashed line shows the mean relation in Illustris.
    The total gas mass excludes cold and multiphase gas as described in Section~\ref{subsec:xray}.
    The observational data (symbols with error bars) are as shown in Fig.~\ref{fig:star_fracs} and \ref{fig:gas_fracs}.
  }
    \label{fig:fracs_models}
\end{figure*}

The `quasar duty cycle' model introduces a quasar-mode duty cycle to the `weak radio' model, which leads to a significant reduction in the abundance of massive galaxies.
This implies that periodic heating is much more effective at suppressing star formation than continuous thermal feedback. Physically, this is because the gas is heated to higher temperatures, reducing cooling losses and slowing the rate at which the gas can condense to form stars.
The GSMF of the `quasar duty cycle' model is similar to Illustris at the high mass end, despite using a far gentler form of radio-mode feedback. In Fig.~\ref{fig:fracs_models} we will see that this change has significant consequences for the gas content of massive haloes.
Our fiducial model combines a quasar-mode duty cycle with the `stronger radio' model to reduce the abundance of massive galaxies even further, in good agreement with the data from \cite{Bernardi2013}.

In the left hand panel of Fig.~\ref{fig:fracs_models} we plot the median stellar mass fraction as a function of halo mass for each of the models at $z=0$. We note that the stellar fractions were not considered during the process of calibrating our fiducial AGN feedback model.
The stellar fractions are largely consistent with the difference in the GSMFs between different models. That is, stronger radio-mode feedback reduces the total stellar mass of massive haloes but the introduction of a quasar-mode duty cycle has a significantly larger effect.
The models converge at $\sim 10^{12} M_{\odot}$ since black holes in these haloes have generally not grown sufficiently in mass for AGN feedback to be effective.
At $M_{500} \sim 7 \times 10^{13} M_{\odot}$ there is a slight overlap with the observational data and we find that the `quasar duty cycle' and fiducial models, which implement a quasar-mode duty cycle, are in agreement with the observed stellar fractions, while the two models with continuous quasar-mode feedback slightly overpredict them.
In Fig.~\ref{fig:star_fracs} we demonstrate that the agreement with observations continues to much higher halo masses with our fiducial model, although we caution that this conclusion is dependent on the question of mass bias in X-ray hydrostatic mass estimates, as we discuss in Section~\ref{sec:discussion}.

In the right hand panel of Fig.~\ref{fig:fracs_models} we compare the median gas mass fraction of each model as a function of halo mass at $z=0$.
The `stronger radio' model yields considerably lower gas fractions than the `weak radio' model. This is owed to its more energetic bubble injections, which are able to eject gas from massive haloes more efficiently.
This also explains why all four models yield much larger gas fractions than the Illustris model, which injected approximately an order of magnitude more energy per bubble feedback event.
The `quasar duty cycle' model yields gas mass fractions in massive haloes almost identical to the `weak radio' model, however, because the `weak radio' run converts significantly more gas into stars, the total mass of baryons accumulated by $z=0$ is lower in the model with a quasar-mode duty cycle.
This is because a quasar-mode duty cycle slows the accumulation of gas onto the virialised region of the halo by heating the ICM to higher temperatures.
This effect is consistent with SPH simulations such as \cite{LeBrun2014}, which show that discontinuous thermal AGN feedback can have a strong impact on the gas content of massive haloes.

In summary, we find that periodic quasar-mode AGN feedback significantly reduces stellar mass buildup in massive haloes compared to the continuous case. This has allowed us to reproduce the observed abundance of galaxies at the high mass end of the present-day GSMF without resorting to extremely strong radio-mode feedback, which was responsible for ejecting too much gas from massive haloes in the Illustris model. Using a far gentler form of radio-mode feedback in combination with a quasar-mode duty cycle allows us to reproduce the $z=0$ GSMF and the local gas fractions of massive haloes simultaneously.

\section{Modelling X-ray luminosities and temperatures}
\label{A:xray}

\subsection{Temperature--density cuts}\label{A:cuts}
In generating synthetic X-ray spectra for our simulated groups and clusters we exclude gas with a temperature less than $3 \times 10^4$~K and gas above the density threshold required for star formation (see Section~\ref{subsec:xray}). These ``fiducial'' temperature--density cuts ensure that the derived X-ray luminosity and spectroscopic temperature are not biased by cold or star-forming gas, which should in reality produce negligible X-ray emission.
In this section we investigate whether the choice of cut can have a significant impact on the derived X-ray properties by comparing our fiducial cuts to a recalibrated version of the method used in \cite{Rasia2012}.

\cite{Rasia2012} (hereafter R12) identify a separated phase of cooling gas in their simulated clusters satisfying the condition
\begin{equation}
T < N \times \rho^{0.25}_{\mathrm{gas}},
\label{eq:cut}
\end{equation}
where $T$ and $\rho_{\mathrm{gas}}$ are the temperature and density of the gas, respectively, and $N$ is a normalisation factor. This relation follows from the polytropic equation for an ideal gas, $T \propto \rho^{\gamma - 1}_{\mathrm{gas}}$, assuming a polytropic index of $\gamma = 1.25$.
R12 consider a small range of cluster masses and assume a fixed normalisation factor, $N$, equal to $3 \times 10^6$ with density in units of g cm$^{-3}$ and temperature in keV. Because our sample covers a much wider range of masses, we scale this normalisation with the virial temperature of the halo, $T_{500}^{\mathrm{vir}} \propto M_{500}/r_{500}$, relative to the mean halo mass of the R12 sample, $\overbar{M}_{500} = 5.61 \times 10^{14} h^{-1} M_{\odot}$.

\begin{figure*}
  \includegraphics[width=0.497\textwidth]{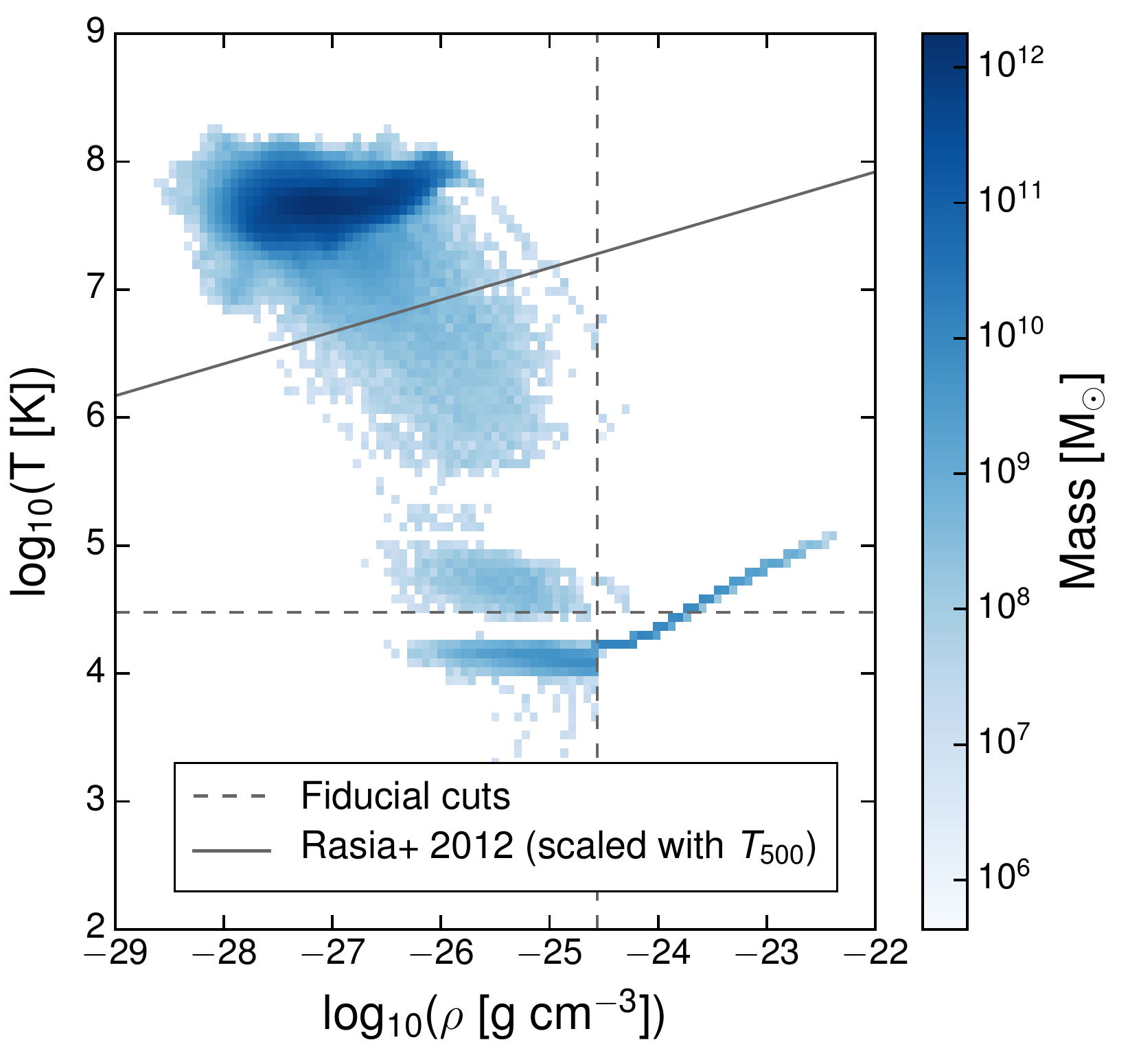}
  \includegraphics[width=0.497\textwidth]{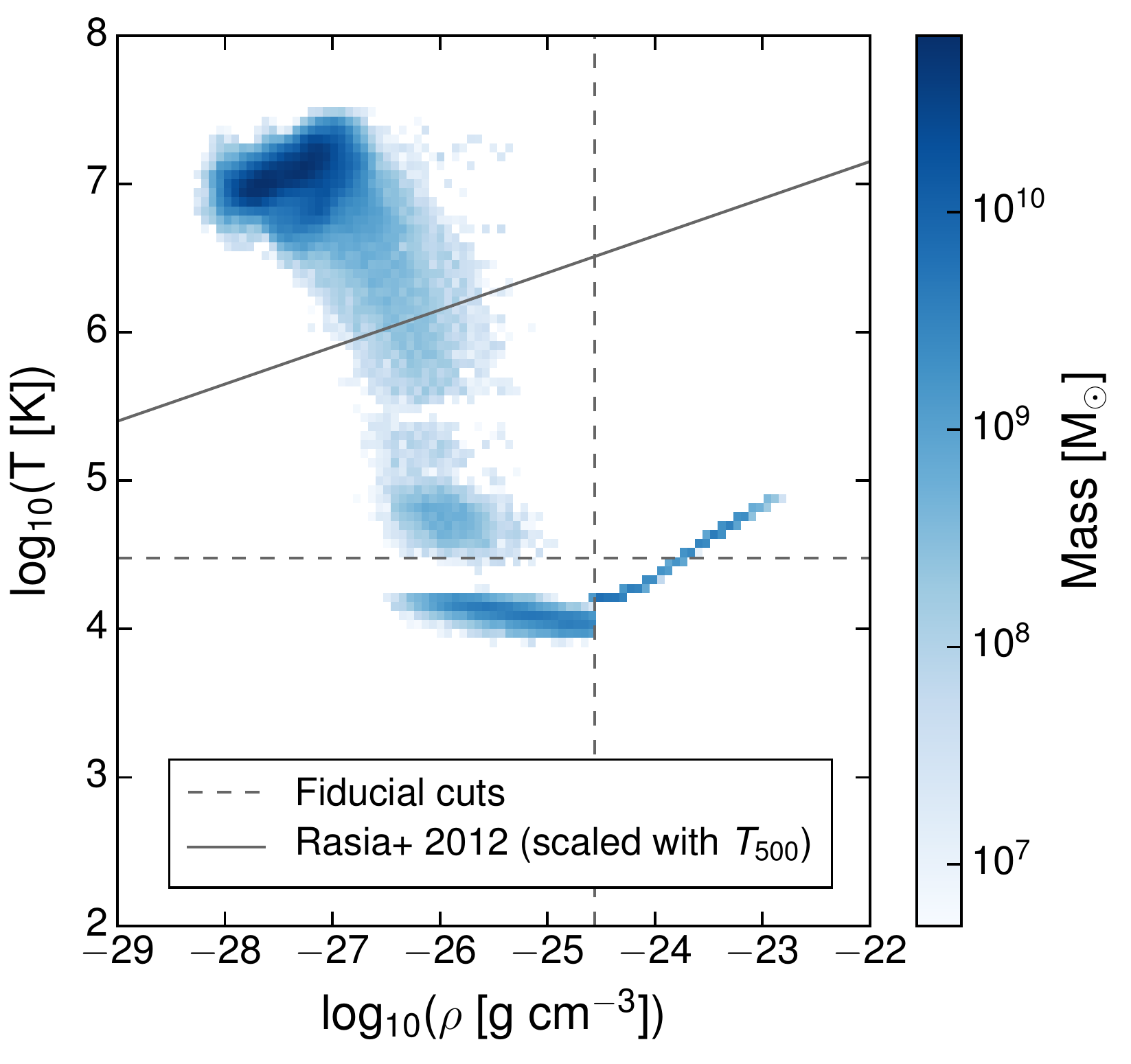}
        \caption{Temperature--density histogram for all gas within $r_{500}$ at $z=0$ in a cluster with $M_{500} = 5.9 \times 10^{14} M_{\odot}$ (left) and a group with $M_{500} = 4.1 \times 10^{13} M_{\odot}$ (right). The colour scale quantifies the total mass of gas in each bin. The solid line shows the temperature--density cut corresponding to the \protect\cite{Rasia2012} method described in the text. Gas below the solid line is excluded using this method and accounts for 0.5 and 7.9 per cent of the total gas mass within $r_{500}$ for the high and low mass system, respectively.
          The horizontal and vertical dashed lines show the fiducial temperature and density cuts used to exclude cold gas and gas followed only with the simple multiphase model for star formation, as discussed in Section~\ref{subsec:xray}. These cuts exclude 0.4 and 6.6 per cent of the total gas mass within $r_{500}$ for the high and low mass system, respectively.
        }
    \label{fig:phase_space}
\end{figure*}

In Fig.~\ref{fig:phase_space} we show the temperature--density distribution for all gas within $r_{500}$ in the case of a high and low mass system: a galaxy cluster with $M_{500} = 5.9 \times 10^{14} M_{\odot}$ and a galaxy group with $M_{500} = 4.6 \times 10^{13} M_{\odot}$.
Dashed lines indicate the fiducial cuts and solid lines the rescaled R12 cut.
The fiducial and R12 cuts exclude 0.4 and 0.5 per cent of the total gas mass in the high mass system, respectively, and 6.6 and 7.9 per cent in the lower mass system, respectively.

\begin{figure*}
	\includegraphics[width=\textwidth]{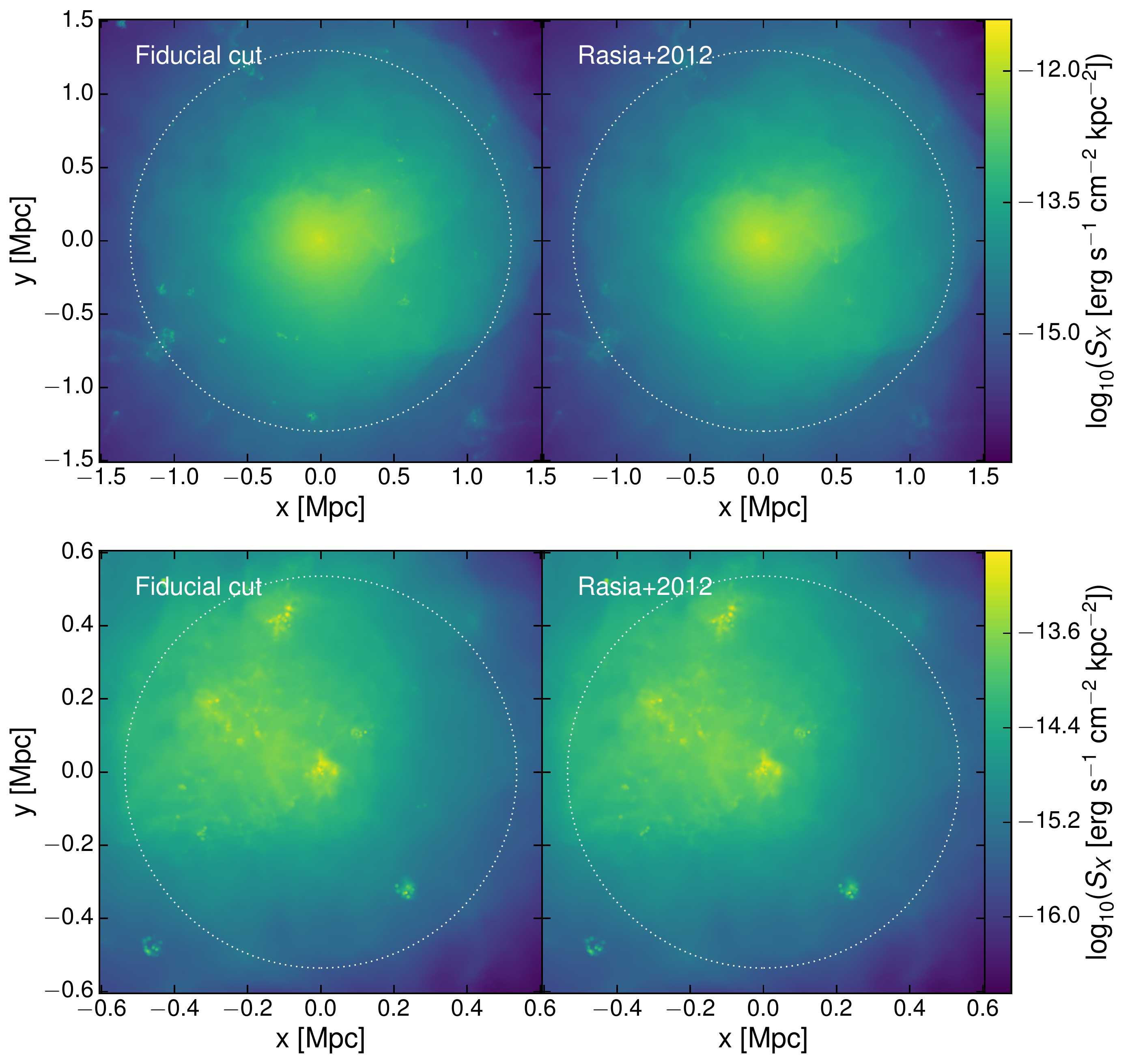}
        \caption{Projected X-ray surface brightness maps in the $0.5-10$~keV energy band for the same systems shown in Fig.~\ref{fig:phase_space} for different temperature--density cuts. The upper panels show a cluster with $M_{500} = 5.9 \times 10^{14} M_{\odot}$ and the lower panels a group with $M_{500} = 4.1 \times 10^{13} M_{\odot}$. The white circle shows $r_{500}$ in each case. In the left hand panels the fiducial temperature and density cuts have been used to exclude cold and star-forming gas. In the right hand panels, gas has been excluded according to the \protect\cite{Rasia2012} method described in the text.
          Maps were derived from X-ray fluxes tabulated using APEC emission models generated for the same temperature bins and metallicity as used in the synthetic X-ray analysis described in Section~\ref{subsec:xray}.
          For a given gas cell we interpolate an X-ray flux from the table and normalise it to the emission measure of the cell.
          We integrate the flux along the full length of the box, although for the more massive object, which was simulated using the zoom-in technique, gas outside the high-resolution region has been excluded from the integration.
        }
    \label{fig:xray_maps}
\end{figure*}

The R12 method was designed to excise compact sources with strong X-ray emission from synthetic X-ray images. In X-ray observations such structures would either be unresolved or would be excised during the analysis. The R12 procedure excludes the cold and dense gas clumps associated with these features without requiring time-consuming post-processing of the synthetic X-ray images.
In Fig.~\ref{fig:xray_maps} we assess the effect of the (rescaled) R12 cut on the X-ray surface brightness maps of the same cluster and group objects shown in Fig.~\ref{fig:phase_space}.
The left and right hand panels correspond to the fiducial and R12 temperature--density cuts, respectively.
We calculate X-ray flux in the $0.5-10$~keV band, which is a typical choice for \textit{Chandra} data analyses and corresponds to the same energy range we use to fit our synthetic spectra (Section~\ref{subsec:xray}).

For the more massive object the R12 cut removes most of the apparent substructure. These structures persist after excluding subhaloes from the analysis, which implies that they are not gravitationally bound to a dark matter substructure. Rather, they belong to a separated cooling phase which is at least partially excluded by the R12 method.
For the lower mass system on the other hand, there is no discernible difference in the X-ray surface brightness distributions between the two cuts. This is in part because the $0.5-10$~keV energy band is fairly insensitive to cool gas.
From Fig.~\ref{fig:phase_space} we see that the gas excluded by either method has temperatures below $0.5$~keV ($\approx 6 \times 10^6$~K). Such gas contributes relatively little to the X-ray surface brightness because much of the X-ray emission falls outside the $0.5-10$~keV energy band.
We caution that this is somewhat dependent on our choice of scaling for the normalisation, $N$, of the R12 cut shown in equation~\ref{eq:cut}. Using a higher normalisation for the same object would exclude gas at higher temperatures, potentially removing some of the X-ray bright substructure.

\begin{figure}
	\includegraphics[width=\columnwidth]{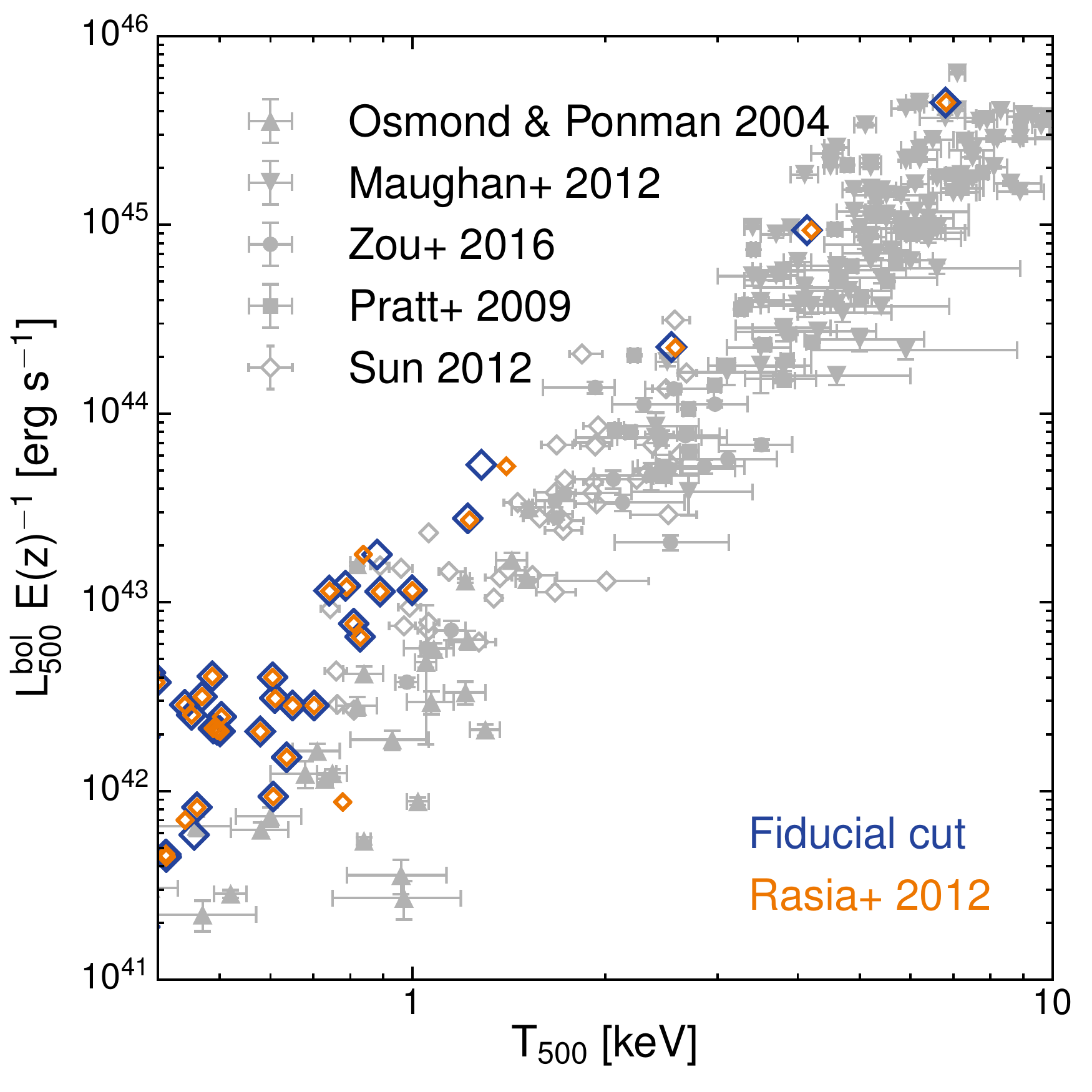}
        \caption{Bolometric X-ray luminosity as a function of spectroscopic temperature measured within $r_{500}$ at $z=0$ for different density--temperature cuts.
          Blue diamonds show the luminosities and temperatures derived when the fiducial cuts are used to exclude cold and star-forming gas.
          Orange diamonds show the same systems when gas is excluded using the \protect\cite{Rasia2012} method described in the text.
          Observational data from \protect\cite{Osmond2004}, \protect\cite{Maughan2012}, \protect\cite{Zou2016}, \protect\cite{Pratt2009} and \protect\cite{Sun2012} are shown for comparison.
        }
    \label{fig:L-T_cuts}
\end{figure}

Fig.~\ref{fig:L-T_cuts} shows the bolometric X-ray luminosity as a function of X-ray spectroscopic temperature when using either the fiducial or R12 temperature--density cuts.
We find that the derived X-ray luminosity and temperature is fairly insensitive to the choice of cut.
At cluster scales ($T \gtrsim 3$~keV) we find little difference between the two methods.
This implies that the X-ray bright structures, which are present in the fiducial case but are excluded with the R12 cut, have a minimal effect on the derived X-ray properties of clusters.
In a handful of group-scale systems the R12 cut produces a notable shift in the $L_{500}-T_{500}$ plane, typically towards higher temperatures.
Unlike the group shown in the lower panel of Fig.~\ref{fig:xray_maps}, these few objects show relatively cold, dense clumps of gas with strong X-ray emission that are largely excluded by the R12 cut.
The lower luminosities and temperatures of these groups compared to clusters means that such clumps can have a more significant impact on the derived X-ray properties.
However, this is true of only a small fraction of objects, while for the majority of groups there is little difference between the fiducial and R12 cuts.

In summary, we find that our derived X-ray luminosities and temperatures are not particularly sensitive to the choice of temperature--density cut. Excluding gas using a rescaled R12 cut does not cause a significant systematic shift in the derived $L_{500}-T_{500}$ relation compared to our fiducial cuts.
In addition, our results suggest that the X-ray properties of clusters are not significantly biased by the presence of X-ray bright substructure.

\subsection{Spectroscopic and Mass-weighted Temperature}\label{A:temp_diff}
A number of studies have demonstrated that there can be significant biases in the temperature of hot gas derived from X-ray analyses compared to the mass-weighted temperature (e.g. \citealt{Mazzotta2004, Rasia2005, Nagai2007a}). Given that the mass-weighted temperature is a direct measure of the thermal energy content of the gas, such a discrepancy could have important consequences for the scaling relations derived from X-ray observations.

\begin{figure}
	\includegraphics[width=\columnwidth]{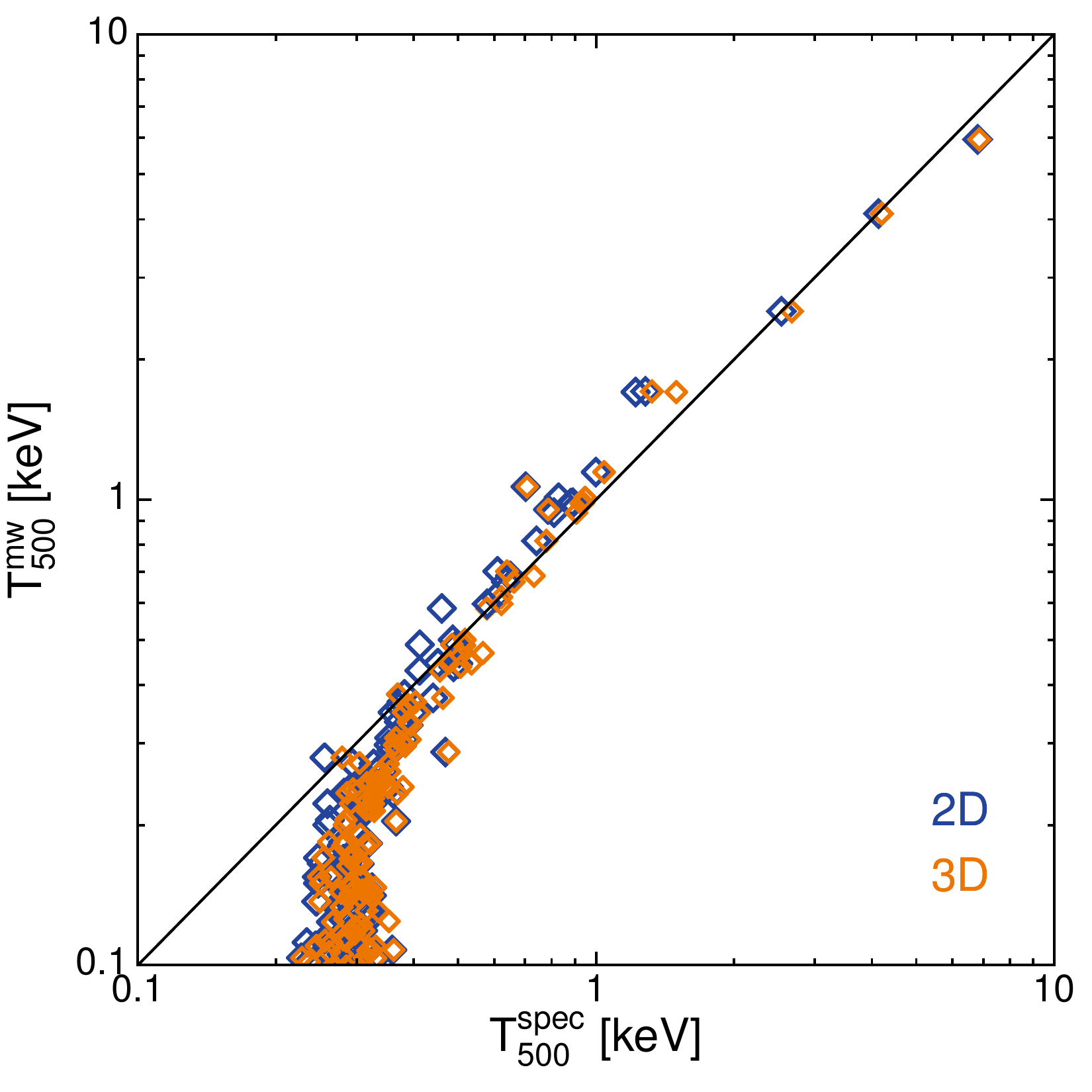}
        \caption{Mass-weighted versus X-ray spectroscopic temperature for \fable\ haloes at $z=0$. The solid line shows equality. The mass-weighted temperature is calculated within the spherical radius $r_{500}$ while the spectroscopic temperature is measured either within the same spherical aperture (3D; orange diamonds) or within a projected aperture of radius $r_{500}$ integrated through the length of the box (2D; blue diamonds). The 2D temperature was used for the X-ray scaling relations presented in this paper.
          The spectroscopic temperature is derived from a single-temperature fit to a synthetic X-ray spectrum convolved with the \textit{Chandra} response function (see Section~\ref{subsec:xray}). Cold and multiphase gas has been excluded from both the mass-weighted and spectroscopic temperature measurements.
          The increase in spectroscopic temperature relative to mass-weighted at low temperatures is largely a result of \textit{Chandra's} effective area, which drops rapidly below $\sim 0.5$~keV.
        }
    \label{fig:temp_diff}
\end{figure}

In Fig.~\ref{fig:temp_diff} we plot the mass-weighted temperature, $T_{500}^{\mathrm{mw}}$, against the X-ray spectroscopic temperature, $T_{500}^{\mathrm{spec}}$, as derived from our synthetic X-ray spectra (see Section~\ref{subsec:xray}) for all haloes with $T_{500}^{\mathrm{spec}} > 0.2$~keV. We calculate the mass-weighted temperature within the spherical radius $r_{500}$ and derive the spectroscopic temperature within either the same spherical aperture of radius $r_{500}$ (3D) or a projected aperture of radius $r_{500}$ integrated along the full length of the box (2D). The latter measure was used in the scaling relations presented in this paper as this is more akin to observations.

For systems with $T_{500}^{\mathrm{spec}} \gtrsim 0.5$~keV the 2D spectroscopic temperature estimate is $\sim 0.1$~dex lower than the mass-weighted temperature.
This is in agreement with other studies using mock X-ray observations of simulated clusters (e.g. \citealt{Mathiesen2001, Biffi2014}).
Unlike these studies, which find a systematic difference across the full $\sim 1-10$~keV range, at $T_{500}^{\mathrm{spec}} > 2$~keV we do not see a systematic offset between mass-weighted and spectroscopic temperatures. However, we only possess three systems at these temperatures and the $T_{500}^{\mathrm{mw}}-T_{500}^{\mathrm{spec}}$ relation shows some object-to-object scatter.
Part of the offset at $0.5~\mathrm{keV} \lesssim T_{500}^{\mathrm{spec}} \lesssim 2~\mathrm{keV}$ can be attributed to contamination of the cluster emission by cooler gas along the line of sight. This is demonstrated by the 3D spectroscopic temperature estimates, which are on average slightly higher than the 2D temperatures.
In some cases, the cold gas clumps discussed in Appendix~\ref{A:cuts} bias the spectroscopic temperature slightly low compared to the mass-weighted temperature, although the difference is notable in only a handful of systems.
The remaining difference is fairly small and may be explained by complex thermal structure in the ICM (e.g. multiple temperature components) that is inadequately modelled by a single-temperature fit to the integrated spectrum.

At temperatures below $\sim 0.5$~keV the spectroscopic temperature rises sharply compared to the mass-weighted temperature.
This is largely due to the effective area of \textit{Chandra}, which drops rapidly at energies below $\sim 0.5$ keV.
It is thus difficult to reliably determine gas temperatures of systems in this regime from \textit{Chandra} observations.
This is not of particular concern for this study, as there are very few global temperature measurements at $\sim 0.5$~keV with which to compare.

\bsp	
\label{lastpage}
\end{document}